\documentclass[12pt]{article}

\usepackage[dvips]{graphicx}	
\usepackage{amsbsy}		
\usepackage{amsfonts}		
\usepackage{amsmath}		
\usepackage{amssymb}		
\usepackage{mciteplus}
\usepackage{url}
\usepackage{cite}
\usepackage{soul}
\usepackage{afterpage}
\usepackage{color}		
\usepackage[dvipsnames]{xcolor}
\usepackage{tabularx}
\usepackage{booktabs}
\usepackage[normalem]{ulem}
\usepackage[utf8]{inputenc}

\include{paperdef}
\graphicspath{{figs/}}
\let\Re\relax
\let\Im\relax

\DeclareMathOperator{\Re}{Re}
\DeclareMathOperator{\Im}{Im}

\newcommand*{\figref}[1]{Fig.~\ref{#1}}

\newenvironment{Eqnarray}%
         {\arraycolsep 0.14em\begin{eqnarray}}{\end{eqnarray}}
\def\beqa{\begin{Eqnarray}}
\def\eeqa{\end{Eqnarray}}
\def\beq{\begin{equation}}
\def\eeq{\end{equation}}
\def\eq#1{Eq.~(\ref{#1})}
\def\eqst#1#2{Eqs.~(\ref{#1})--(\ref{#2})}
\def\eqs#1#2{Eqs.~(\ref{#1}) and (\ref{#2})}

\def\nn{\nonumber}
\def\ls#1{\ifmath{_{\lower1.5pt\hbox{$\scriptstyle #1$}}}}
\def\sbma{s_{\beta-\alpha}}
\def\cbma{c_{\beta-\alpha}}
\def\sbmaii{s^2_{\beta-\alpha}}
\def\cbmaii{c^2_{\beta-\alpha}}

\def\ifmath#1{\relax\ifmmode #1\else $#1$\fi}
\def\half{\tfrac{1}{2}}
\def\quarter{\tfrac14}
\def\vev#1{\langle#1\rangle}

\begin{document}
\mbox{}\hfill
{\tt IFT-UAM/CSIC-17-074}, 
{\tt SCIPP 17/09}

\def\thefootnote{\fnsymbol{footnote}}

\begin{center}
\Large\bf\boldmath
\vspace*{1cm} 
The Impact of Two-Loop Effects on the Scenario of\\[0.3cm] MSSM Higgs Alignment without Decoupling 
\unboldmath
\end{center}
\vspace*{0.2cm}
\begin{center}
Howard~E.~Haber$^{1}$
, Sven~Heinemeyer$^{2,3,4}$
, Tim~Stefaniak$^{1,}$\footnote{Electronic addresses: 
haber@scipp.ucsc.edu,
Sven.Heinemeyer@cern.ch, 
tistefan@ucsc.edu} \\[0.4cm] 
\vspace{0.2cm}
{\small
 {\sl $^1$ Santa Cruz Institute for Particle Physics (SCIPP) and Department of Physics\\
 University of California, Santa Cruz, 1156 High Street, Santa Cruz, CA 95060, USA} \\[0.1cm]
{\sl$^2$Campus of International Excellence UAM+CSIC, Cantoblanco, E--28049 Madrid, Spain}\\[0.1cm]  
{\sl$^3$Instituto de F\'isica Te\'orica, (UAM/CSIC), Universidad
  Aut\'onoma de Madrid,\\ Cantoblanco, E-28049 Madrid, Spain
}\\[0.1cm]
{\sl$^4$Instituto de F\'isica de Cantabria (CSIC-UC), E-39005 Santander,
  Spain}\\[0.1cm]
  }
\end{center}
\vspace{0.2cm}

\renewcommand{\thefootnote}{\arabic{footnote}}
\setcounter{footnote}{0}

\begin{abstract}
In multi-Higgs models, the properties of one neutral scalar state approximate those of the Standard Model (SM) Higgs boson in a limit where the corresponding scalar field is roughly aligned in field space with the scalar doublet vacuum expectation value.  
In a scenario of alignment without decoupling, a SM-like Higgs boson can be accompanied by additional scalar states
whose masses are of a similar order of magnitude.  In the Minimal Supersymmetric Standard Model (MSSM), alignment without decoupling can be achieved due to an accidental cancellation of tree-level and radiative loop-level effects.
In this paper we assess the impact of the leading two-loop $\order{\alpha_s h_t^2}$ corrections on the Higgs alignment condition in the MSSM. These corrections are sizable and important in the relevant regions of parameter space and furthermore give rise to solutions of the alignment condition that are not present in the approximate one-loop description. We provide a comprehensive numerical comparison of the alignment condition obtained in the approximate one-loop and two-loop approximations, and discuss its implications for phenomenologically viable regions of the MSSM parameter space.

\end{abstract}
\newpage



\section{Introduction}
\label{sec:Introduction}

Since the initial discovery of a new scalar particle with mass of about $125\gev$~\cite{ATLASDiscovery,CMSDiscovery}, detailed studies of the data from Run 1 and 2 of the Large Hadron Collider (LHC) at CERN are beginning to establish the phenomenological profile of what appears to be the Higgs boson associated with electroweak symmetry breaking. Indeed, the measurements of Higgs production cross sections times decay branching ratios into a variety of final states appear to be consistent with the Higgs boson predicted by the Standard Model (SM)~\cite{Aad:2015zhl}. One can now say with some confidence that a ``SM-like'' Higgs boson has been discovered. Nevertheless, the limited precision of the current Higgs data from the LHC still allows for deviations from SM expectations.  If such deviations were to be confirmed,  new physics beyond the SM would be required.

Deviations from the SM Higgs behavior can be accommodated by introducing additional Higgs scalars to the electroweak model. Typically, the SM Higgs sector is extended by adding additional electroweak scalar doublets and/or singlets in
order to avoid deviations of the approximate relation between the $W$ and $Z$ boson mass, $M_W\simeq M_Z\cos\theta_W$, where $\theta_W$ is the weak mixing angle.
However, the existence of a SM-like Higgs boson already imposes significant constraints on any extended Higgs sector.  
We can always define a neutral Higgs field that points in the direction of the scalar doublet vacuum expectation value (vev)  in field space. The tree-level couplings of such a scalar field to the SM gauge bosons and fermions are precisely those of the SM Higgs boson.  However in general, this \textit{aligned} scalar field is not a mass eigenstate field, since it will mix with other neutral scalar fields of the extended Higgs sector.   Thus, the current Higgs data is consistent with an extended Higgs sector only if the observed scalar particle with mass $125\gev$ is approximately aligned in field space with the doublet vev.  This so-called \textit{alignment limit}~\cite{Gunion:2002zf,Craig:2013hca,Haber:2013mia,Asner:2013psa,Carena:2013ooa} is either the result of some symmetry of the scalar sector~\cite{Dev:2014yca,Pilaftsis:2016erj}, or it is the result of some special choice of the scalar sector parameters.  

An example of the latter is the decoupling regime of the extended Higgs sector~\cite{Haber:1989xc,Gunion:2002zf}.  The scalar potential typically contains a number of mass parameters.    One of those mass parameters is fixed by the doublet scalar vev, which must be set to $v=246$~GeV to explain the observed value of the Fermi constant $G_F$.  If other scalar sector mass parameters are characterized by a scale $M$ that is significantly larger than $v$, then one of the neutral scalar mass eigenstates will be of $\mathcal{O}(v)$, whereas all other scalar masses will be of order $M\gg v$.   One can then integrate out the heavy scalar states below the mass scale $M$.  The resulting effective scalar theory will be that of the SM with a single Higgs doublet, which will yield one neutral Higgs boson state whose couplings are approximately those of the SM Higgs boson.  
Of course, in such a scenario, additional scalar states would be quite heavy and may be difficult to discover at the LHC.  

One can also achieve alignment independently of the masses of the non-SM like Higgs bosons.  Generically, the aligned scalar field (which possesses the couplings of the SM Higgs boson) is not a mass eigenstate.  However,  
if the parameters of the scalar sector (either accidentally or due to a
symmetry) yield suppressed mixing between the aligned scalar field and the
other neutral scalar interaction eigenstates, then approximate alignment is
realized.   In any multi-Higgs doublet model, an \textit{exact alignment
  condition} can be specified, in which the aligned scalar field is a mass
eigenstate (and thus its mixing with all other scalar eigenstate fields
vanishes).  Hence, if this alignment condition is approximately fulfilled, it
is possible to have a SM-like Higgs boson along with additional scalar states
with masses that are not significantly larger  than the electroweak scale and
thus more amenable to discovery in future LHC runs.  We denote this latter
scenario as \textit{alignment without decoupling}~\cite{Gunion:2002zf,Craig:2013hca,Haber:2013mia,Asner:2013psa,Carena:2013ooa,Carena:2014nza,Bernon:2015qea,Bernon:2015wef}.

Extended Higgs sectors in isolation suffer from the same problem as the SM
Higgs sector, namely there is no natural explanation for the origin of the
electroweak  scale.  There have been numerous attempts in the literature to
devise models of new physics beyond the SM (BSM) that can provide a natural
explanation of the electroweak scale, either via new dynamics or a new
symmetry.  All such approaches invoke new fundamental degrees of freedom, and
many models of BSM physics incorporate enlarged scalar sectors.    One of the
most well studied models of this type is the minimal supersymmetric extension
of the SM
(MSSM)~\cite{Nilles:1983ge,Barbieri:1987xf,Haber:1984rc,Gunion:1984yn}, which
requires a second Higgs doublet in order to avoid 
anomalies associated with the supersymmetric fermionic partners of the SM
Higgs doublet.  In light of the fact that no supersymmetric particles have yet
been discovered, it follows that the scale of supersymmetry (SUSY)-breaking,
$M_{\rm S}$, must lie somewhat above the electroweak scale.  This already leads
to some tension with the requirements of a natural explanation of the
electroweak scale (sometimes called the little hierarchy
problem~\cite{Giusti:1998gz,Cheng:2003ju,Harnik:2003rs,Cheng:2004yc}). Nevertheless,
if supersymmetric particles are ultimately discovered at the LHC, it would
provide a significant amelioration of the large hierarchy problem associated
with the fact that the electroweak scale is 17 orders of magnitude smaller
than the Planck scale. 

Numerous searches for supersymmetric particles at the LHC (as well as at
previous lower energy colliders such as LEP and Tevatron) provide important
constraints on the allowed MSSM parameter space~\cite{ATLAS-SUSY,CMS-SUSY},
with additional constraints 
from considerations of virtual supersymmetric particle contributions to SM
processes (see, e.g., \citere{Heinemeyer:2004gx} for a review).  Finally, due to the
enlarged Higgs sector of the MSSM, the 
properties of the observed Higgs boson and the absence of evidence for
additional Higgs scalars yield additional constraints.   In particular, given
that the observed Higgs boson appears to be SM-like, it follows that the Higgs
sector of the MSSM must be close to the alignment limit.   In the MSSM, the
scale of the non-SM-like Higgs boson is governed by a SUSY-breaking mass
parameter.   Although this mass parameter is logically distinct from the mass
parameter $M_{\rm S}$ that governs the mass scale of the heavy supersymmetric
particles, one might expect these two parameters to be of a similar order of
magnitude.  If this is the case, then the approximate alignment limit of the
MSSM Higgs sector is a result of the decoupling of heavy Higgs states.  On the other hand, one
may wonder whether the approximate alignment limit of the MSSM Higgs sector
can be achieved outside of the decoupling limit, in which case one might
expect the possibility that additional non-SM like Higgs scalars could soon be
discovered in future LHC running.

The possibility of alignment without decoupling has been analyzed in detail in
Refs.~\cite{Gunion:2002zf,Craig:2013hca,Haber:2013mia,Asner:2013psa,Carena:2013ooa,Carena:2014nza,Bernon:2015qea,Bernon:2015wef}.\footnote{It is noteworthy that a number of benchmark scenarios (e.g. the ``$\tau$-phobic'' and low-MH scenarios) proposed for different reasons in \citere{Carena:2013qia} also provide parameter regimes in which approximate alignment without decoupling is achieved.}
More recently, the connection of Higgs alignment without decoupling in the MSSM with dark matter has been investigated in
Ref.~\cite{Profumo:2016zxo}. 
In Ref.~\cite{Bechtle:2016kui},  a parameter scan of the phenomenological MSSM (pMSSM) with eight parameters was performed, taking into account the experimental Higgs boson results from Run I of the LHC and further low-energy observables.  One of the central questions considered in Ref.~\cite{Bechtle:2016kui} was whether parameter regimes with approximate Higgs alignment without decoupling are still allowed in light of the current LHC data.  Two separate cases were considered in which either the lighter or the heavier of the two CP-even neutral Higgs bosons of the MSSM is identified with the observed Higgs boson of mass 125 GeV.   In the first case, we identified allowed regions of the MSSM parameter space in which the non-SM-like Higgs bosons could be as light as 200 GeV.  In the second case,  we demonstrated that the heavy CP-even Higgs boson is still a viable candidate to explain the Higgs signal --- albeit only in a highly constrained parameter region.   Both cases correspond to parameter regimes of approximate alignment without decoupling.

In the MSSM, alignment without decoupling arises due to an approximate accidental cancellation between tree-level and loop-level effects.   Given the current precision of the Higgs data, we concluded in Ref.~\cite{Bechtle:2016kui} that this region of approximate cancellation, while accidental in nature, does not require an extreme fine-tuning of the MSSM parameters.  Indeed, such regions must appear in any comprehensive scan of the MSSM parameter space.
In Ref.~\cite{Bechtle:2016kui}, we showed that the result of our numerical scans could be understood using simple analytical expressions in which the leading one-loop and two-loop radiative corrections to the MSSM Higgs sector are included.  In this paper, we provide a detailed treatment of this analytic approximation and demonstrate the importance of the leading two-loop radiative effects in determining the allowed parameter regions for approximate alignment without decoupling.  

The remainder of this paper is structured as follows. In Section~\ref{Sec:MSSMHiggs}, we review the alignment limit at tree-level in the context of the general CP-conserving two Higgs doublet model 2HDM).   Both the decoupling limit and the limit of alignment without decoupling are discussed.   We can apply these results to the MSSM by treating the MSSM Higgs sector as an  effective non-supersymmetric 2HDM at tree-level, obtained by integrating out heavier supersymmetric particles.  The effects of the SUSY-breaking lead to corrections that are logarithmic in the supersymmetry breaking scale, $M_S$, as well as finite threshold corrections that can be of $\mathcal{O}(1)$.   The leading one-loop corrections to the exact alignment condition are treated in Section~\ref{Sec:alignment1L}.
However, it is known that the two-loop corrections to the MSSM Higgs sector can be phenomenologically relevant.   Employing a procedure first introduced in Ref.~\cite{Haber:1996fp} and later extended in Ref.~\cite{Carena:2000dp}, the one-loop results of Section~\ref{Sec:alignment1L} are modified to obtain the leading two-loop corrections to the exact alignment condition in
Section~\ref{Sec:alignment2L}. In Section~\ref{onevstwo}, a numerical comparison of the impact of the corresponding leading one-loop and two-loop corrections is given.  In addition, we discuss the $M_S$ values required to achieve a SM-like Higgs boson mass of $125\gev$, and give a criterion on the CP-odd Higgs mass, $M_A$, that determines whether the lighter or the heavier CP-even Higgs boson is aligned with the SM Higgs vev. In the latter scenario in which the heavier of the two CP-even Higgs bosons is identified with the observed Higgs scalar at 125 GeV, a new decay mode $H\to hh$ is possible if $m_H>2m_h$. We discuss the magnitude of the relevant triple Higgs coupling and the resulting branching fraction for this decay in Section~\ref{sec:Hhh}.  Finally, we present our conclusions and outlook in Section~\ref{conclusions}.

\section{The alignment limit in the two Higgs doublet model}
\label{Sec:MSSMHiggs}

In light of the LHC Higgs data, which strongly suggests that the properties of the
observed Higgs boson are SM-like~\cite{Aad:2015zhl}, we seek to explore the region of the MSSM parameter space that yields a SM-like Higgs boson.   
Since the Higgs sector of the MSSM is a constrained $\mathcal{CP}$-conserving 2HDM, we first review the limit of the 2HDM that yields a SM-like Higgs boson. In a multi Higgs doublet model, a SM-like Higgs boson arises in the \textit{alignment limit}, in which one of the neutral Higgs mass eigenstates is approximately aligned with the direction of the Higgs vacuum expectation value (vev) in field space.

The 2HDM contains two hypercharge-one weak SU(2)$_L$ doublet scalar fields, $\Phi_1$ and $\Phi_2$.
 By an appropriate rephasing of these two fields, one can choose their vevs, $\vev{\Phi_1^0}\equiv v_1/\sqrt{2}$ and $\vev{\Phi_2^0}\equiv v_2/\sqrt{2}$, to be real and non-negative.  In this convention,
$\tan\beta\equiv v_2/v_1$, with $0\leq\beta\leq\half\pi$.  Note that $v\equiv (v_1^2+v_2^2)^{1/2}=(2G_F^2)^{-1/4}\simeq 246$~GeV is fixed by the value of the Fermi constant,~$G_F$.

It is convenient to introduce the following linear combinations of Higgs doublet fields,
\begin{align}
\cHe = \begin{pmatrix} H_1^+\\H_1^0 \end{pmatrix} \equiv \frac{v_1 \Phi_1 + v_2\Phi_2}{v}, \qquad \cHz = \begin{pmatrix} H_2^+\\H_2^0 \end{pmatrix} \equiv \frac{-v_2 \Phi_1 + v_1\Phi_2}{v},
\end{align}
such that $\langle H_1^0 \rangle = v/\wz$ and $\langle H_2^0\rangle = 0$, which defines the \textit{Higgs basis}~\cite{Georgi:1978ri,Branco:1999fs,Davidson:2005cw}.
The most general 2HDM scalar potential, expressed in terms of the Higgs basis fields $\mathcal{H}_1$ and $\mathcal{H}_2$, is given by
\beqa \mathcal{V}&=& Y_1  \cHe^\dagger\cHe + Y_2 \cHz^\dagger\cHz +[Y_3
\cHe^\dagger\cHz +{\rm h.c.}
+\half Z_1(\cHe^\dagger\cHe)^2+\half Z_2(\cHz^\dagger\cHz)^2
+Z_3(\cHe^\dagger\cHe)(\cHz^\dagger\cHz)\nn\\
&&\qquad
+Z_4(\cHe^\dagger\cHz)(\cHz^\dagger\cHe)
+\left\{\half Z_5 (\cHe^\dagger\cHz)^2 +\big[Z_6 (\cHe^\dagger\cHe)
+Z_7 (\cHz^\dagger\cHz)\big] \cHe^\dagger\cHz+{\rm
h.c.}\right\}.
\eeqa
The Higgs basis is uniquely defined up to a rephasing of the Higgs basis field $\mathcal{H}_2$.  If the tree-level Higgs scalar potential and vacuum is $\mathcal{CP}$-conserving, then it is possible to rephase the Higgs basis field $\mathcal{H}_2$ such that all the scalar potential parameters are real.  

The scalar potential minimum conditions determine the values of $Y_1$ and $Y_3$, 
\beq \label{potmin}
Y_1=-\half Z_1 v^2\,,\qquad\quad Y_3=-\half Z_6 v^2\,.
\eeq
The tree-level squared masses of the charged Higgs boson and the CP-odd neutral Higgs boson are given by
\beqa
M^2_{H^\pm}&=&Y_2+\half v^2 Z_3\,,\label{mhpm}\\
M^2_{A}&=&Y_2+\half v^2(Z_3+Z_4-Z_5)\,.\label{ma}
\eeqa
In particular, the squared mass parameter $Y_2$ can be eliminated in favor of $M_A^2$.

One can then evaluate the squared-mass matrix of the neutral $\cp$-even Higgs bosons, with respect to the neutral Higgs basis states, $\{\sqrt{2}\,{\rm Re}~H^0_1-v$\,,\,$\sqrt{2}\,{\rm Re}~H^0_2\}$.  After employing \eqs{potmin}{ma}, the $\mathcal{CP}$-even neutral Higgs squared-mass matrix takes the following simple form, 
\begin{align}
  \cM^2 = \ML Z_1 v^2 &\quad Z_6 v^2\,, \\
        Z_6 v^2 & \quad \MA^2 + Z_5 v^2 \MR\,.
\label{HiBa-massmatrix}
\end{align}
If $\sqrt{2}\,{\rm Re}~H^0_1-v$ were a Higgs mass eigenstate, then its tree-level couplings to SM particles would be precisely those of the SM Higgs boson.  This would correspond to the exact alignment limit.  To achieve a SM-like Higgs boson,
it is sufficient for one of the neutral Higgs mass eigenstates to be approximately given by $\sqrt{2}\,{\rm Re}~H^0_1-v$, with a corresponding squared-mass $\simeq Z_1 v^2$.  The observed Higgs mass implies that $Z_1\simeq 0.26$.

The $\mathcal{CP}$-even neutral Higgs squared-mass matrix given by \eq{HiBa-massmatrix} is controlled by two independent mass scales, $v\simeq 246$~GeV and $Y_2$, where the latter enters via the parameter $M_A^2$ [cf.~\eq{ma}].   In addition, the scalar potential parameters $Z_1$, $Z_5$ and $Z_6$ are typically of $\mathcal{O}(1)$ or less (in the MSSM, they are of order the square of a gauge coupling).
Consequently, a SM-like neutral Higgs boson can arise in two different ways: 
\begin{enumerate}
\item
$\MA^2\gg (Z_1-Z_5)v^2$.  This corresponds to the so-called \textit{decoupling limit}, where $h$ is SM-like and $\MA\sim\MH\sim M_{H^\pm}\gg\Mh$.
\item
$|Z_6|\ll 1$.  In this case $h$ is SM-like if $\MA^2+(Z_5-Z_1)v^2>0$ and $H$ is SM-like if $\MA^2+(Z_5-Z_1)v^2<0$.
\end{enumerate}
In particular, one can achieve alignment \textit{without} decoupling if $|Z_6|\ll 1$, independently of the value of the non-SM-like Higgs states $H$, $A$ and $H^\pm$.  Indeed, if the heavier of the two neutral $\mathcal{CP}$-even Higgs states is SM-like, then one must have $|Z_6|\ll 1$ in a non-decoupling parameter regime.

After diagonalizing the $\mathcal{CP}$-even neutral Higgs squared-mass matrix, one obtains the $\mathcal{CP}$-even Higgs mass eigenstates $h$ and $H$ (where $m_h<m_H$),
\beq \label{mixing}
\begin{pmatrix} H\\ h\end{pmatrix}=\begin{pmatrix} \cba & \,\,\, -\sba \\
\sba & \,\,\,\phantom{-}\cba\end{pmatrix}\,\begin{pmatrix} \sqrt{2}\,\,{\rm Re}~H_1^0-v \\ 
\sqrt{2}\,{\rm Re}~H_2^0
\end{pmatrix}\,,
\eeq
where $\cba\equiv\cos(\be - \al)$ and $\sba\equiv\sin(\be - \al)$ are defined in terms of the mixing angle $\alpha$ that diagonalizes the CP-even Higgs squared-mass matrix when expressed in the original basis of scalar fields, $\{\sqrt{2}\,{\rm Re}~\Phi_1^0-v_1\,,\,\sqrt{2}\,{\rm Re}~\Phi_2^0-v_2\}$.
Since the SM-like Higgs field must be approximately $\sqrt{2}\,\Re~H_1^0-v$, it follows that
$h$ is SM-like if $|\cba|\ll 1$ and $H$ is SM-like if $|\sba|\ll 1$.

We can now apply the above results to the MSSM Higgs sector.    In the usual treatment of the MSSM, one introduces 
two Higgs doublets, $H_U$ and $H_D$ of hypercharge $Y=+1$ and $Y=-1$, respectively.\footnote{The notation derives from the fact that the MSSM superpotential is a holomorphic gauge-invariant function of the corresponding superfields $\hat{H}_U$ and $\hat{H}_D$.   As a consequence, $\hat{H}_U$ couples exclusively to the up-type SU(2)$_{\rm L}$ singlet quark superfield $\hat{U}$ and
$\hat{H}_D$ couples exclusively to the down-type SU(2)$_{\rm L}$ singlet quark superfield $\hat{D}$.} 
To make contact with the notation of the 2HDM presented above, we can relate these fields to the hypercharge $Y=+1$ scalar fields,
\beq
(\Phi_1)^i=\epsilon_{ij}(H_D^*)^j\,,\qquad\quad (\Phi_2)^i=(H_U)^i\,,
\eeq
where $\epsilon_{12}=-\epsilon_{21}=1$ and $\epsilon_{11}=\epsilon_{22}=0$,
and there is an implicit sum over the repeated SU(2)$_L$ index $j=1,2$.  
The tree-level quartic couplings $Z_i$ can be expressed in terms of the electroweak SU(2)$_L$ and U(1)$_{\rm Y}$ gauge couplings $g$ and $g'$, respectively,
\beqa
Z_1&=&Z_2=\quarter(g^2+g^{\prime\,2}) c_{2\beta}^2\,,\qquad Z_5=\quarter(g^2+g^{\prime\,2})s_{2\beta}^2\,,\qquad
Z_7=-Z_6=\quarter(g^2+g^{\prime\,2}) s_{2\beta}c_{2\beta}\,,\nn \\
Z_3&=&Z_5+\quarter(g^2-g^{\prime\,2})\,,\qquad\quad\,\,
Z_4=Z_5-\half g^2\,,\label{zsusy} 
\eeqa
where $c_{2\beta}\equiv\cos 2\beta$ and $s_{2\beta}\equiv \sin 2\beta$.   We have already noted that the squared-mass of the SM-like Higgs boson 
is approximately given by $Z_1 v^2$, which is equal to $M_Z^2 c_{2\beta}$ at tree-level in the MSSM, and thus incompatible with the observed Higgs mass of 125~GeV.  Moreover, if the existence of a SM-like Higgs boson is due to alignment without decoupling, then the relation $Z_6=0$ must be approximately fulfilled, which implies that $\sin 4\beta=0$ (i.e.~$\beta=0,\tfrac14\pi$ or $\half\pi$).   Of course, the extreme values of $\beta=0$ or $\beta=\half\pi$ are not phenomenologically realistic, whereas $\beta=\tfrac14\pi$ would yield a massless $\textit{CP}$-even Higgs boson at tree-level.

In order to achieve a realistic MSSM Higgs sector, radiative corrections must
be incorporated~\cite{Haber:1990aw,Okada:1990vk,Ellis:1990nz} (see, e.g.,
\citeres{Heinemeyer:2004gx,Djouadi:2005gj,Heinemeyer:2004ms,Draper:2016pys} for reviews).
It is well-known that the observed Higgs mass of 125~GeV is
compatible with a radiatively-corrected Higgs sector in certain regions of the
MSSM parameter space~\cite{Heinemeyer:2011aa}.   Moreover, a SM-like Higgs
state is easily achieved  
in the decoupling limit where $M_A^2\gg v^2$, where $h$ is identified as the
observed Higgs boson.  In this paper, we focus on the alternative scenario in
which a SM-like Higgs boson is a consequence of approximate alignment without
decoupling.   When loop corrections are taken into account, the possibility of
alignment without decoupling must be reconsidered.

\section{Alignment without decoupling at the one-loop level}
\label{Sec:alignment1L}

In the MSSM, exact alignment via $Z_6 = 0$ can only happen through an
accidental cancellation of the tree-level terms with contributions arising
at the one-loop level (or higher). In this case the Higgs alignment is
independent of the values of $M_A^2$, $Z_1$ and $Z_5$. 
The leading one-loop contributions to $Z_1$, $Z_5$ and $Z_6$ proportional to $h_t^2 m_t^2$, where $m_t$ is the top quark mass and
\beq \label{ht}
h_t=\frac{\sqrt{2}m_t}{v s_\beta}
\eeq
is the top quark Yukawa coupling, have been obtained in Ref.~\cite{Carena:2014nza} in the limit $M_Z, M_A \ll M_S$ (using results from Ref.~\cite{Haber:1993an}):
\begin{align}
Z_1v^2 &= M_Z^2 c_{2\beta}^2 + \frac{3m_t^4}{2\pi^2v^2} \left[\ln\left(\frac{M_S^2}{m_t^2}\right) +\frac{X_t^2}{M_S^2} \left(1 - \frac{X_t^2}{12M_S^2}\right) \right],\label{zone}\\
Z_5v^2 &=s_{2\beta}^2 \left\{ M_Z^2  + \frac{3 m_t^4}{8\pi^2v^2 s_\beta^4} \left[\ln\left(\frac{M_S^2}{m_t^2}\right) +\frac{X_tY_t}{M_S^2} \left(1 - \frac{X_tY_t}{12M_S^2}\right) \right]\right\},\label{Eq:Z5v2}\\
Z_6v^2 &= -s_{2\beta} \left\{M_Z^2c_{2\beta}  - \frac{3 m_t^4 }{4\pi^2v^2s_\beta^2} \left[\ln\left(\frac{M_S^2}{m_t^2}\right) +\frac{X_t(X_t+Y_t)}{2 M_S^2} - \frac{X_t^3Y_t}{12M_S^4} \right]\right\},
\label{Eq:Z6v2}
\end{align}
where $s_\beta\equiv \sin\beta$, $s_{2\beta}\equiv\sin 2\beta$, $c_{2\beta}\equiv\cos 2\beta$, 
$M_S\equiv \sqrt{m_{\tilde{t}_1}m_{\tilde{t}_2}}$ denotes the SUSY-breaking mass scale that governs the top-squark (stop) sector, given by the geometric mean of the light and heavy stop masses, 
and\footnote{The elements of the top-squark squared-mass matrix are governed by the supersymmetric higgsino mass parameter $\mu$ and the soft-SUSY-breaking trilinear $H^0_U \widetilde t_L\widetilde t^*_R$ coupling $A_t$.  For simplicity, we ignore potential CP-violating effects by taking $\mu$ and $A_t$ to be real parameters in this work.}
\begin{align} 
X_t \equiv A_t - \mu/\tan\beta,\qquad\qquad Y_t \equiv A_t + \mu\tan\beta. \label{XY}
\end{align}

The approximate expression for $Z_6 v^2$ given in \refeq{Eq:Z6v2} depends only on the unknown parameters $\mu$, $A_t$, $\tan\beta$ and $M_S$.    Exact alignment arises if $Z_6=0$.
Note that $Z_6=0$ is trivially satisfied if $\beta=0$ or $\tfrac12 \pi$ (corresponding to the vanishing of either $v_1$ or $v_2$).  However, this choice of parameters is not relevant for phenomenology as it leads to a massless $b$ quark or $t$ quark, respectively, at tree-level.\footnote{A potential loophole to this last remark arises in models, dubbed ``uplifted supersymmetry'', in which down-type fermion masses are absent at tree-level but are generated radiatively by loop-induced couplings to the up-type Higgs doublet, $H_U$.  Further details are described in Ref.~\cite{Dobrescu:2010mk}.}   
Henceforth, we assume that $\tan\beta$ is finite and non-zero; by convention, we take $\tan\beta$ to be positive.  
Regarding the other parameters, $\mu$, $A_t$ and $M_S$, we generously allow for rather large parameter values in this work. However, one should keep in mind that parameter points with $|\mu/M_S|$ and $|A_t/M_S|$ larger than about 3 are often severely restricted by vacuum (meta-)stability requirements, in particular the absence of a color and/or electric charge-breaking global minimum of the full MSSM scalar potential~\cite{Frere:1983ag,Claudson:1983et,Kounnas:1983td,Gunion:1987qv,Casas:1995pd,Langacker:1994bc,Strumia:1996pr} (for recent analyses, see also Refs.~\cite{Chowdhury:2013dka,Bagnaschi:2015pwa,Hollik:2016dcm}.) Furthermore, for values of $|X_t/M_S|\gsim 3$, the theoretically predicted loop-corrected Higgs squared-mass decreases rapidly from its maximal value (which at one-loop is achieved at $X_t/M_S\simeq \sqrt{6}$), and is ultimately driven to negative values.  
In our numerical analysis, we consider only $|A_t/M_S|$ values up to about $3$, and highlight regions of the parameter space that exhibit $|X_t/M_S|\ge 3$ where our analysis is untrustworthy. 

We simplify the analysis by solving \eq{zone} for $\ln(M_S^2/m_t^2)$ and inserting the result back into \eq{Eq:Z6v2}.
The resulting expression for $Z_6$ now depends on $Z_1$, $\tan\beta$, and the dimensionless
ratios
\beq
\widehat{A}_t\equiv \frac{A_t}{M_S}\,,\qquad \quad \widehat{\mu}\equiv\frac{\mu}{M_S}\,.
\eeq
Using \eq{XY} to rewrite the final expression in terms of $\widehat{A}_t$ and $\widehat{\mu}$, we obtain
\beq \label{z6zero}
Z_6 v^2=-\cot\beta\biggl\{m_Z^2 c_{2\beta}-Z_1 v^2 +\frac{3m_t^4\widehat{\mu}(\widehat{A}_t\tan\beta-\widehat{\mu})}{4\pi^2 v^2 s_\beta^2}
\bigl[\tfrac{1}{6}(\widehat{A}_t-\widehat{\mu}\cot\beta)^2-1\bigr]\biggr\}\,.
\eeq
Setting $Z_6 = 0$, we can identify $Z_1 v^2$ with the mass of the observed (SM-like) Higgs boson (which may be either $h$ or $H$ depending on whether $\sba$ is close to 1 or 0, respectively).   
We can then numerically solve for $\tan\beta$ for given values of  $\widehat{A}_t$ and $\widehat{\mu}$. 
Indeed, $t_\beta\equiv \tan\beta$ is the solution to a seventh order polynomial equation,
\beq \label{poly}
M_Z^2 t_\beta^4(1-t_\beta^2)-Z_1 v^2 t_\beta^4(1+t_\beta^2)+\frac{3m_t^4\widehat{\mu}(\widehat{A}_t t_\beta
-\widehat{\mu})(1+t_\beta^2)^2}{4\pi^2 v^2}\bigl[\tfrac16(\widehat{A}_t t_\beta-\widehat{\mu})^2-t_\beta^2\bigr]=0\,.
\eeq
A seventh order polynomial has either one, three, five or seven real roots.  In light of the comments below \eq{XY}, we are only interested in real positive solutions of \eq{poly}; i.e., we exclude the possibility of $t_\beta=0$. Moreover, we can interpret the negative $t_\beta$ solutions at the point $(\widehat\mu\,,\,\widehat{A}_t)$ as corresponding to positive $t_\beta$ solutions at the point $(-\widehat\mu\,,\,\widehat{A}_t)$.\footnote{In light of the interactions of the Higgs bosons with quarks and with squarks, if we were to adopt an alternative convention in which both signs of $t_\beta$ were allowed, then under $t_\beta\to -t_\beta$ one must also transform $\mu\to -\mu$ and $h_t\to -h_t$ (note that the signs of $X_t$ and $Y_t$ are unaffected).   In this alternative convention, the points $(\widehat{\mu}\,,\,\widehat{A}_t\,,\,t_\beta)$ and $(-\widehat{\mu}\,,\,\widehat{A}_t\,,\,-t_\beta)$ would be physically equivalent.}      Finally, the solution to \eq{poly} is invariant under the simultaneous inversion of $\widehat{\mu}\to -\widehat{\mu}$ and $\widehat{A}_t\to-\widehat{A}_t$ (keeping the sign of $\tan\beta$ fixed).  The latter is a consequence of the symmetry properties of the approximate one-loop expressions for $Z_1$, $Z_5$ and $Z_6$.  It therefore follows that a negative $t_\beta$ solution at the point $(\widehat\mu\,,\,\widehat{A}_t)$ corresponds to a positive $t_\beta$ solution at the point $(\widehat\mu\,,\,-\widehat{A}_t)$. 

\begin{figure}[t!]
\centering
\includegraphics[width=0.48\textwidth]{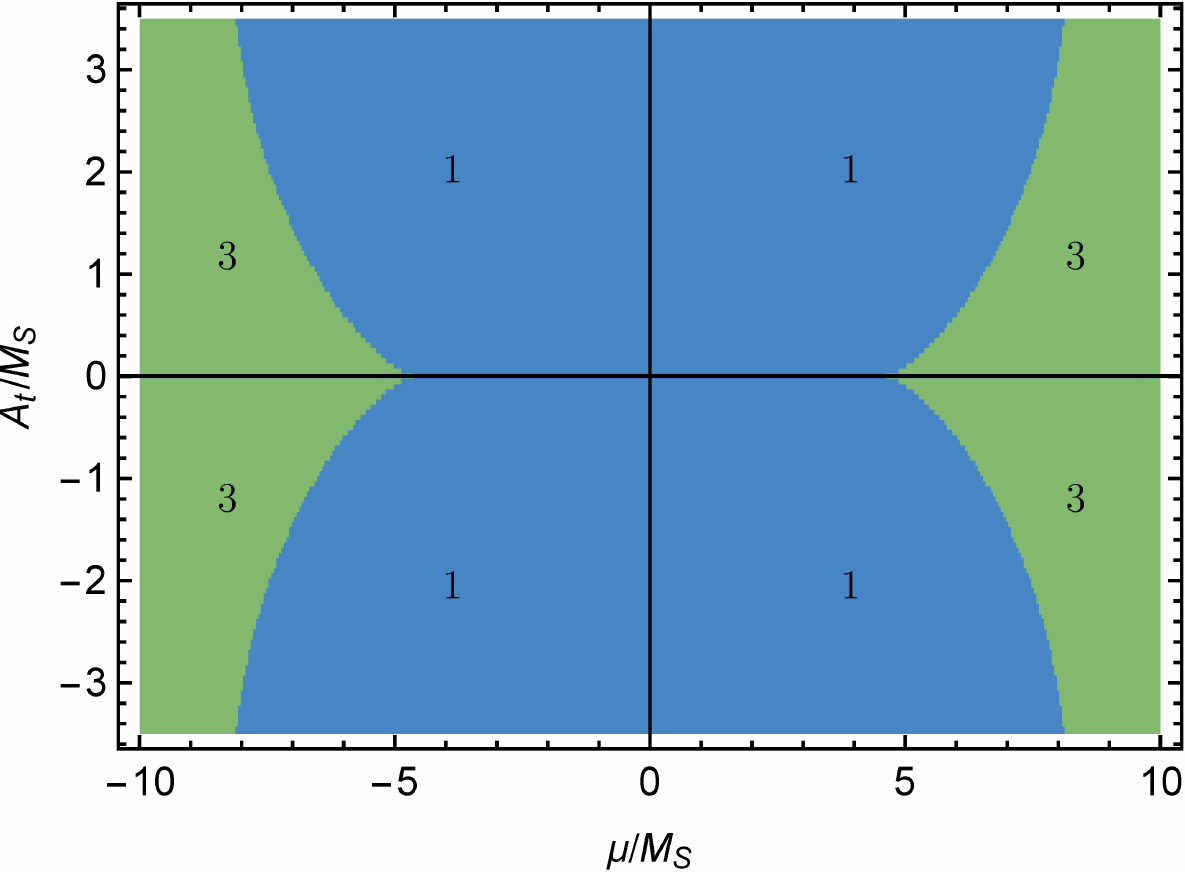}\hfill
\includegraphics[width=0.48\textwidth]{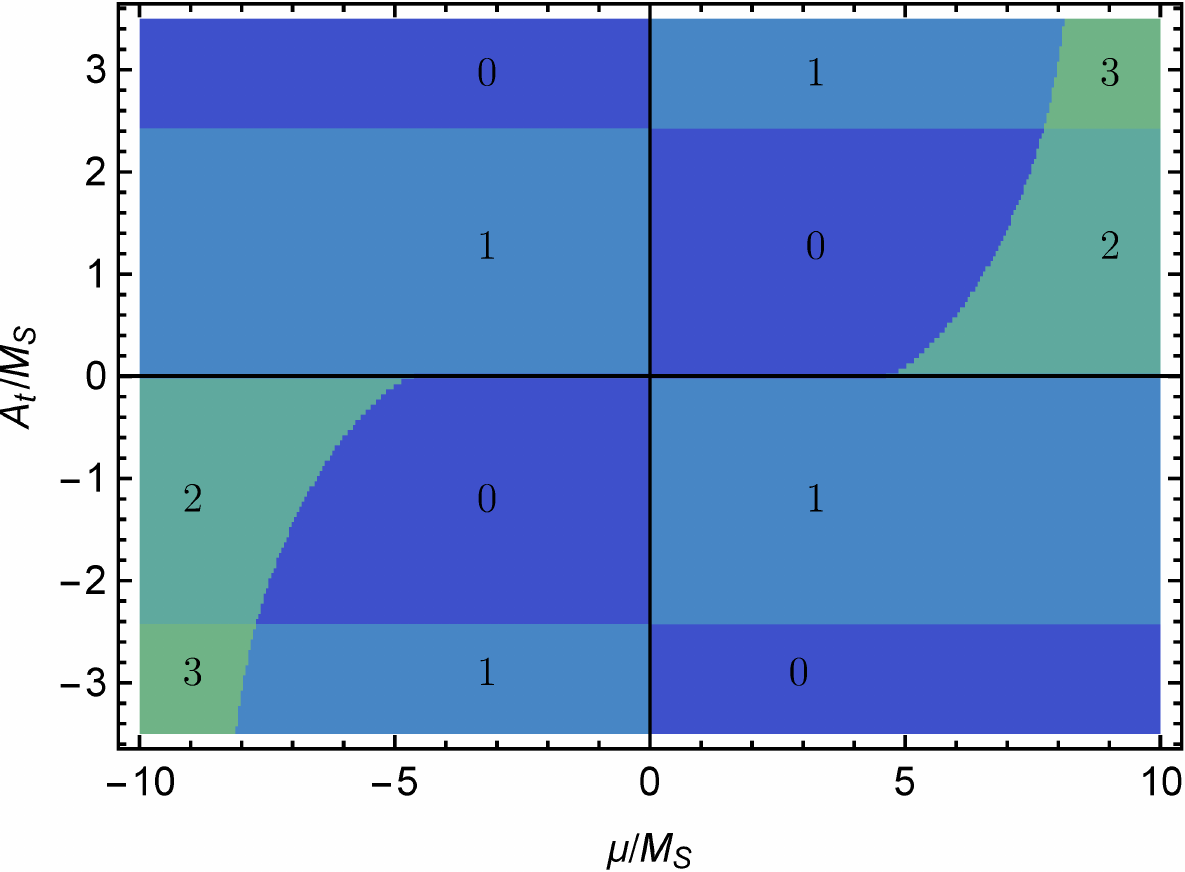}
\caption{Number of real solutions to the one-loop alignment condition, \refeq{poly}. We set $Z_1v^2 = 125\gev$. \emph{Left:} All solutions with real $\tan\beta$; \emph{Right:} Real, positive $\tan\beta$ solutions.}
\label{Fig:Nsol1L}
\end{figure}

In the left [right] panel of Fig.~\ref{Fig:Nsol1L} we show the number of real [positive] solutions to the above polynomial, \refeq{poly}, in the $(\widehat\mu\,,\,\widehat{A}_t)$ plane. We observe that  there is one real root of \eq{poly} for $|\widehat{\mu}|\lsim 5$ to $8$ (depending on the value of $\widehat{A}_t \ne 0$).  For larger values of $|\widehat{\mu}|$ in the $(\widehat\mu\,,\,\widehat{A}_t)$ plane, there are three real roots.
The transition between these two regions occurs when two of the three roots coalesce (yielding a degenerate real root) and then move off the real axis to form a complex conjugate pair. In the quadrants with $\widehat\mu \widehat{A}_t > 0 $ ($\widehat\mu \widehat{A}_t < 0 $) with large $|\widehat{\mu}|$, these two real roots are always positive (negative), whereas the sign of the third root, which also exists at smaller $|\widehat{\mu}|$, depends on the value of $\widehat{A}_t$: if $|\widehat{A}_t| \ge \sqrt{6}$, this root is positive (negative) in the quadrant with $\widehat\mu \widehat{A}_t > 0 $ ($\widehat\mu \widehat{A}_t < 0 $), whereas if $|\widehat{A}_t| < \sqrt{6}$, it is of the opposite sign.

To see how the roots evolve in a continuous manner in the ($\widehat{\mu}$, $\widehat{A}_t$) parameter plane,
consider a path in the left panel of \figref{Fig:Nsol1L} that begins at $\widehat{A}_t\sim 3$ and $\widehat{\mu}\sim -9$.  For these values, \eq{poly} possesses three negative roots and no positive roots.  Keeping $\widehat{\mu}$ fixed and reducing $\widehat{A}_t$, one of the negative roots decreases without bound until it reaches $-\infty$ at $\widehat{A}_t\sim\sqrt{6}$.  Taking $\widehat{A}_t$ below $\sqrt{6}$, the root switches over to $+\infty$, and then steadily decreases.
When $\widehat{A}_t$ crosses from positive to negative values, all positive and negative roots interchange.  Consequently, when we enter the quadrant where  $\widehat{A}_t$ is negative, we now have two positive roots and one negative root.  
This happens because one of the negative roots goes to $-\infty$ as $\widehat{A}_t$ approaches zero from above, and then switches over to $+\infty$ after crossing $\widehat{A}_t=0$. Finally, the remaining negative root goes to $-\infty$ as $\widehat{A}_t$ approaches $-\sqrt{6}$ from above, and then switches over to $+\infty$.  For values of $\widehat{A}_t < -\sqrt{6}$,  there is now a third positive root.   For smaller values of $|\widetilde{\mu}|$, \eq{poly} possesses only one real root, since the two other roots that were real at larger values of $|\widetilde{\mu}|$ are now complex, as noted above.   Finally, the two panels of 
Fig.~\ref{Fig:Nsol1L} are symmetric under  $\widehat{\mu}\to -\widehat{\mu}$ and $\widehat{A}_t\to-\widehat{A}_t$, reflecting the symmetry of \eq{poly}.
We shall discuss the $\tb$ values of all these roots in greater detail in Section~\ref{onevstwo}, when we present the results of our numerical analysis.

It is instructive to obtain an approximate analytic expression for the value of the largest real root. Assuming $\widehat{\mu}\widehat{A}_t\tan\beta\gg1$ the following approximate alignment condition, first written in Ref.~\cite{Carena:2014nza}, is obtained,
\begin{align}
\tb &\simeq  \frac{M_{h/H}^2 + \MZ^2 +\displaystyle\frac{3 \mt^4 \widehat{\mu}^2}{8\pi^2 v^2}(\widehat{A}_t^{\,2}-2)}
            {\displaystyle\frac{m_t^4 \widehat{\mu}\widehat{A}_t}{8\pi^2 v^2}(\widehat{A}_t^{\,2}-6)}
            \simeq \frac{127+3\widehat{\mu}^2(\widehat{A}_t^{\,2}-2)}
            {\widehat{\mu}\widehat{A}_t(\widehat{A}_t^{\,2}-6)}\,,
\label{Eq:alignmentcondition}
\end{align}
where $M^2_{h/H}\simeq Z_1 v^2$ denotes the (one-loop) mass of the SM-like Higgs boson obtained from \eq{zone}, which could be either the light or heavy $\cp$-even Higgs boson.  It is clear from Eq.~\eqref{Eq:alignmentcondition} that a positive $\tan\beta$ solution exists if either $\widehat{\mu} \widehat{A}_t (\widehat{A}_t^2 - 6) > 0$ and $\widehat{A}_t^2>2$, or if 
$\widehat{\mu} \widehat{A}_t (\widehat{A}_t^2 - 6) <0$, $\widehat{A}_t^2<2$ and $|\widehat\mu|$ is sufficiently large such that the numerator of 
Eq.~\eqref{Eq:alignmentcondition} is negative.  Keeping in mind that Eq.~\eqref{Eq:alignmentcondition} was derived under the assumption that $\widehat{\mu}\widehat{A}_t\tan\beta\gg1$, we have observed in our numerical evaluation in Section~\ref{onevstwo} that the largest of the three roots of \eq{poly} always satisfies the stated conditions above.
Another consequence of Eq.~\eqref{Eq:alignmentcondition} is that  by increasing the value of $|\widehat{\mu}\widehat{ A}_t|$ (in the region where $2<\widehat{A}^2_t<6$), it is
possible to lower the $\tb$ value at which alignment occurs.

If $|\widehat{A}_t|\ll 1$, then \eq{Eq:alignmentcondition} is no longer a good approximation.   Returning to \eq{poly}, we set $\widehat{A}_t=0$ and again assume that $\tan\beta\gg 1$. We can then solve approximately for $\tan\beta$,
\beq
\label{Eq:alignmentcondition_Atsmall}
\tan^2\beta\simeq \frac{M_Z^2-M^2_{h/H}+\displaystyle\frac{3m_t^4\widehat{\mu}^2}{4\pi^2 v^2}\bigl(\tfrac16 \widehat{\mu}^2-2\bigr)}{M_Z^2+M^2_{h/H}+\displaystyle\frac{3m_t^4\widehat{\mu}^2}{4\pi^2 v^2}}\simeq \frac{-39+\widehat{\mu}^2(\widehat{\mu}^2-12)}{126+6\widehat{\mu}^2}\,.
\eeq
For example, in the parameter regime where $\widehat{A}_t\simeq 0$ and $|\widehat{\mu}|\gg 1$, we obtain
$\tan\beta\simeq |\widehat{\mu}|/\sqrt{6}$.

Once the value of $t_\beta$ corresponding to exact alignment is known at a specific point in the
$(\widehat\mu\,,\,\widehat{A}_t)$ plane, we can use \eq{zone} to
determine the value of the SUSY mass scale, $M_S$, such that $Z_1 v^2=(125~{\rm GeV})^2$ is
the observed Higgs squared mass. We shall explore the numerical values of $M_S$ in Section~\ref{onevstwo} for each of the physical solutions of the alignment condition.

The question of whether the light or the heavy $\cp$-even Higgs boson possesses SM-like Higgs couplings in the alignment without decoupling regime depends on the relative size of $Z_1v^2$ and $Z_5v^2 + M_A^2$.   Combining \eqs{Eq:Z5v2}{Eq:Z6v2}, it follows that in the limit of exact alignment where $Z_6=0$, we can identify $Z_1 v^2$ as the squared mass of the observed SM-like Higgs boson and
\beq \label{zfive}
Z_5 v^2=M_Z^2(1+ c_{2\beta}) +\frac{3m_t^4 \widehat{\mu}(\widehat{A}_t-\widehat{\mu}\cot\beta)}{8\pi^2 v^2 s_\beta^4}
\biggl\{s_{2\beta}-\tfrac16\bigl[(\widehat{A}_t^{\,2}-\widehat{\mu}^2)s_{2\beta}-2\widehat{A}_t\widehat{\mu} c_{2\beta}\bigr]\biggr\}\,,
\eeq
We define a critical value of $M^2_A$,
\beq \label{macrit}
M_{A,c}^2\equiv{\rm max}\bigl\{(Z_1-Z_5)v^2\,,\,0\bigr\}\,,
\eeq
where $Z_1 v^2=(125~{\rm GeV})^2$ and $Z_5 v^2$ is given by \eq{zfive}.   Note further that the squared-mass of the non-SM-like CP-even Higgs boson in the exact alignment limit, $M_A^2+Z_5 v^2$, must be positive, which implies that the minimum value possible for the squared-mass of the CP-odd Higgs boson is
\beq \label{mamin}
M_{A,m}^2\equiv{\rm max}\bigl\{-Z_5 v^2\,,\,0\bigr\}\,.
\eeq
That is, if $Z_5$ is negative, then the minimal allowed value of $M_A^2$ is non-zero and positive.

If we compute $Z_5$ from \eq{zfive} using the value of $\tan\beta$ obtained
by setting $Z_6=0$ in \eq{z6zero},  the value
of $M_{A,c}^2$ for each point in the $(\widehat{\mu}\,,\widehat{A}_t$) plane can be determined.  The interpretation of
$M_{A,c}^2$ is as follows.  If $M^2_A> M^2_{A,c}$, then $h$ can be identified as the SM-like Higgs boson with 
$M_h\simeq 125$~GeV.  If $M_{A,m}^2<M^2_A<M^2_{A,c}$ [where $M^2_{A,m}$ is the minimal allowed value of $M_A^2$ given in \eq{mamin}], then $H$ can be identified as the SM-like Higgs boson with $M_H\simeq 125$~GeV.  We shall exhibit numerical results for $M_{A,c}$ in Section~\ref{onevstwo} for each of the realistic $\tb$ solutions.

Finally, we note that using the same one-loop approximations employed in this section, the leading contribution to the squared-mass splitting of the charged Higgs boson and CP-odd Higgs boson is given by~\cite{Haber:1993an},
\beq \label{mhpm1L}
M_{H^\pm}^2-M_A^2\simeq \MW^2-\frac{3\mu^2 m_t^4}{16\pi^2 v^2 s_\beta^4 M_{\rm SUSY}^2}\simeq \MW^2\left(1-\frac{0.035\widehat{\mu}^2}{s_\beta^4}\right)\,.
\eeq
In particular, in the parameter regime in which $H$ is identified as the SM-like Higgs boson, there is an upper bound on the charged Higgs mass obtained by inserting $M_A=M_{A,c}$ into \eq{mhpm1L}. In this case, collider and flavor constraints relevant to such a light charged Higgs boson 
can significantly reduce the allowed MSSM parameter space~\cite{Bechtle:2016kui,Arbey:2017gmh}.

\section{Alignment without decoupling at the two-loop level} 
\label{Sec:alignment2L}

As previously noted, the analysis above was based on approximate one-loop formulae given in \eqst{zone}{Eq:Z6v2}, where only the leading terms proportional to $m_t^2 h_t^2$ are included.   In the exact alignment limit, we identify $Z_1 v^2$ given by \eq{zone} as  the squared-mass of the observed SM-like Higgs boson.  However, it is well known that \eq{zone} overestimates the value of the radiatively corrected Higgs mass.  Remarkably, one can obtain a significantly more accurate result simply by including the leading two-loop radiative corrections proportional to $\alpha_s m_t^2 h_t^2$.   

In Ref.~\cite{Carena:2000dp}, it was shown that the dominant part of these two-loop corrections can be obtained from the corresponding one-loop formulae with the following very simple two step prescription.   First, we replace
\beq \label{step1}
m_t^4\ln\left(\frac{M_S^2}{m_t^2}\right)\longrightarrow m_t^4(\lambda)\ln\left(\frac{M_S^2}{m_t^2(\lambda)}\right)\,,\qquad
\text{where $\lambda\equiv\bigl[m_t(m_t)M_S\bigr]^{1/2}$}\,,
\eeq
where $m_t(m_t)\simeq 163.6$~GeV is the $\overline{\rm MS}$ top quark mass~\cite{Marquard:2015qpa}, and the running top quark mass in the one-loop approximation is given by
\beq \label{mtmu}
m_t(\lambda)=m_t(m_t)\left[1+\frac{\alpha_s}{\pi}\ln\left(\frac{m_t^2(m_t)}{\mu^2}\right)\right]\,.
\eeq
In our numerical analysis, we take $\alpha_s = \alpha_s(m_t(m_t)) \simeq 0.1088$~\cite{Marquard:2015qpa}.
Second, when $m_t^4$ multiplies the threshold corrections (i.e., the one-loop terms proportional to $X_t$ and $Y_t$), then
we make the replacement,
\beq
m_t^4\longrightarrow m_t^4(M_S)\,,
\eeq
where
\beq
\label{mtms}
m_t(M_S)=m_t(m_t)\left[1+\frac{\alpha_s}{\pi}\ln\left(\frac{m_t^2(m_t)}{M_S^2}\right)+\frac{\alpha_s}{3\pi}\,\frac{X_t}{M_S}\right]\,.
\eeq
Note that the running top-quark mass evaluated at $M_S$ includes a threshold correction at the SUSY-breaking scale that is proportional to $X_t$.
Here, we only keep the leading contribution to the threshold correction under the assumption that $m_t\ll M_S$ (a more precise formula can be found in Appendix B of Ref.~\cite{Carena:2000dp}).
The above two step prescription can now be applied to \eqst{zone}{Eq:Z6v2}, which yields a more accurate expression for the radiatively corrected Higgs mass and the condition for exact alignment without decoupling.

In applying the prescription outlined above, we formally work to $\mathcal{O}(\alpha_s)$  while dropping terms of $\mathcal{O}(\alpha_s^2)$ and higher.  For example,
\beq
\ln\left(\frac{M_S^2}{m_t^2(\mu)}\right)\simeq \left[1+\frac{\alpha_s}{2\pi}\right]\ln\left(\frac{M_S^2}{m_t^2(m_t)}\right)\,.
\eeq
The end results are the following approximate expressions for $Z_1$, $Z_5$ and $Z_6$ that incorporate the leading two-loop $\mathcal{O}(\alpha_s m_t^2 h_t^2)$ effects,
\beqa
Z_1 v^2&=&M_Z^2 c^2_{2\beta}+CL\bigl(1-2\overline{\alpha}_s L+\overline{\alpha}_s\bigr)+CX_1\bigl(1-4\overline{\alpha}_sL+\tfrac43\overline{\alpha}_s x_t\bigr)\,,\label{zeeone}\\[6pt]
Z_5 v^2 &=& s_{2\beta}^2\biggl[M_Z^2+\frac{CL}{4s_\beta^4}\bigl(1-2\overline{\alpha}_s L+\overline{\alpha}_s\bigr)+\frac{C}{4s_\beta^4}X_5\bigl(1-4\overline{\alpha}_sL+\tfrac43\overline{\alpha}_s x_t\bigr)\biggr]\,,\label{zeefive}\\[6pt]
Z_6 v^2&=&-s_{2\beta}\biggl[M_Z^2 c_{2\beta}-\frac{CL}{2s_\beta^2}\bigl(1-2\overline{\alpha}_s L+\overline{\alpha}_s\bigr)-\frac{C}{2s_\beta^2}X_6\bigl(1-4\overline{\alpha}_sL+\tfrac43\overline{\alpha}_s x_t\bigr)\biggr]\,,\label{zeesix}
\eeqa
where we have defined,
\beqa
\label{Eq:definequantities}
\!\!\!\!\!\!\!\!
C\equiv \frac{3m_t^4(m_t)}{2\pi^2 v^2}\,,\qquad \overline{\alpha}_s\equiv\frac{\alpha_s}{\pi}\,,\qquad
x_t\equiv X_t/M_S\,,\qquad y_t\equiv Y_t/M_S\,,\qquad L\equiv\ln\left(\frac{M_S^2}{m_t^2(m_t)}\right),
\eeqa
and
\beqa
X_1\equiv x_t^2\bigl(1-\tfrac{1}{12}x_t^2\bigr)\,,\qquad X_5\equiv x_t y_t\bigl(1-\tfrac{1}{12}x_t y_t\bigr)\,,
\qquad
X_6\equiv \tfrac12 x_t(x_t+y_t)-\tfrac{1}{12}x_t^3 y_t\,.
\eeqa
In the above equations, $m_t\equiv m_t(m_t)$ is the $\overline{\rm MS}$ top quark mass.
Note that the approximate loop-corrected formulae for $Z_1$, $Z_5$ and $Z_6$ are no longer invariant under $X_t\to -X_t$, $Y_t\to -Y_t$ (or equivalently $A_t\to -A_t$, $\mu\to -\mu$) due to the asymmetry introduced by \eq{mtms} at $\mathcal{O}(\alpha_s)$.

We can now derive analogous expressions to Eqs.~\eqref{poly} and~\eqref{zfive} that incorporate the leading two-loop effects at $\mathcal{O}(\alpha_s m_t^2 h_t^2)$.  First, we note that \eq{zeeone} yields
\beq \label{L1}
L= C^{-1}\bigl(Z_1 v^2 -M_Z^2 c^2_{2\beta}\bigr)-X_1+\overline{\alpha}_s B_1\,,
\eeq
where $B_1$ is to be determined.  Inserting \eq{L1} into \eq{zeeone}, the $\mathcal{O}(1)$ terms cancel exactly.   Keeping only terms of $\mathcal{O}(\overline{\alpha}_s)$, we end up with the following expression for $B_1$,
\beq \label{B1}
B_1=2C^{-2}\bigl(Z_1 v^2-M_Z^2 c^2_{2\beta}\bigr)^2-C^{-1}\bigl(Z_1 v^2-M_Z^2 c^2_{2\beta}\bigr)+X_1\left(1-2X_1-\tfrac43 x_t\right)\,.
\eeq

Now we substitute \eq{B1} back into \eq{L1} to obtain
\beqa 
L&=&C^{-1}(1-\overline{\alpha}_s)\bigl(Z_1 v^2 -M_Z^2 c^2_{2\beta}\bigr)
+2\overline{\alpha}_s C^{-2}\bigl(Z_1 v^2 -M_Z^2 c^2_{2\beta}\bigr)^2
-X_1\bigl[1-\overline{\alpha}_s\bigl(1-2X_1-\tfrac43 x_t\bigr)\bigr]\,.\nn \\
&&\phantom{line}
\label{LL1}
\eeqa
Finally, we insert \eq{LL1} into \eq{zeesix} and set $Z_6=0$ to obtain,
\beq \label{tanbeq}
2M_Z^2 s_\beta^2 c_{2\beta}-(Z_1 v^2-M_Z^2 c^2_{2\beta})\bigl[1+4\overline{\alpha}_s(X_1-X_6)\bigr]
+C(X_1-X_6)\bigl[1+\overline{\alpha}_s(4X_1+\tfrac43 x_t)\bigr]=0\,.
\eeq
That is,  $t_\beta\equiv \tan\beta$ is the solution to a 11th order polynomial equation,
\beqa \label{poly2}
&&M_Z^2 t_\beta^8(1-t_\beta^2)-Z_1 v^2 t_\beta^8(1+t_\beta^2)+\frac{3m_t^4\widehat{\mu}(\widehat{A}_t t_\beta
-\widehat{\mu})t_\beta^4(1+t_\beta^2)^2}{4\pi^2 v^2}\bigl[\tfrac16(\widehat{A}_t t_\beta-\widehat{\mu})^2-t_\beta^2\bigr]\nn \\[6pt]
&& \qquad
+2\overline{\alpha}_s t_\beta^4\bigl[M_Z^2 (1-t_\beta^2)^2-Z_1 v^2 (1+t_\beta^2)^2\bigr]\widehat{\mu}(\widehat{A}_t t_\beta
-\widehat{\mu})\bigl[\tfrac16(\widehat{A}_t t_\beta-\widehat{\mu})^2-t_\beta^2\bigr]\nn \\[6pt]
&& \qquad
+\frac{\overline{\alpha}_s m_t^4 \widehat{\mu}(\widehat{A}_t t_\beta-\widehat{\mu})^2(1+t_\beta^2)^2}{\pi^2 v^2}\bigl[\tfrac16(\widehat{A}_t t_\beta-\widehat{\mu})^2-t_\beta^2\bigr]\bigl[t_\beta^3+3t_\beta^2(\widehat{A}_t t_\beta-\widehat\mu)-\tfrac14(\widehat{A}_t t_\beta-\widehat\mu)^3\bigr]=0.\nn \\
&&
\phantom{line}
\eeqa
As previously noted, solutions to this equation for negative $\tan\beta$ at a point in the $(\widehat{\mu}\,,\,\widehat{A}_t)$ plane can be reinterpreted as positive $\tan\beta$ solutions at the point $(-\widehat{\mu}\,,\,\widehat{A}_t)$.

In order to obtain two-loop improved versions of $M^2_{A,c}$ and $M^2_{A,m}$ [cf.~\eqs{macrit}{mamin}], we need to impose the alignment limit condition, $Z_6=0$, on the two-loop expression for $Z_5$ given by \eq{zeefive}.  Our strategy is similar to the one employed above in deriving \eq{tanbeq}.  First, we derive another expression for $L$ based on \eq{zeefive}; the steps leading to \eq{LL1} are modified by the following substitutions,
\begin{align}
M_Z^2 c^2_{2\beta}\to M_Z^2 s^2_{2\beta},\quad C\to C/t_\beta^2,\quad Z_1\to Z_5,\quad\text{and}\quad X_1\to X_5\,.
\end{align}
The end result is,
\beqa 
L&=&C^{-1}(1-\overline{\alpha}_s)\bigl(Z_5 v^2 -M_Z^2 s^2_{2\beta}\bigr)t_\beta^2
+2\overline{\alpha}_s C^{-2}\bigl(Z_5 v^2 -M_Z^2 s^2_{2\beta}\bigr)^2 t_\beta^4
-X_5\bigl[1-\overline{\alpha}_s\bigl(1-2X_5-\tfrac43 x_t\bigr)\bigr]\,.\nn \\
&&\phantom{line}
\label{LL5}
\eeqa
Finally, we insert \eq{LL5} into \eq{zeesix} and set $Z_6=0$ to obtain,
\beq
2M_Z^2 s_\beta^2 c_{2\beta}-(Z_5 v^2-M_Z^2 s^2_{2\beta})t_\beta^2\bigl[1+4\overline{\alpha}_s(X_5-X_6)\bigr]
+C(X_5-X_6)\bigl[1+\overline{\alpha}_s(4X_5+\tfrac43 x_t)\bigr]=0\,.
\eeq
Solving for $Z_5$, and again expanding out in $\alpha_s$ and dropping terms of $\mathcal{O}(\alpha_s^2)$ and higher,
\beq
Z_5 v^2=M_Z^2(1+c_{2\beta})+\frac{C(X_5-X_6)}{t_\beta^2}\biggl\{1+4\overline{\alpha}_s\bigl(X_6+\tfrac13 x_t-2s_\beta^2 c_{2\beta}C^{-1}M_Z^2\bigr)\biggr\}\,,
\eeq
which yields the $\mathcal{O}(\alpha_s)$ correction to \eq{zfive}.  One can now define the two-loop improved versions of $M^2_{A,c}$ and $M^2_{A,m}$ via \eqs{macrit}{mamin}.   Likewise, the two-loop improved formula for the charged Higgs mass is obtained by replacing $m_t$ in \eq{mhpm1L} by $m_t(M_S)$ according to \eq{mtms}.  The end result is 
\beq \label{mhpm2L}
M_{H^\pm}^2\simeq M_A^2+\MW^2-\frac{3\mu^2 m_t^4(m_t)}{16\pi^2 v^2 s_\beta^4 M_{\rm SUSY}^2}\
\left[1+\frac{4\alpha_s}{\pi}\ln\left(\frac{m_t^2(m_t)}{M_S^2}\right)+\frac{4\alpha_s}{3\pi}\,\frac{X_t}{M_S}\right]\,.
\eeq

\begin{figure}[b!]
\centering
\includegraphics[width=0.44\textwidth]{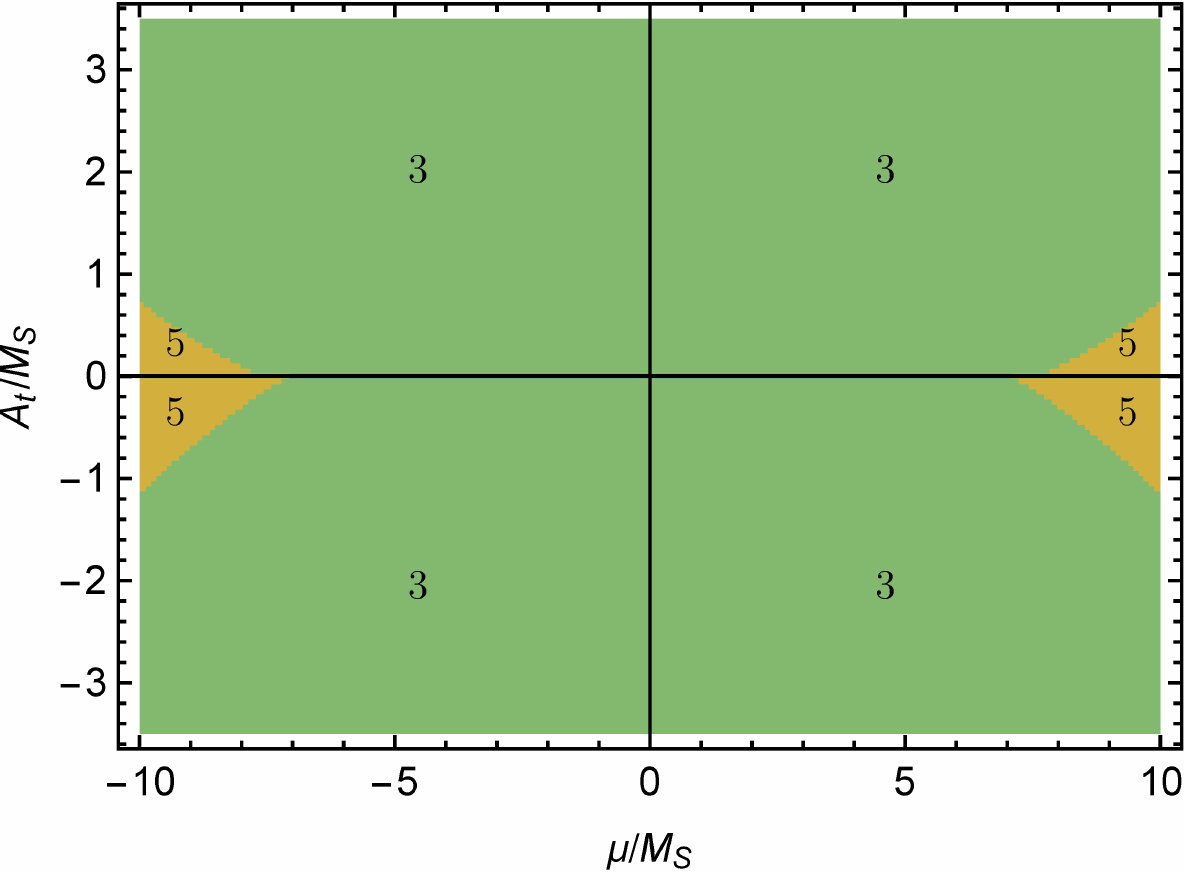}\hfill
\includegraphics[width=0.44\textwidth]{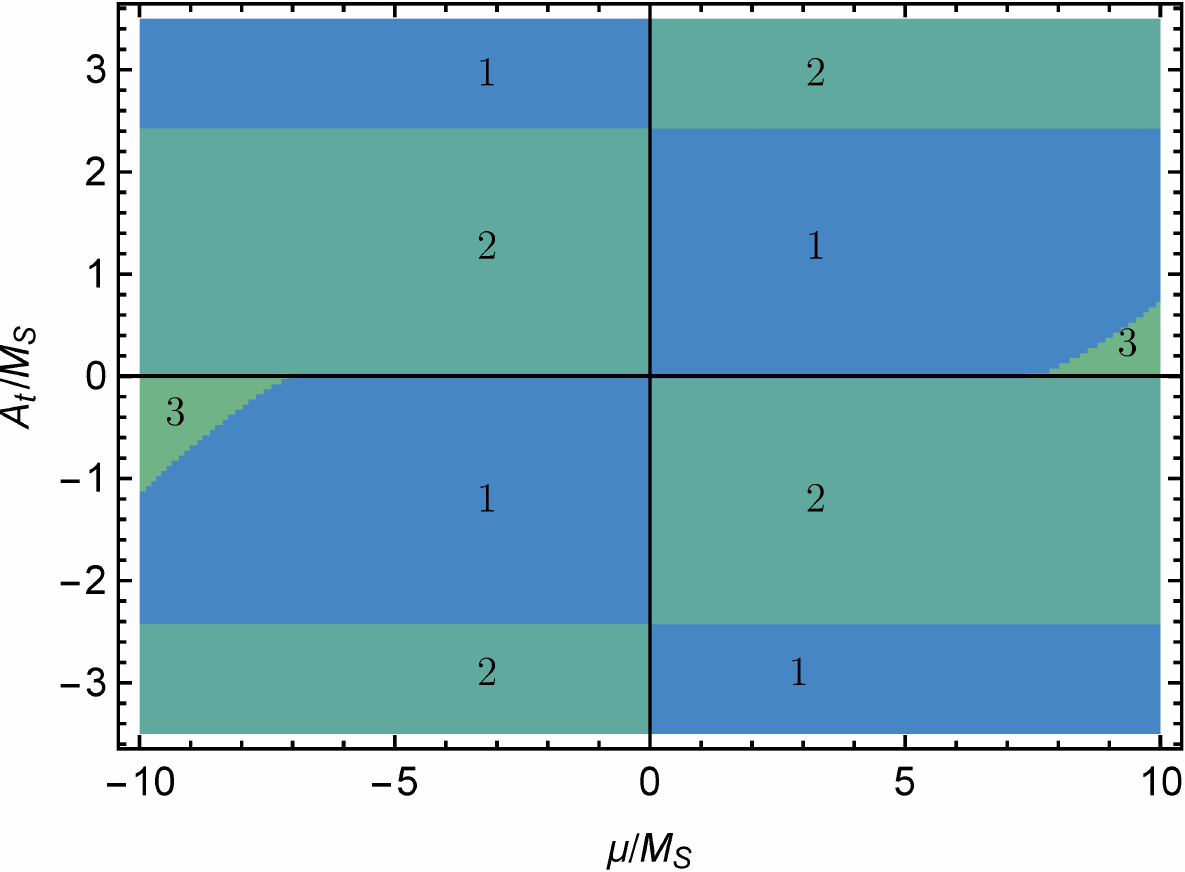}
\caption{Number of real solutions to the two-loop alignment condition, \refeq{poly2}. We set $Z_1v^2 = 125\gev$. \emph{Left:} All solutions with real $\tan\beta$; \emph{Right:} Real, positive $\tan\beta$ solutions.}
\label{Fig:Nsol2L}
\end{figure}

In the left [right] panel of Fig.~\ref{Fig:Nsol2L} we show the number of real [positive] solutions to the polynomial given in \refeq{poly2}, 
corresponding to the two-loop condition for alignment without decoupling which determines $\tan\beta$ as a function of $\widehat\mu$ and $\widehat{A}_t$.  Compared to the one-loop results of Section~\ref{Sec:alignment1L}, there are a few notable changes, which we now discuss.  First, 
in our scan of the $(\widehat\mu\,,\,\widehat{A}_t)$ plane, we have observed numerically that there are three real roots of \eq{poly2} for $|\widehat{\mu}|\lsim 8$--$10$ (depending on the value of $\widehat{A}_t$), whereas for larger values of $|\widehat{\mu}|$, a region opens up in which there are five real roots.  As previously discussed,
the transition between these two regions occurs when two of the real roots in the large $|\widehat{\mu}|$ regime coalesce (yielding a degenerate real root) and then move off the real axis to form a complex conjugate pair as the value of $|\widehat{\mu}|$ is reduced.  
Comparing with the roots of \eq{poly}, we see that two new roots have come into play.  We have analyzed these two roots and find that one is positive and one is negative.  However, the positive root always corresponds to a value of $|{X}_t|>3M_S$, which lies outside our region of interest.  Henceforth, we simply discard this possibility.  What remains then are at most two real roots at a given point in the  $(\widehat\mu\,,\,\widehat{A}_t)$ plane that can be identified as the two-loop corrected versions of the corresponding one-loop results obtained earlier.

We can now see the effects of including the leading $\mathcal{O}(h_t^2\alpha_s)$ corrections.   The regions where positive solutions to \eq{poly2} exist, shown in the right panels of Fig.~\ref{Fig:Nsol2L} (excluding the positive solution corresponding to $|{X}_t|>3M_S$ as noted above),
have shrunk considerably in the two quadrants where $\widehat{\mu}\widehat{A}_t>0$, 
as compared to the corresponding positive solutions to \eq{poly} shown in the right panel of Fig.~\ref{Fig:Nsol1L}.  In contrast, in
the two quadrants where $\widehat{\mu}\widehat{A}_t<0$, the respective sizes of the regions where positive solutions to \eq{poly} and \eq{poly2} exist are comparable.

One new feature of the two-loop approximation not yet emphasized is that we must now carefully define the input parameters $\mu$ and $A_t$.  In the formulae presented in this section, we interpret these parameters as $\overline{\rm MS}$ parameters.  However, it is often more convenient to re-express these parameters in terms of on-shell parameters.  In Ref.~\cite{Carena:2000dp}, the following expression was obtained for the on-shell squark mixing parameter $X_t^{\rm OS}$ in terms of the $\overline{\rm MS}$ squark mixing parameter $X_t$, where only the leading $\mathcal{O}(\alpha_s)$ corrections are kept,
\beq
X_t^{\rm OS}=X_t-\frac{\alpha_s}{3\pi}M_S\left[8+\frac{4X_t}{M_S}-\frac{X_t^2}{M_S^2}-\frac{3X_t}{M_S}\ln\left(\frac{m_t^2}{M_S^2}\right)\right]\,.
\eeq
Since the on-shell and $\overline{\rm MS}$ versions of $\mu$ are equal at this level of approximation, we also have
\beq
A_t^{\rm OS}=X_t^{\rm OS}+\frac{\mu}{\tan\beta}\,.
\eeq

The approximations employed in the section capture some of the most important radiative corrections relevant for analyzing the alignment limit of the MSSM.   However, it is important to appreciate what has been left out.  The analysis of this section ultimately corresponds to a renormalization of $\cos(\beta-\alpha)$, which governs the couplings of the Higgs boson in the effective 2HDM theory below the SUSY-breaking scale
and its departure from the alignment limit.  However, radiative corrections also contribute other effects that modify Higgs production cross-sections and branching ratios.  It is well-known that for $M_A\ll M_S$, the effective low-energy theory below the scale $M_S$ is a general two Higgs doublet model with the most general Higgs-fermion Yukawa couplings.  These include the so-called wrong-Higgs couplings of the MSSM~\cite{Haber:2007dj}, which ultimately are responsible for the $\Delta_b$ and $\Delta_\tau$ corrections that can significantly modify the coupling of the Higgs boson to bottom quarks and tau leptons.\footnote{For a review of these effects and a guide to the original literature, see Ref.~\cite{Carena:2002es}. \label{fn}}  In addition, integrating out heavy SUSY particles at the scale $M_S$ can generate higher dimensional operators that can also modify Higgs production cross-sections and branching ratios~\cite{Dine:2007xi}.  None of these effects are accounted for in the analysis presented in this section.

\section{Numerical results}
\label{onevstwo}

In this section we present the numerical results for the physical (i.e.\ real positive) $\tb$ solutions of the alignment condition, and, in particular, compare the results obtained in the one-loop and two-loop approximations given in Sections~\ref{Sec:alignment1L} and \ref{Sec:alignment2L}, respectively. Moreover, we shall discuss for each of these solutions their implications for the correlated parameters, i.e.~the SUSY-breaking mass scale, $M_S$, and the critical $M_A$ value, $M_{A,c}$, which determines whether the light or the heavy CP-even Higgs boson is the one aligned with the SM Higgs vev in field space.

\begin{figure}[t!]
\centering
\includegraphics[width=0.48\textwidth]{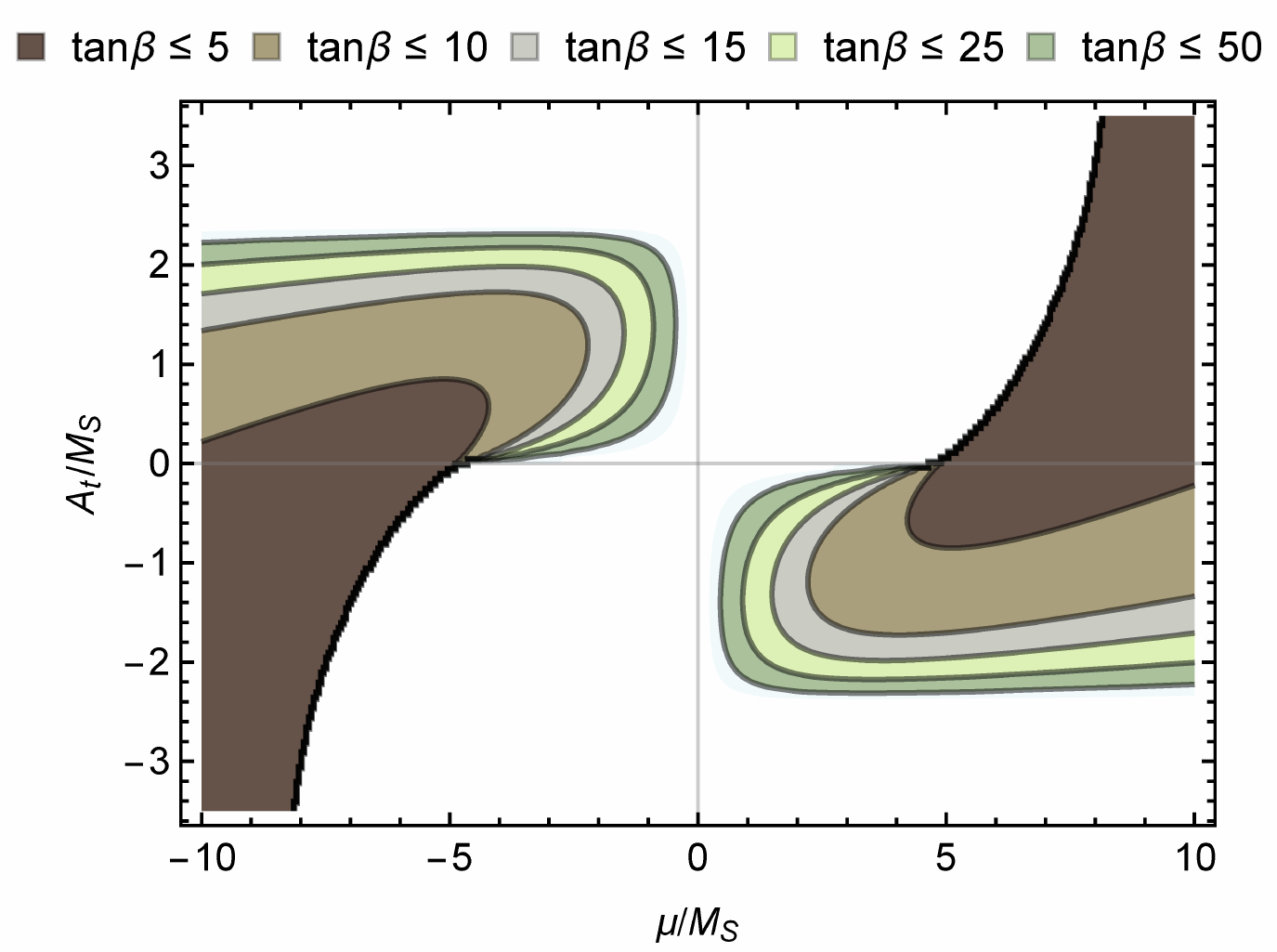}\hfill
\includegraphics[width=0.48\textwidth]{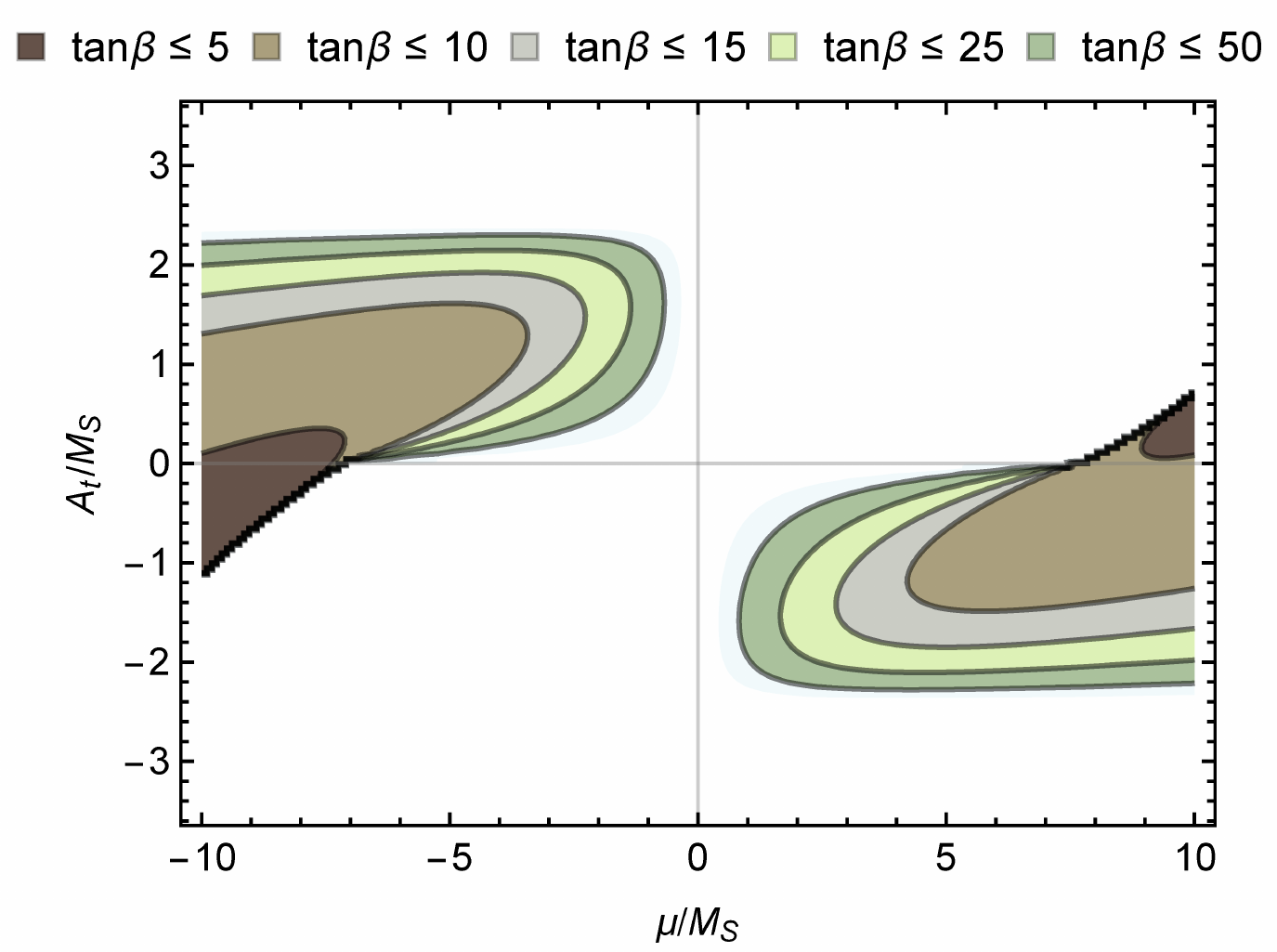}
\includegraphics[width=0.48\textwidth]{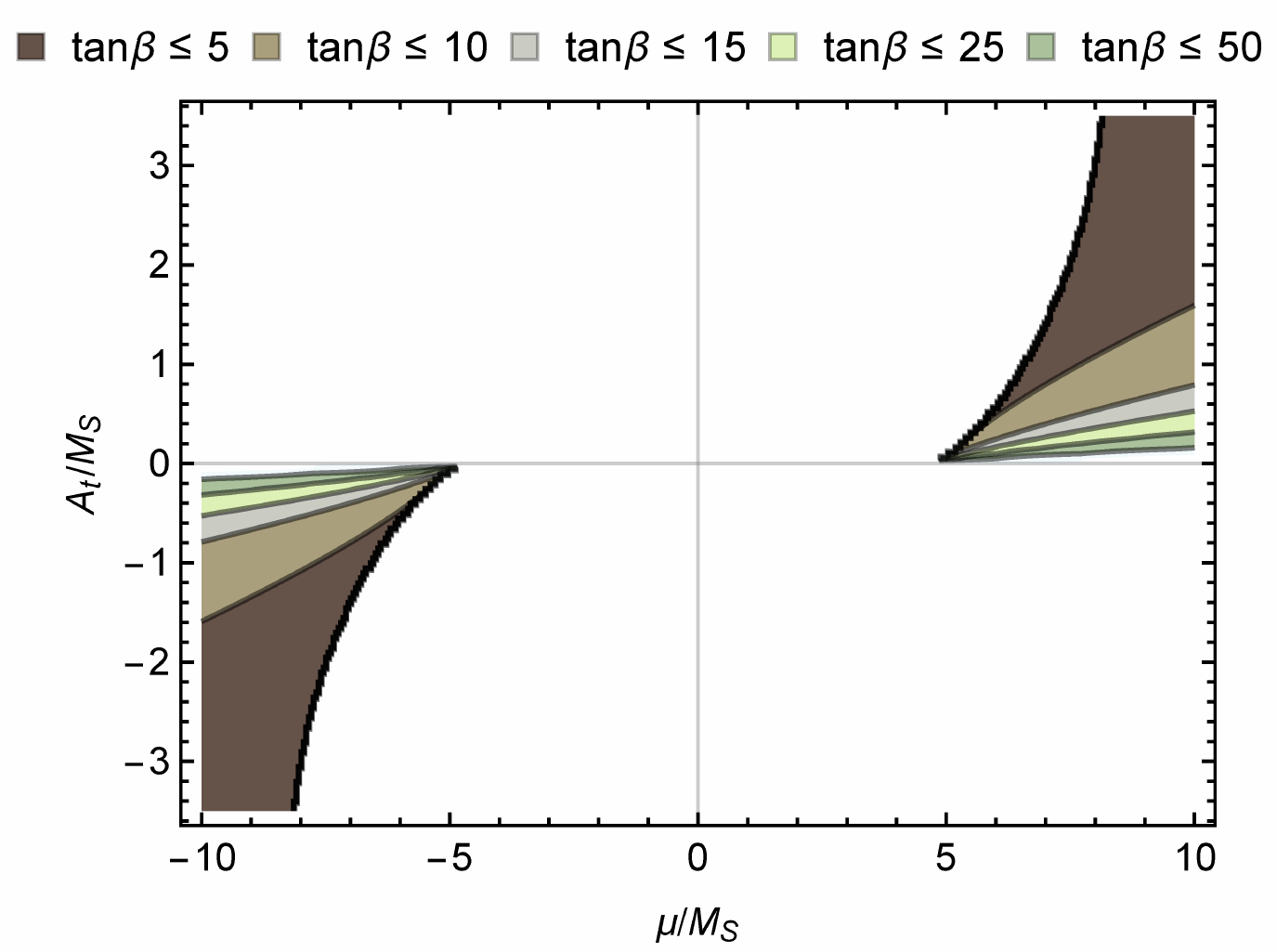}\hfill
\includegraphics[width=0.48\textwidth]{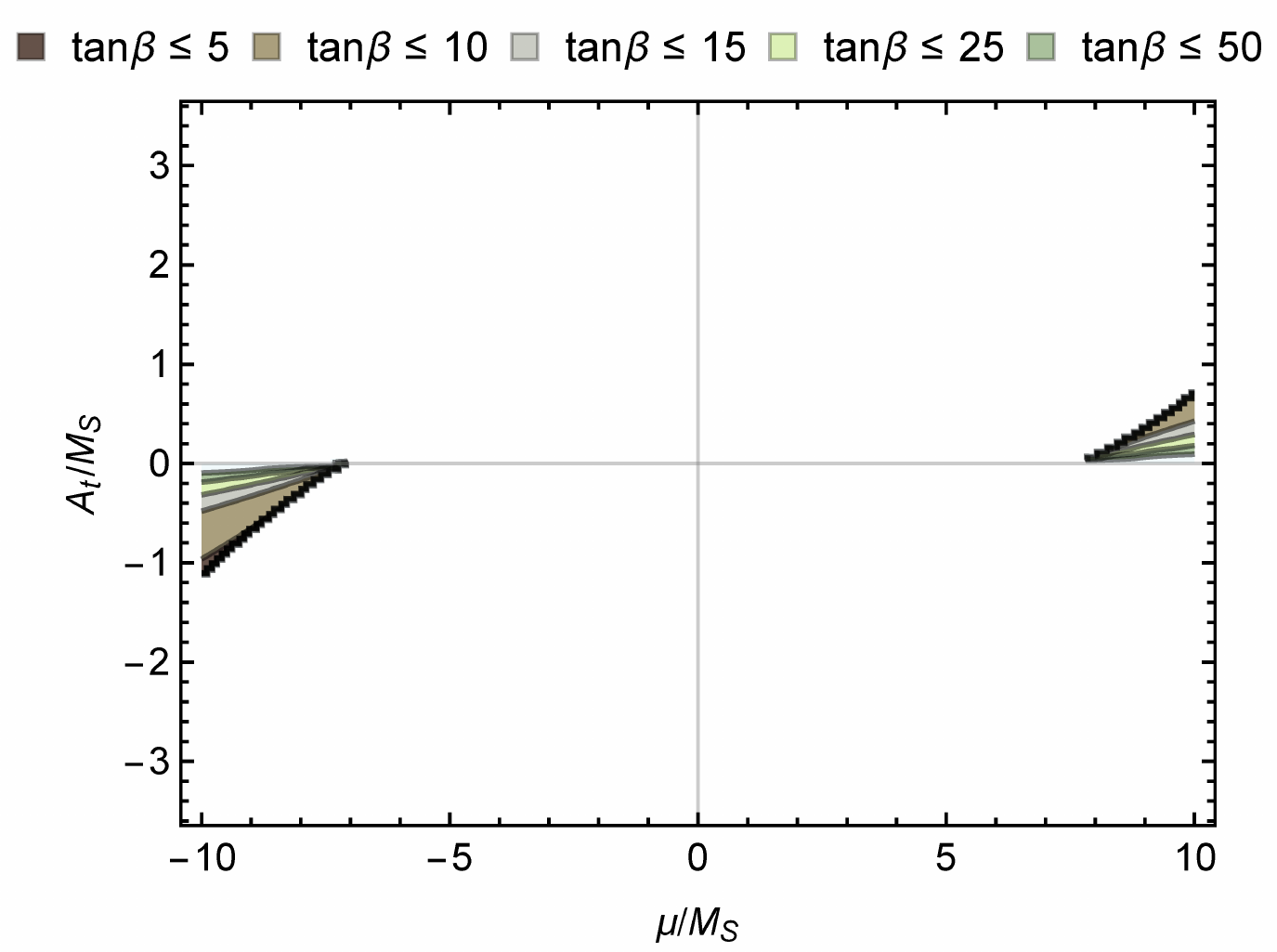}
\includegraphics[width=0.48\textwidth]{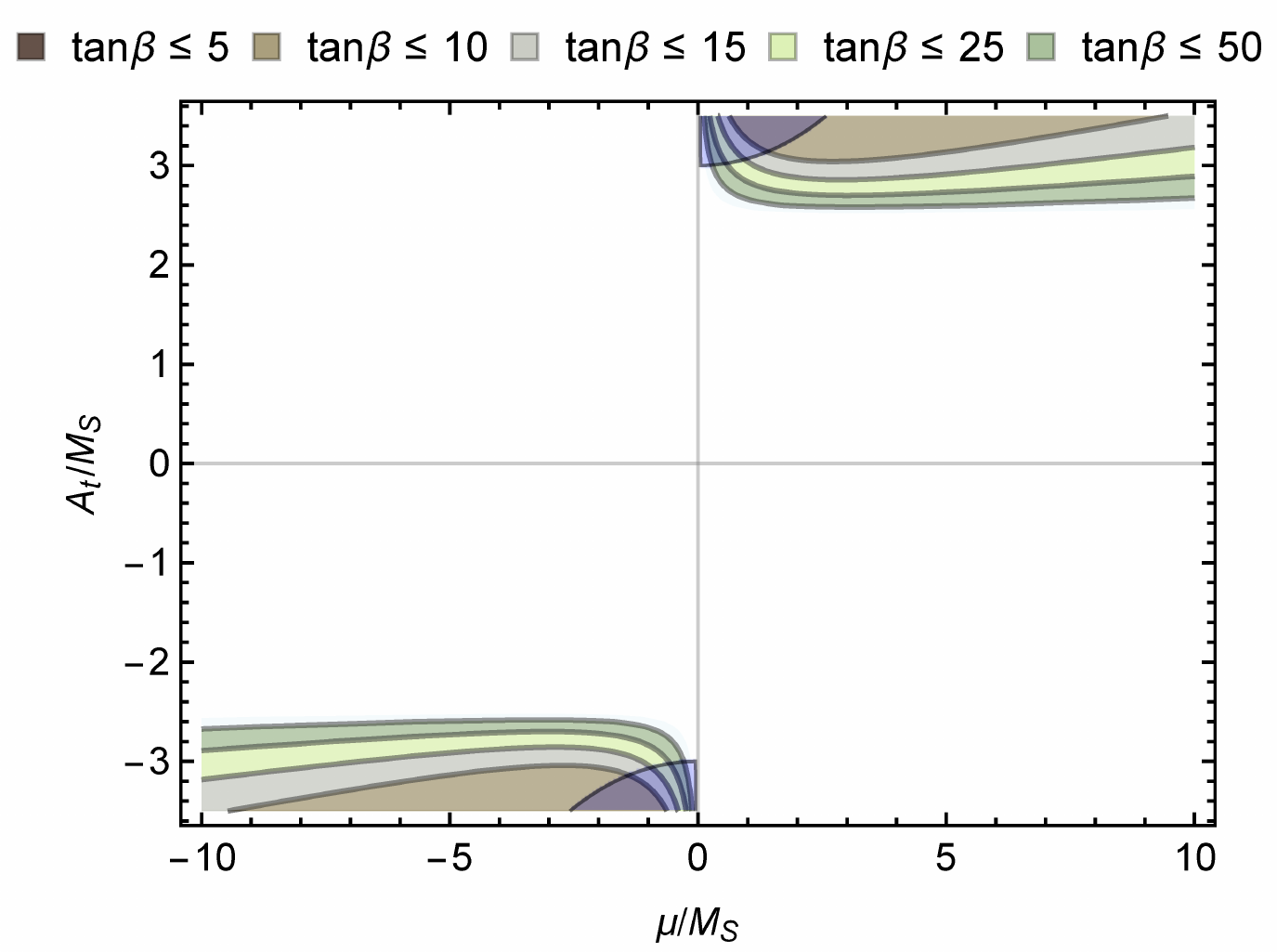}\hfill
\includegraphics[width=0.48\textwidth]{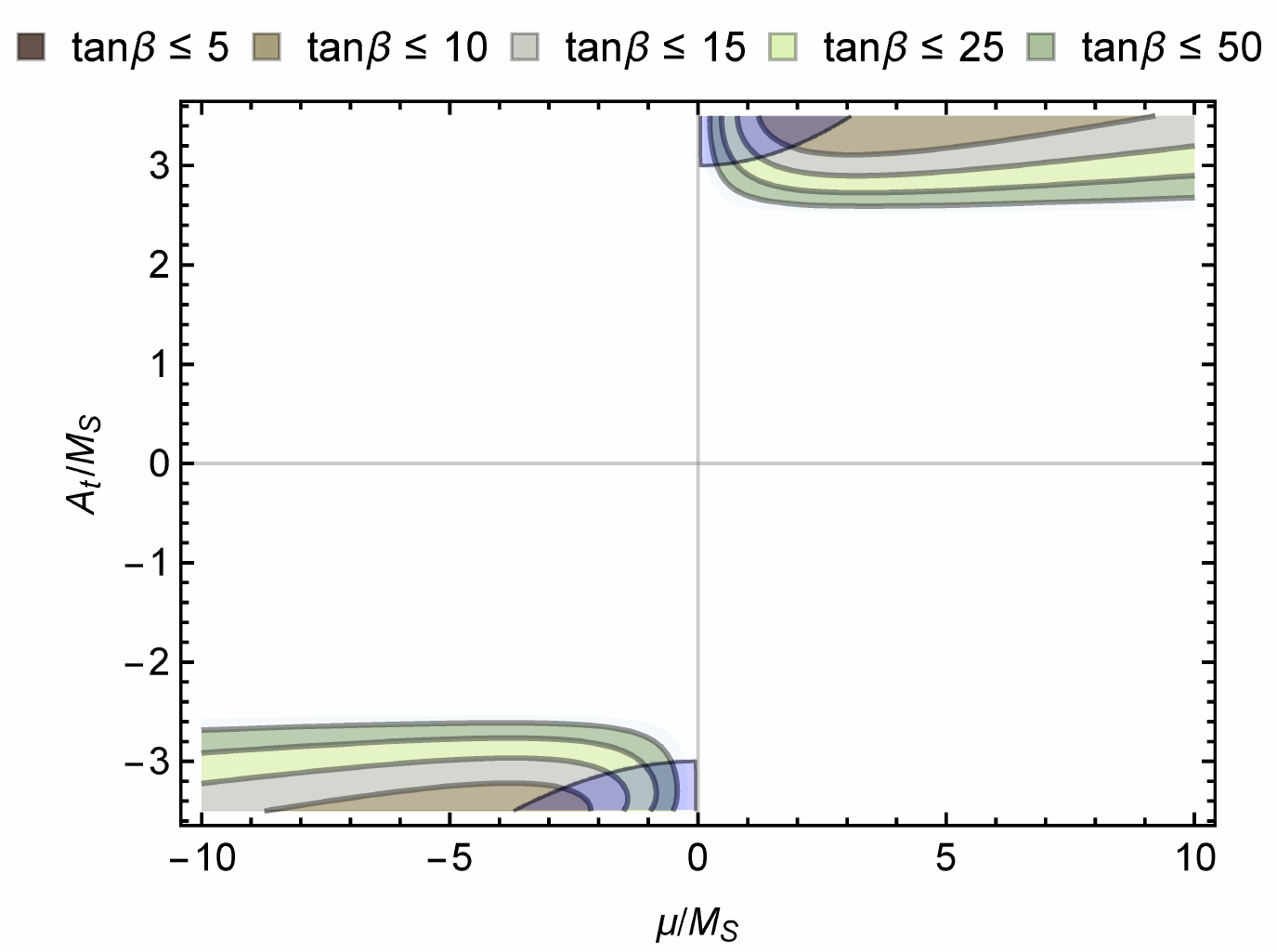}
\caption{Contours of $\tan\beta$ corresponding to exact alignment, $Z_6= 0$,
in the $(\mu/M_S, A_t/M_S)$ plane.
$Z_1$ is adjusted to give the correct Higgs mass.
Left: Approximate one-loop result; Right: Two-loop improved result.
Taking the three panels on each side together, one can immediately discern the regions of zero, one, two and three values of $\tan\beta$ in which exact alignment is realized. In the overlaid blue regions we have (unstable) values of $|X_t/M_S| \ge 3$.
}
\label{Fig:alignment_numerical_TB}
\end{figure}

As shown in Figs.~\ref{Fig:Nsol1L} and \ref{Fig:Nsol2L}, there may be more than one value of $\tan\beta$ corresponding to exact alignment for a given $\widehat{\mu}$ and $\widehat{A}_t$.
In the left and right panels of \figref{Fig:alignment_numerical_TB} these $\tan\beta$ solutions in the one-loop [\refeq{poly}] and two-loop [\refeq{poly2}] approximation, respectively, are
displayed as filled contours in the $(\widehat{\mu}, \widehat{A}_t)$ parameter plane. The three panels from top to bottom of \figref{Fig:alignment_numerical_TB} correspond to three different roots, with the respective $\tb$ values being the smallest in the top panel and the largest in the bottom panel. Taking the top, middle and bottom panel together, one can immediately discern the regions of zero, one, two and three positive roots of \eq{poly} and \eq{poly2}, and their corresponding values.

\afterpage{\clearpage}

Previous works on Higgs alignment without decoupling in the MSSM~\cite{Carena:2013ooa,Carena:2014nza,Bechtle:2016kui} have largely focused on the $\tb$ solution displayed in the two bottom panels of \figref{Fig:alignment_numerical_TB}. This value of $\tb$, which can appear already at moderately large $|\widehat\mu|$ values,  
has also been employed in the definition of MSSM benchmark scenarios with Higgs alignment without decoupling~\cite{Carena:2014nza,Bechtle:2015pma}.
However,  this solution is associated with a large trilinear scalar coupling, $\widehat{A}_t$, and thus part of the parameter space exhibiting this solution 
may yield a color or electric charge breaking vacuum and/or feature an unreliable theoretical prediction of the Higgs mass. In order to highlight this, we overlay the region where $|X_t|/M_S \ge 3$ with a blue shading in Fig.~\ref{Fig:alignment_numerical_TB}. Since for the relevant parameter space, $\widehat{\mu}\widehat{A}_t \tb \gg 1$, this $\tb$ solution is approximated by \refeq{Eq:alignmentcondition}, as first employed in Ref.~\cite{Carena:2014nza}. Alignment without decoupling at moderately small values of $\tb\lesssim10$, as suggested by constraints from LHC $H/A\to \tau^+\tau^-$ searches~\cite{CMSMSSMHiggs,ATLASMSSMHiggs},
can be found for $|\widehat{\mu}| \sim 2$--$3$ and $|\widehat{A}_t|\sim 3$. Comparing the numerical values of this $\tan\beta$ solution obtained in the approximate one-loop and two-loop descriptions, we observe that the improved description at the two-loop level yields rather small corrections, which slightly increase the value of $\tb$. Furthermore, note the small asymmetry between the two sectors ($\widehat{\mu}> 0$ and $\widehat{A}_t>0$ vs.\ $\widehat{\mu}<0$ and $\widehat{A}_t<0$) introduced by the finite threshold correction proportional to $X_t$ entering at the two-loop level.

The smallest of the $\tb$ solutions, displayed in the top panel of Fig.~\ref{Fig:alignment_numerical_TB}, was only briefly mentioned in Ref.~\cite{Carena:2013ooa,Carena:2014nza}, but was subject to detailed discussions in Ref.~\cite{Bechtle:2016kui} in the context of scenarios where the observed SM-like Higgs boson was interpreted in terms of the heavy CP-even Higgs boson. In fact, such scenarios were found viable in this parameter region at $|\widehat\mu|\sim 6$--$8$, partly because for such large $\widehat{\mu}$ values, large $\Delta_b$ corrections suppress the light charged Higgs contribution to the rare flavor physics decays $B\to X_s\gamma$. In the top panel of Fig.~\ref{Fig:alignment_numerical_TB}, we observe that this $\tb$ solution extends over all four sectors of the $(\widehat{\mu}, \widehat{A}_t)$ parameter space, however, with the restriction that for $|\widehat{\mu}|\lesssim 5~(7)$, the parameters $\widehat{\mu}$ and $\widehat{A}_t$ have to be of opposite sign in the one-loop (two-loop) description. In the latter case, and as long as $|\widehat{\mu}| |\widehat{A}_t| \tb \gg 1$, the $\tb$ alignment solution derived at the one-loop level [top left panel of Fig.~\ref{Fig:alignment_numerical_TB}] is approximately described by \refeq{Eq:alignmentcondition}. The impact of the two-loop improved calculation on the numerical values of this solution is significant and again shifts the $\tb$ values towards larger values. Whereas alignment without decoupling at moderately small values of $\tb\lesssim 10$ is achieved in the one-loop description 
for $\widehat{\mu} \gtrsim 2.2$, $\widehat{A}_t \sim -1.3$ and for $\widehat{\mu} \lsim -2.2$, $\widehat{A}_t \sim 1.3$
(due to the $\widehat{\mu}\to - \widehat{\mu}$, $\widehat{A}_t \to - \widehat{A}_t$ symmetry), the two-loop description pushes these results to higher absolute values of $\widehat{\mu} \gsim 4.2$ and $\widehat{\mu}\lsim -3.4$, respectively. Even lower $\tb$ values $\lesssim 5$ can be obtained by allowing even larger $\widehat{\mu}$ values, as can be seen in \figref{Fig:alignment_numerical_TB}. However, with further increasing $\widehat{\mu}\gg1$, this turns over into the approximate behavior $\tb \simeq |\widehat{\mu}|/\sqrt{6}$, found in the limit $\widehat{\mu}\gg1$ and small $\widehat{A}_t$ of \refeq{Eq:alignmentcondition_Atsmall}, and thus $\tb$ starts to increase with $\widehat{\mu}$. At such large $\widehat{\mu}$ values alignment solutions are also found in the parameter regions with $\widehat{\mu}$ and $\widehat{A}_t$ having the same sign, which feature small values of $\tb\lesssim 5$ (for the $\widehat{\mu}$ range considered here).

The remaining $\tan\beta$ solution, displayed in the middle panels in Fig.~\ref{Fig:alignment_numerical_TB}, has not been discussed previously in the literature (except for some brief comments in our previous work~\cite{Bechtle:2016kui}). It occurs only in the regions where $\widehat{\mu}$ and $\widehat{A}_t$ have the same sign, and only for very large $|\widehat{\mu}| \gtrsim 5~(7-8)$ in the one-loop (two-loop) description. The $\tb$ value of this solution is small at large $|\widehat{A}_t|$, and approaches $+\infty$ as $|\widehat{A}_t|\to 0$. This solution is only found in regions of the parameter space that also exhibit the solution shown in the top panels of Fig.~\ref{Fig:alignment_numerical_TB}, and its $\tb$ values are always larger. Therefore, and because of the very large $\widehat{\mu}$ values required especially after taking into account the two-loop corrections, this alignment solution is phenomenologically not relevant.

\begin{figure}[t!]
\centering
\includegraphics[width=0.44\textwidth,height = 0.29\textheight]{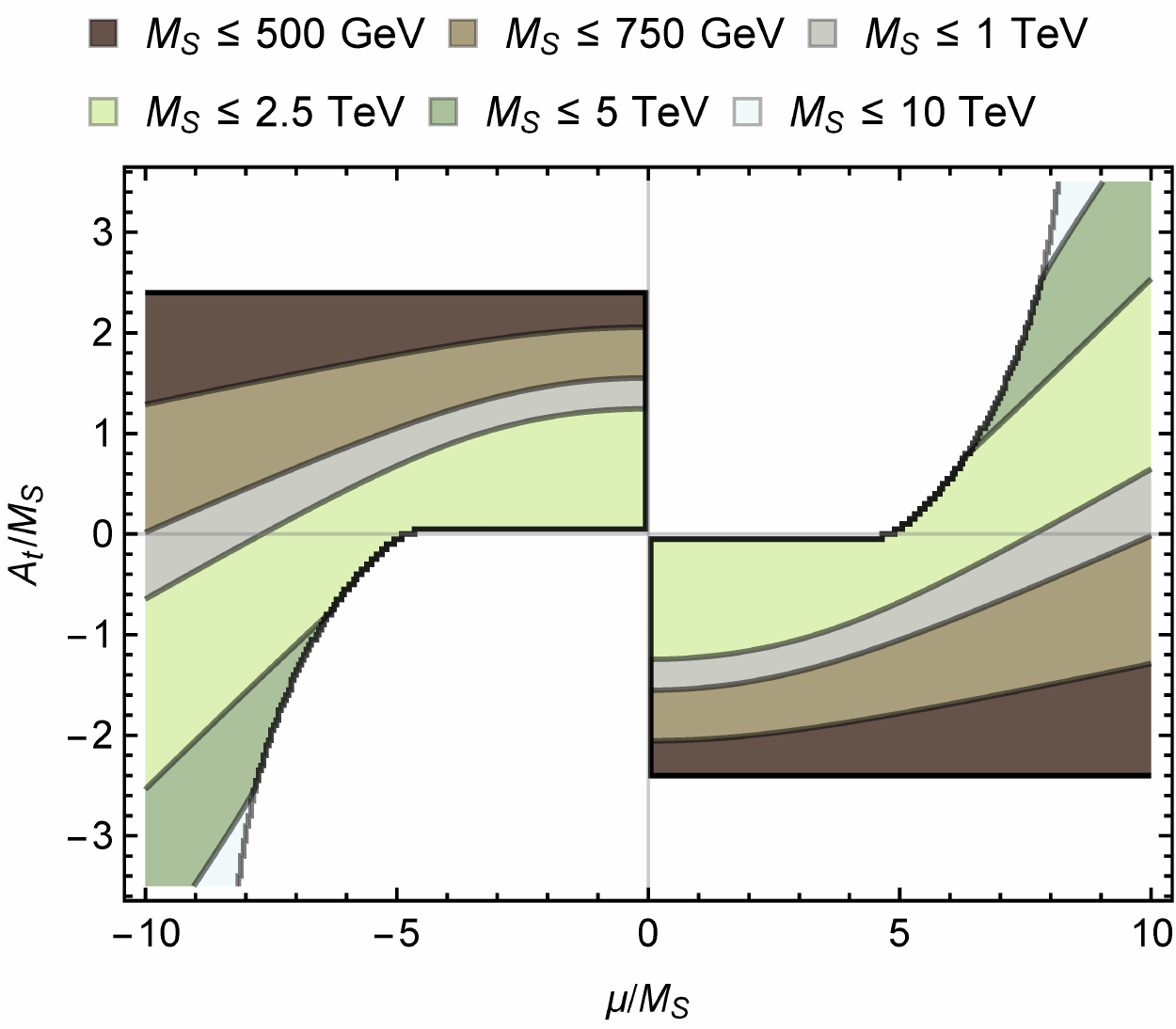}\hfill
\includegraphics[width=0.44\textwidth,height = 0.29\textheight]{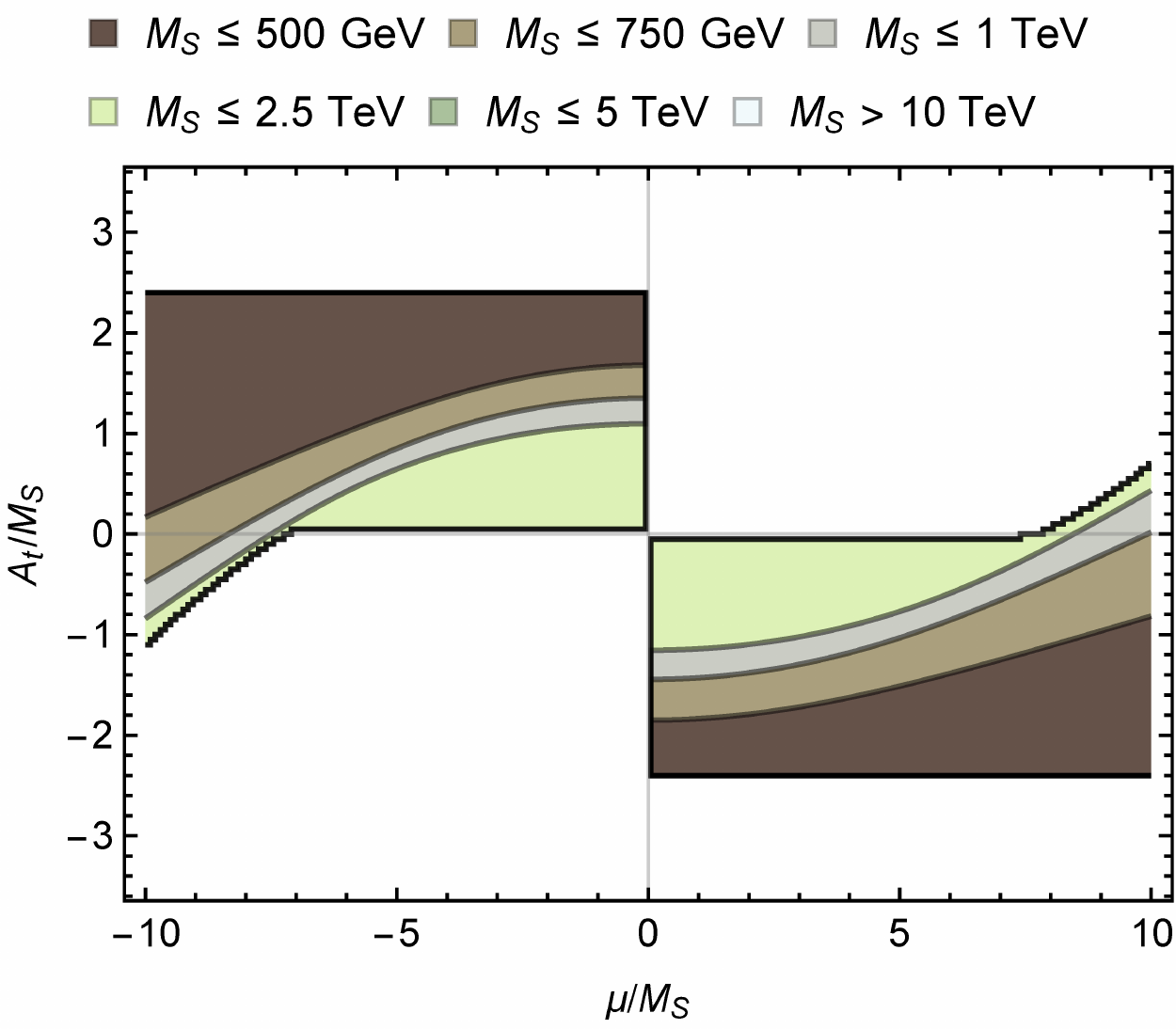}
\includegraphics[width=0.44\textwidth,height = 0.29\textheight]{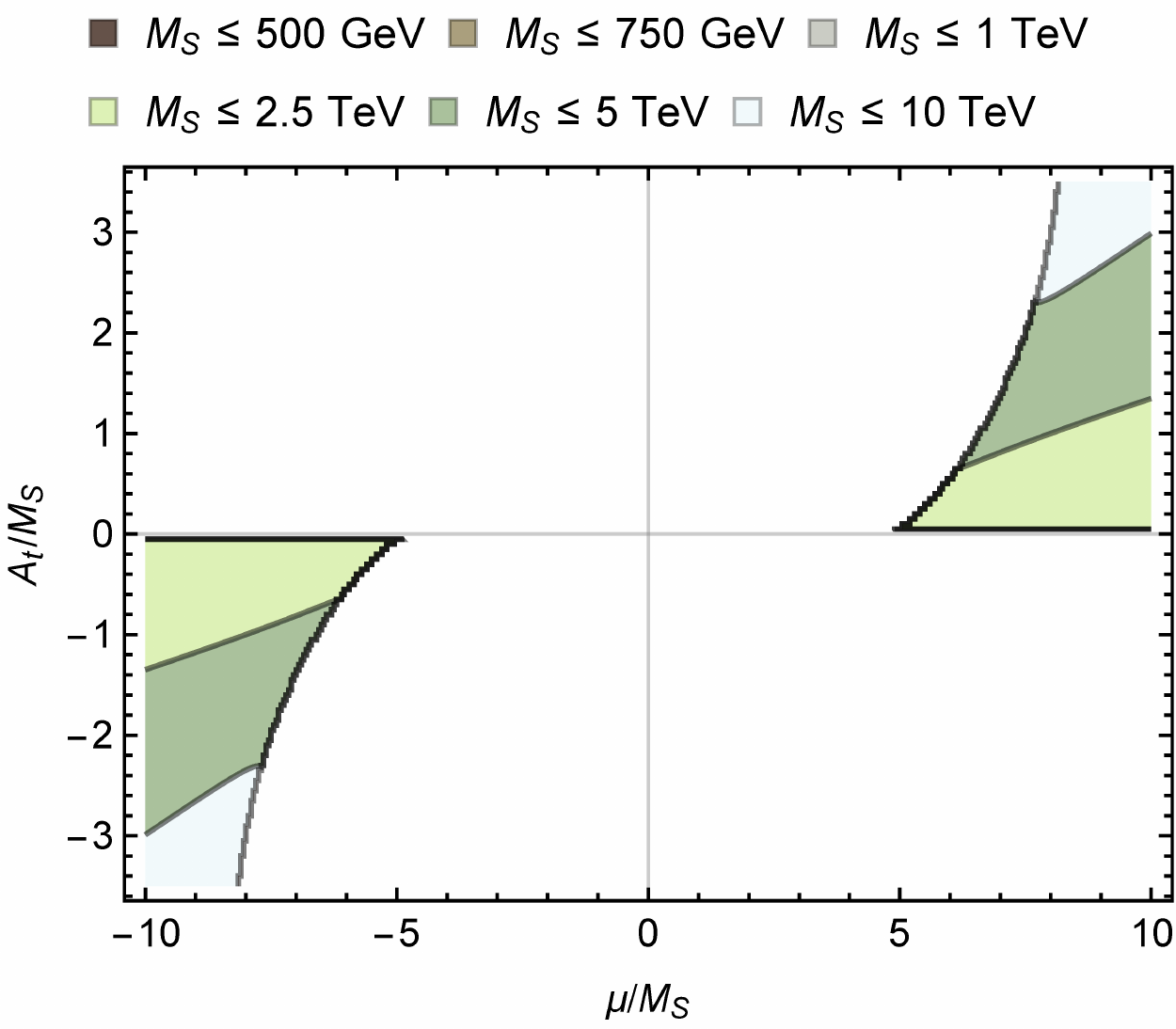}\hfill
\includegraphics[width=0.44\textwidth,height = 0.29\textheight]{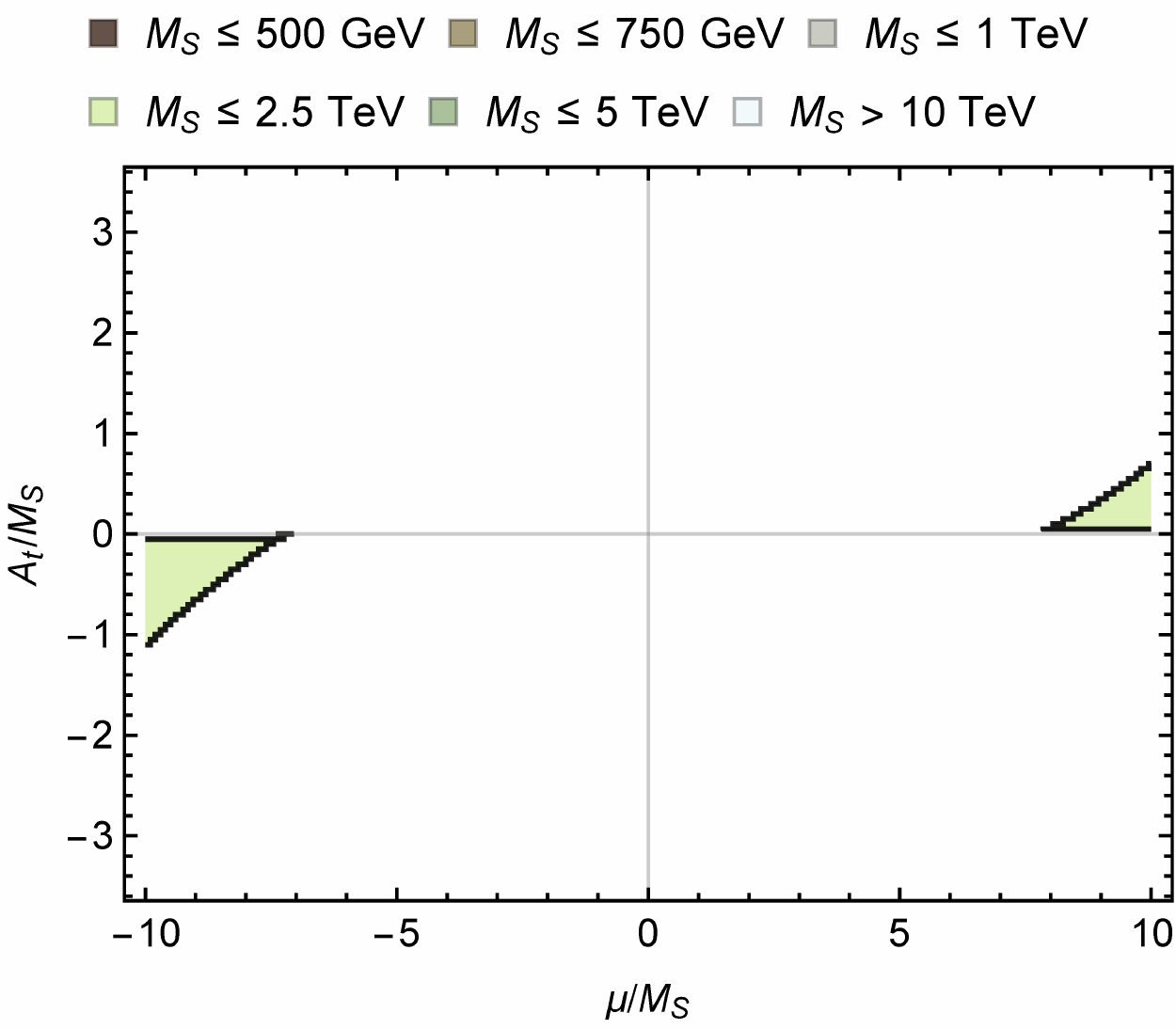}
\includegraphics[width=0.44\textwidth,height = 0.29\textheight]{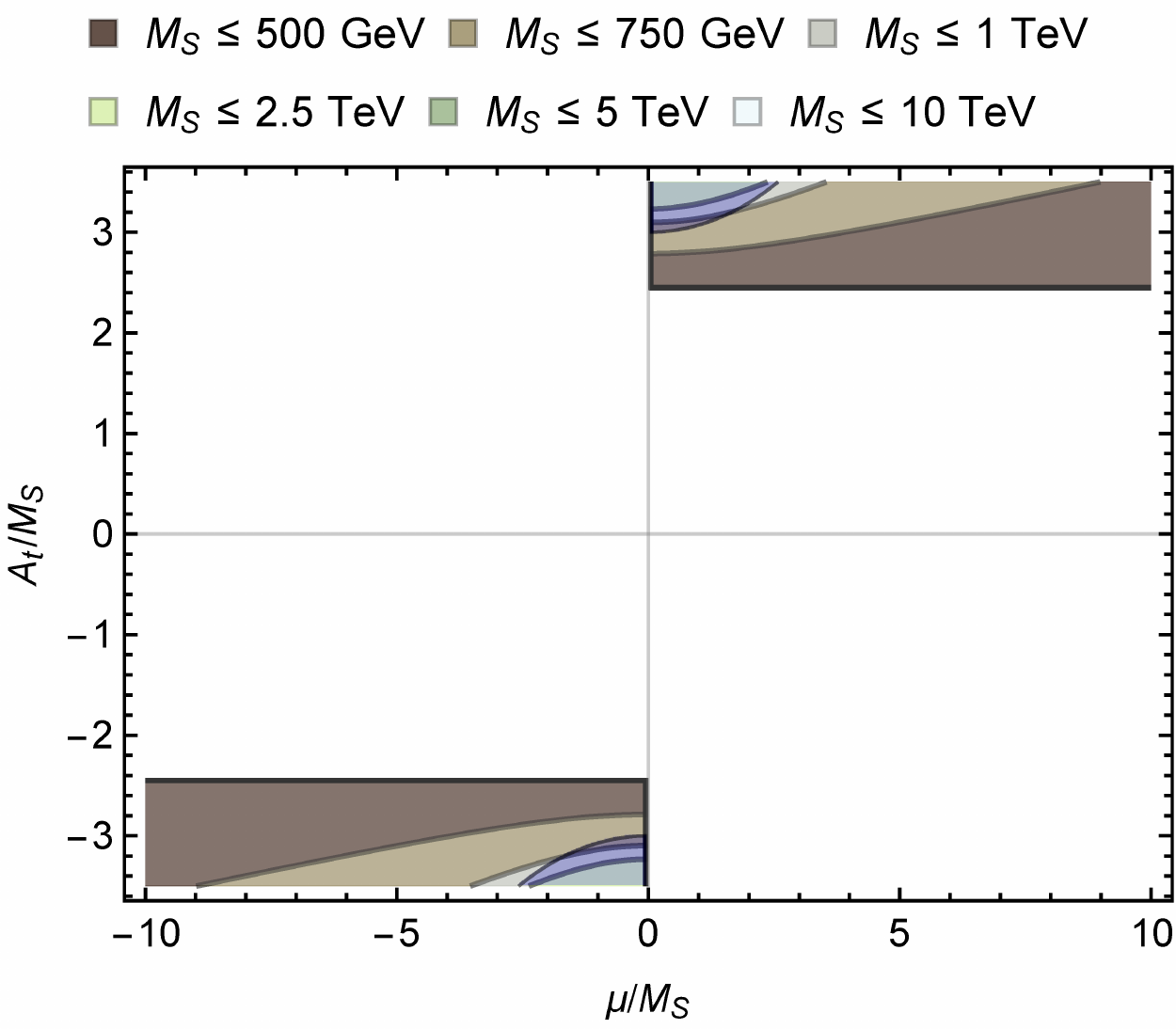}\hfill
\includegraphics[width=0.44\textwidth,height = 0.29\textheight]{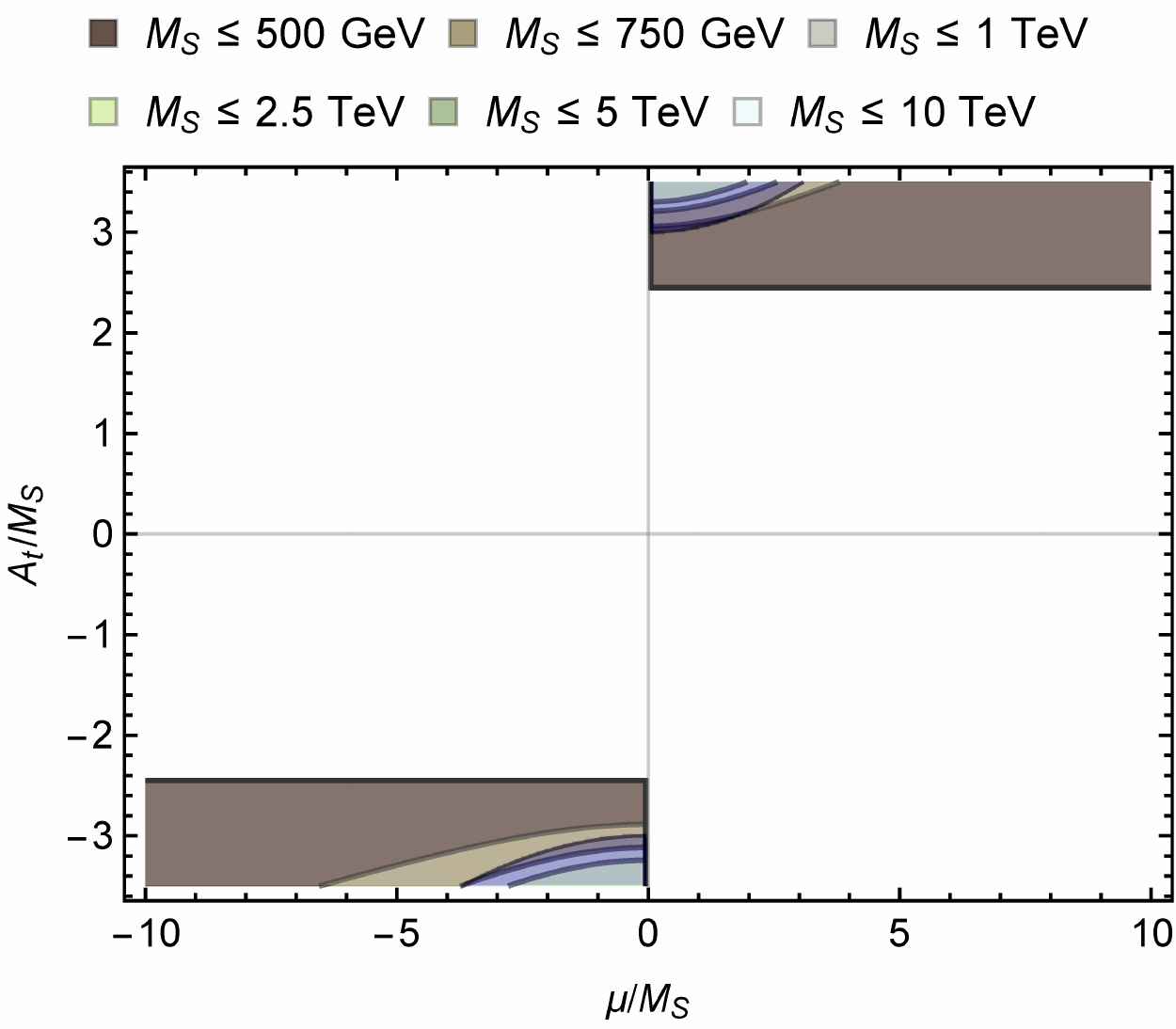}
\caption{$M_S$ value needed to obtain the correct Higgs mass in the limit of exact alignment, corresponding to the solutions found in~\figref{Fig:alignment_numerical_TB} in the $(\mu/M_S, A_t/M_S)$ plane.
Left: Approximate one-loop result; Right: Two-loop improved result. In the overlaid blue regions we have (unstable) values of $|X_t/M_S| \ge 3$.}
\label{Fig:alignment_numerical_MS}
\end{figure}

In our numerical scans in the $(\widehat\mu\,,\,\widehat{A}_t)$ plane, the size of the SUSY-breaking mass scale, $M_S$, varies as required by the condition of exact alignment ($Z_6=0$) such that the SM-like Higgs mass is fixed to its observed value of $125\gev$.
That is, given the value of $t_\beta$ for exact alignment at a point in the $(\widehat\mu\,,\,\widehat{A}_t)$ plane, one can use \eq{zone} [\eq{zeeone}] in the one-loop [two-loop] approximation to determine the value of $M_S$ such that $Z_1 v^2=(125~{\rm GeV})^2$. These $M_S$ values are exhibited in the three rows of \figref{Fig:alignment_numerical_MS}, which are in one-to-one correspondence with the three rows of \figref{Fig:alignment_numerical_TB}, i.e.~each row shows a different solution of the alignment condition, and on the left [right] we show the one-loop [two-loop] result.  We define \textit{maximal mixing} in the top squark sector to correspond to the value of $X_t/M_S$ that maximizes the value of $Z_1 v^2$ given by  \eq{zone} [\eq{zeeone}] in the one-loop [two-loop] approximation, prior to fixing the Higgs mass at its observed value of 125 GeV. In the one-loop approximation, maximal mixing occurs at $X_t/M_S=\sqrt{6}$, and the maximal value of the Higgs mass corresponds to maximal mixing with $\tan\beta\gg 1$.  Turning this around, if we fix the value of the Higgs mass to be its
observed value of 125 GeV, then the minimal value of $M_S$ occurs at maximal mixing with $\tan\beta\gg 1$.  This can be seen in the top panels of \figref{Fig:alignment_numerical_MS}, where the minimal $M_S$ contour is located in the region of $\widehat{A}_t=\sqrt{6}$ and $\widehat{\mu}\widehat{A}_t<0$.  In this region, $\tan\beta\gg 1$ so that
$\widehat{A}_t\simeq X_t/M_S$, corresponding to the region of maximal mixing at large $\tan\beta$.
Maximal mixing at large $\tan\beta$ is also evident in the bottom panels of \figref{Fig:alignment_numerical_MS}.
At smaller values of $|\widehat{A_t}|$, it is still possible to reach maximal mixing at large values of $|\widehat{\mu}|$, albeit with smaller values of $\tan\beta$ shown in \figref{Fig:alignment_numerical_TB}.  In contrast, maximal mixing is never reached in the middle panels of \figref{Fig:alignment_numerical_MS}, as the corresponding $X_t/M_S$ values are closer to the minimal mixing value of $X_t=0$.  In this case, the smaller values of $M_S$ are associated with the larger values of $\tan\beta$, which occur when $|\widehat{A}_t|\to 0$.

\afterpage{\clearpage}

Lastly, we turn to the question whether the light or the heavy neutral CP-even Higgs boson is the state that is aligned with the SM Higgs vev. Recall that the answer depends on the relative size of $Z_1 v^2$ and $Z_5 v^2 + M_A^2$. In the end of Section~\ref{Sec:alignment1L} we defined a critical and a minimal $M_A$ value, $M_{A,c}$ and $M_{A,m}$ [see \refeq{macrit} and \refeq{mamin}], respectively, such that $h$ is SM-like for the parameter points with $M^2_A> M^2_{A,c}$ and $H$ is SM-like for the parameter points with $M_{A,m}^2<M^2_A<M^2_{A,c}$. 

\begin{figure}[t!]
\centering
\includegraphics[width=0.44\textwidth,height = 0.29\textheight]{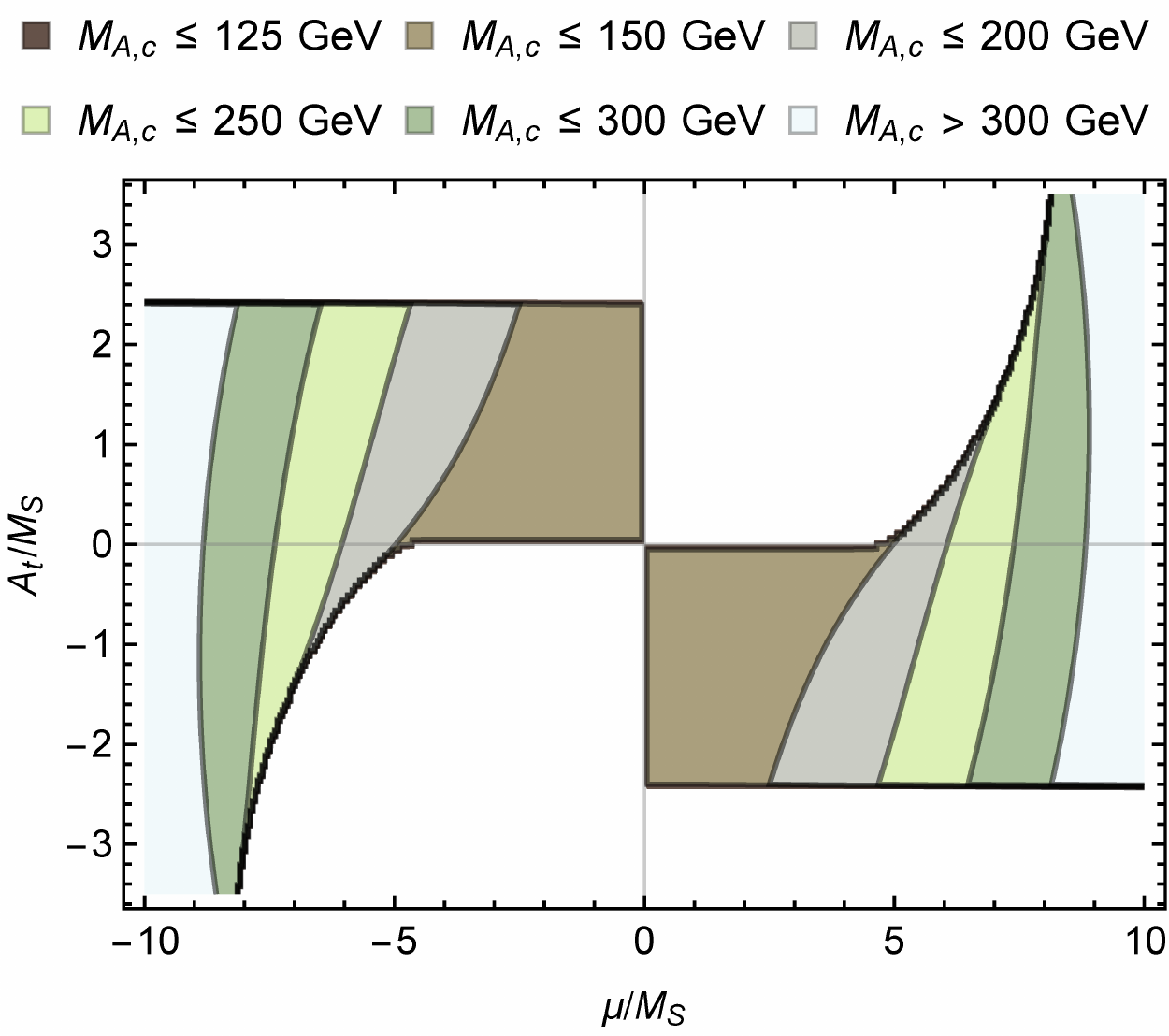}\hfill
\includegraphics[width=0.44\textwidth,height = 0.29\textheight]{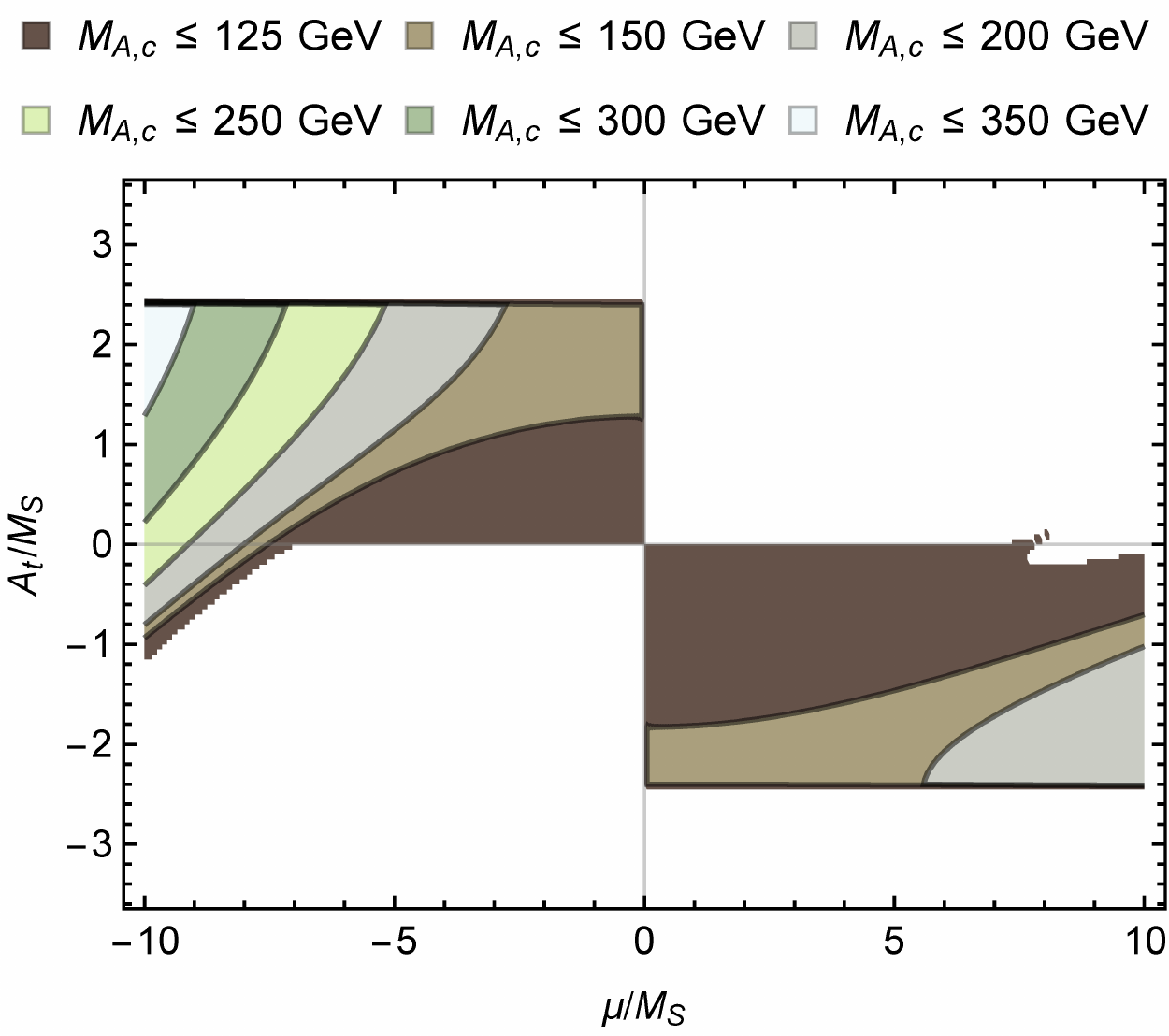}
\includegraphics[width=0.44\textwidth,height = 0.29\textheight]{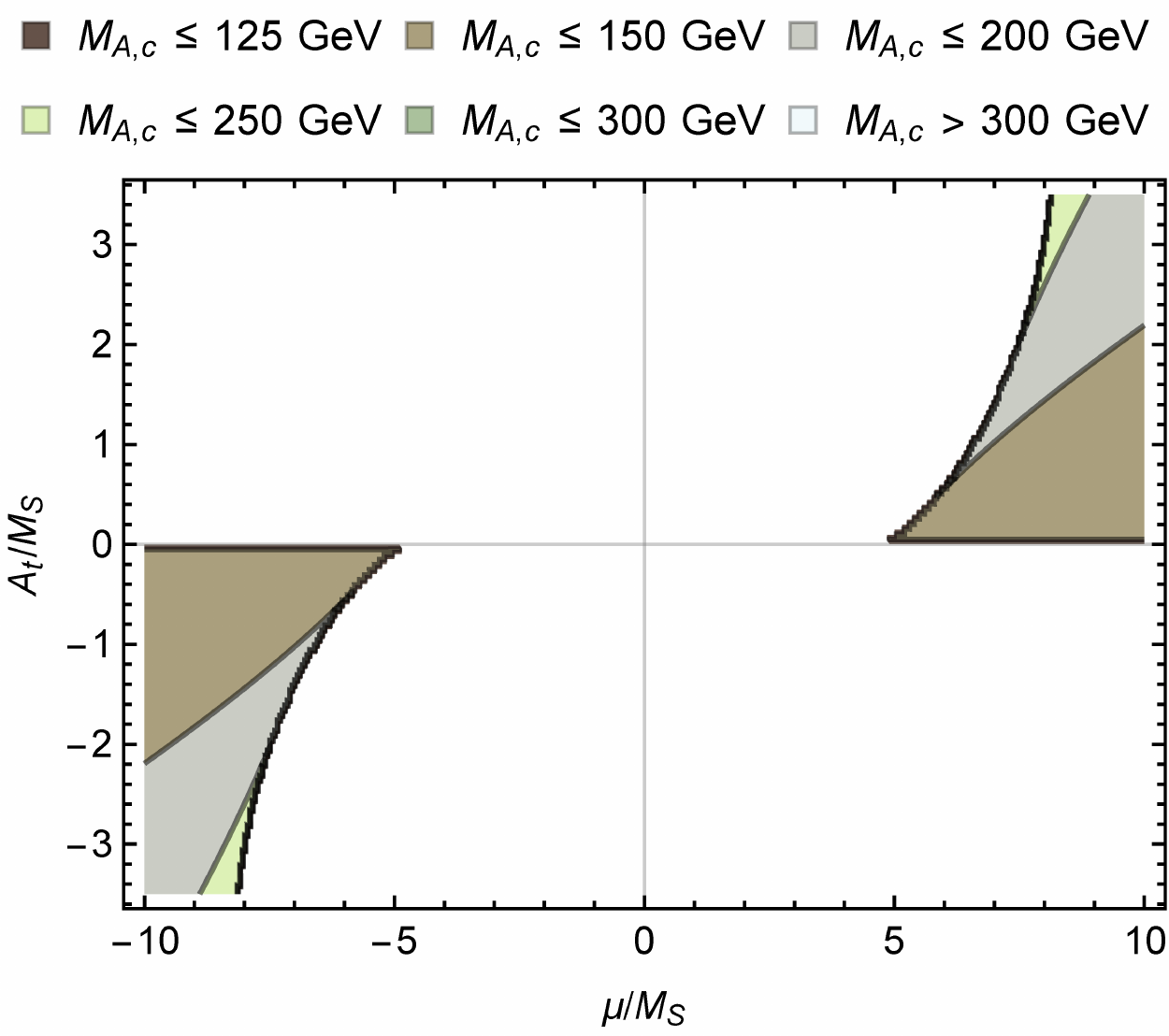}\hfill
\includegraphics[width=0.44\textwidth,height = 0.29\textheight]{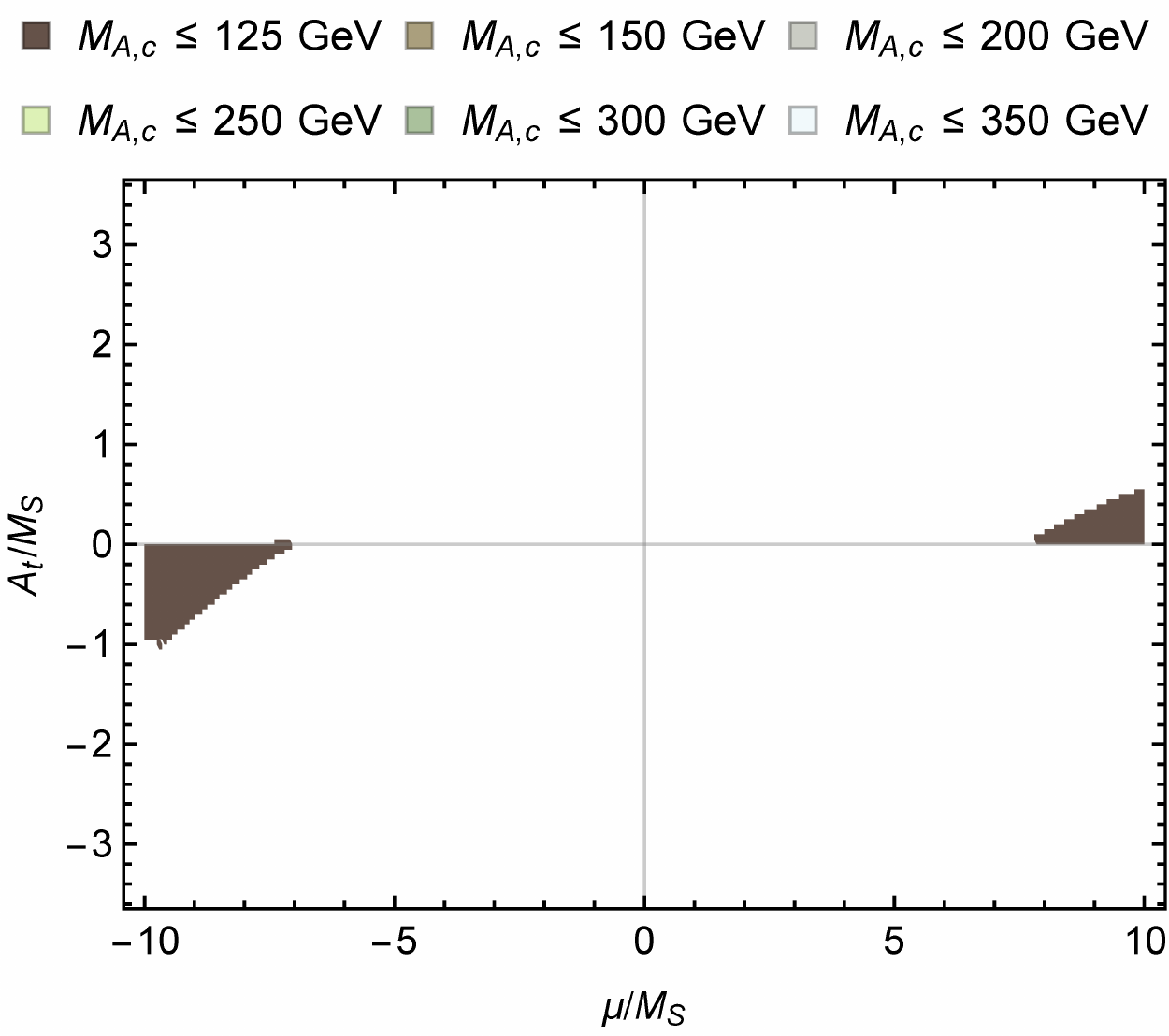}
\includegraphics[width=0.44\textwidth,height = 0.29\textheight]{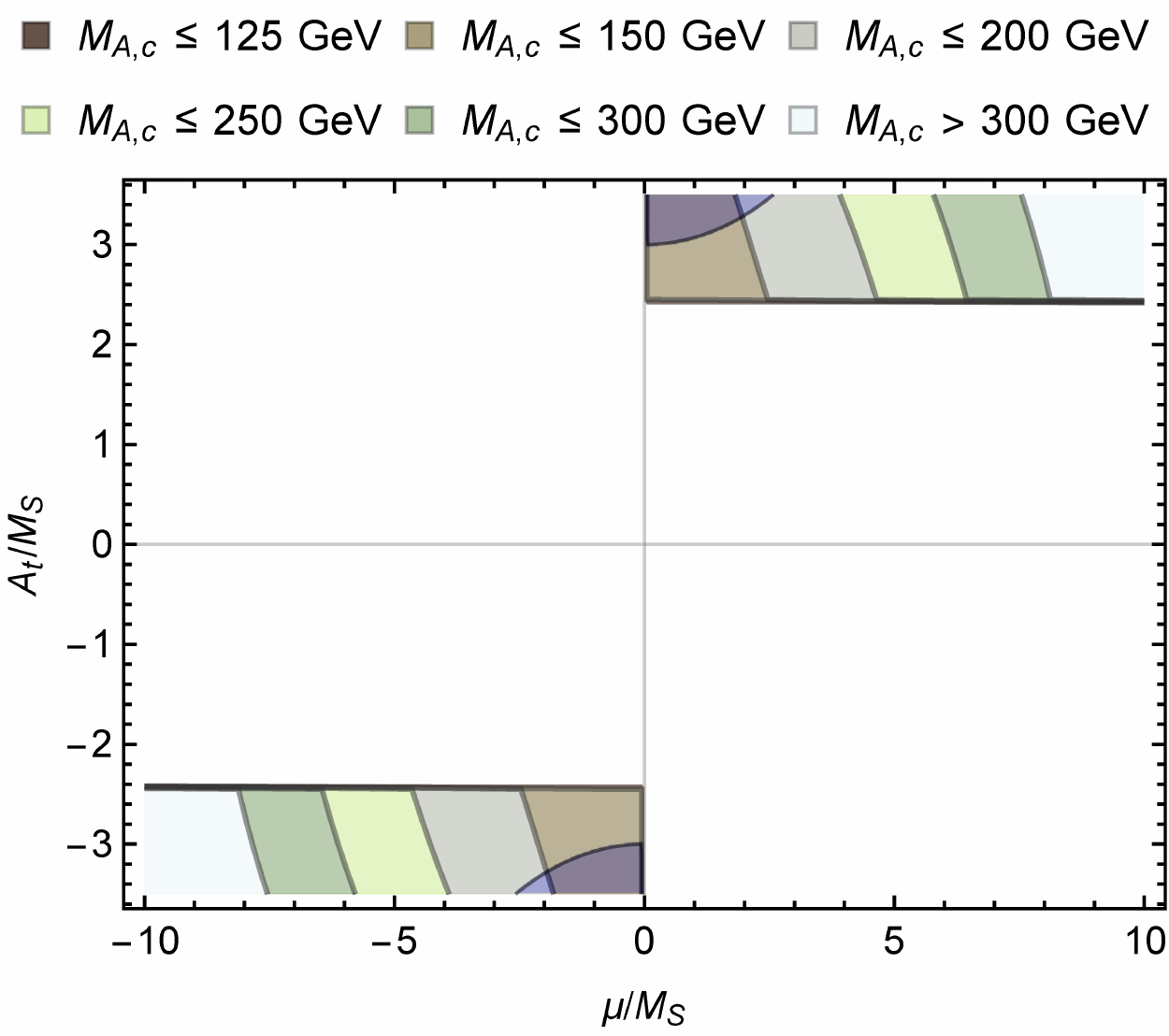}\hfill
\includegraphics[width=0.44\textwidth,height = 0.29\textheight]{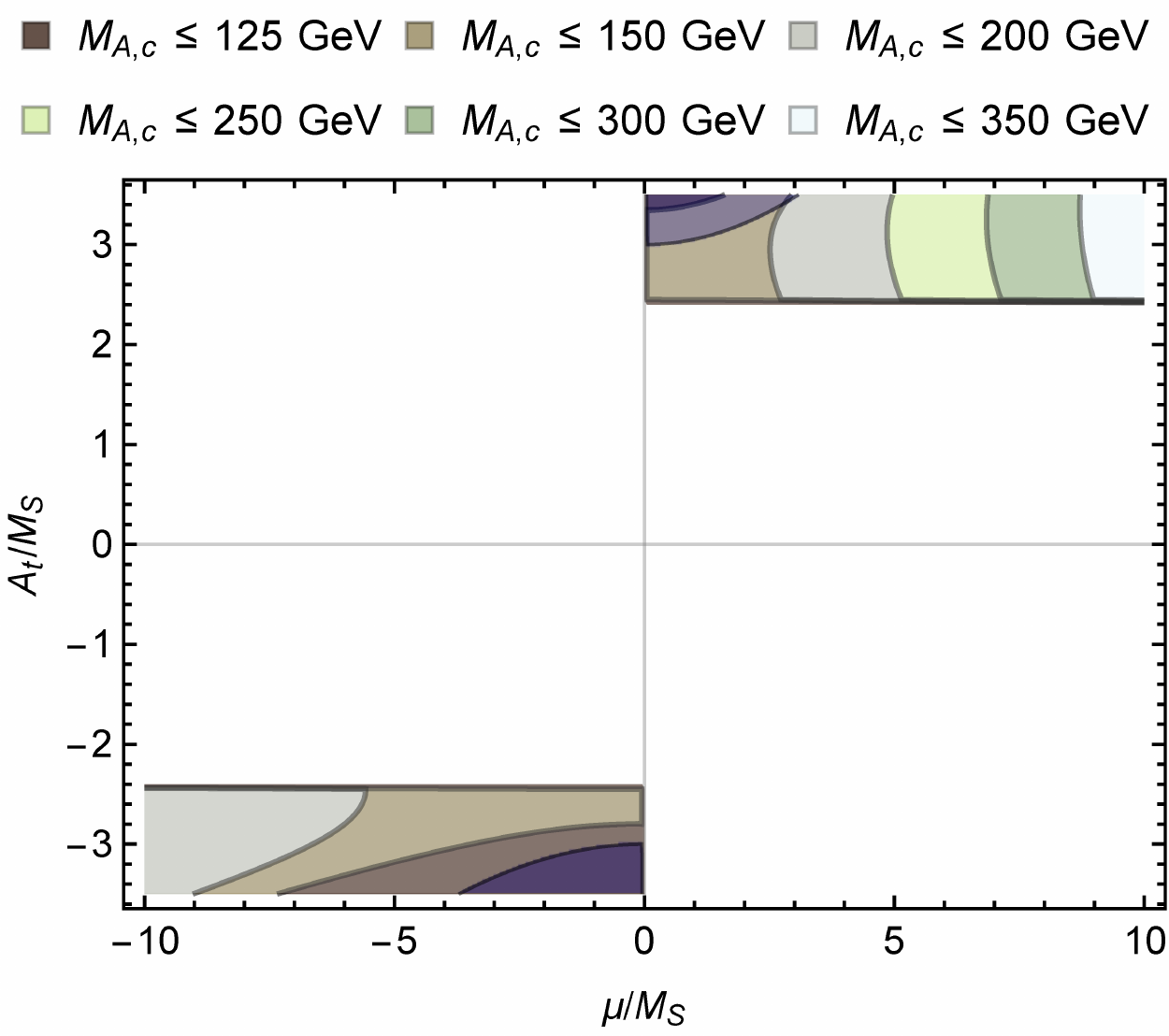}
\caption{Critical $M_A$ value, $M_{A,c}$, in the exact alignment, indicating the maximal $M_A$ value for which the mass hierarchy of the heavy Higgs interpretation is obtained, corresponding to the solutions found in~\figref{Fig:alignment_numerical_TB} in the $(\mu/M_S, A_t/M_S)$ plane.
Left: Approximate one-loop result; Right: Two-loop improved result. In the overlaid blue regions we have (unstable) values of $|X_t/M_S| \ge 3$.
}
\label{Fig:alignment_numerical_MAc}
\end{figure}

We can compute $Z_5$ at one-loop [two-loop] accuracy from \eq{zfive} [\eq{zeefive}] using the value of $\tb$ for which exact alignment without decoupling occurs. This allows us to determine the value of $M_{A,c}^2$ for each point in the
$(\widehat{\mu}\,,\widehat{A}_t$) plane. The corresponding contours of $M_{A,c}$ are exhibited in the three rows of \figref{Fig:alignment_numerical_MAc}, which are in one-to-one correspondence with the three rows of Figs.~\ref{Fig:alignment_numerical_TB} and \ref{Fig:alignment_numerical_MS}. Again, we show the one-loop [two-loop] results on the left [right]-hand side of \figref{Fig:alignment_numerical_MAc}. For the two phenomenologically relevant alignment solutions, displayed in the top and bottom panels in \figref{Fig:alignment_numerical_MAc}, we observe that $M_{A,c}$ generally increases with $|\widehat{\mu}|$. In the alignment solution shown in the top panel, a slight increase of $M_{A,c}$ can also be noted with $|\widehat{A}_t|$. At the two-loop level, the asymmetry between the relative signs of $\widehat{\mu}$ and $\widehat{A}_t$ introduced by the finite threshold corrections proportional to $X_t$ are quite noticeable in these figures. Furthermore, the two-loop corrections lead to a sizable shift of $M_{A,c}$ towards lower values in the entire parameter space, thus narrowing the available parameter space that can feature a heavy SM-like Higgs boson at $125\gev$.

Under the assumption that the heavy CP-even Higgs boson $H$ is identified with the observed Higgs boson at $125\gev$, the $M_{A,c}$ values can easily be translated into upper bounds on the charged Higgs boson mass, $M_{H^+}$, according to Eqs.~\eqref{mhpm1L} and \eqref{mhpm2L} in the leading one- and two-loop description, respectively. Here one should keep in mind that the leading radiative corrections are negative and proportional to $\widehat{\mu}^2$. Consequently, at large $\widehat{\mu}$, these radiative corrections can substantially decrease the $M_{H^+}$ prediction with respect to its tree-level prediction, $M_{H^+}^\text{tree} = \left( M_{A}^2 + M_{W}^2\right)^{1/2}$. Collider and flavor constraints on such scenarios arising from a light charged Higgs boson have been extensively discussed in Ref.~\cite{Bechtle:2016kui} (see also Ref.~\cite{Arbey:2017gmh} for a similar analysis in the framework of the 2HDM).

We close this section with a few comments on how these results compare with the numerical fit results found in Ref.~\cite{Bechtle:2016kui}. In the global fit, Ref.~\cite{Bechtle:2016kui} identified two distinct parameter regions with phenomenologically viable points near the limit of alignment without decoupling. These regions resemble the parameter regions that are exhibited in the top and bottom panels in Figs.~\ref{Fig:alignment_numerical_TB}-\ref{Fig:alignment_numerical_MAc}. Specifically, under the assumption that $h$ is identified as the SM-like Higgs boson at $125\gev$, the preferred points with low $M_A$ (i.e.~in the non-decoupling regime) found in Ref.~\cite{Bechtle:2016kui} were located near the alignment solution displayed in the bottom panels. The main reason for this, however, is the restriction $|\mu|/M_S \le 3$ imposed in the fit for the light Higgs interpretation, which essentially excludes the other possible alignment solutions identified in this work. In contrast, assuming that $H$ is identified as the SM-like Higgs boson at $125\gev$, Ref.~\cite{Bechtle:2016kui} found viable points only near the parameter regions displayed in the top panels of Figs.~\ref{Fig:alignment_numerical_TB}-\ref{Fig:alignment_numerical_MAc}. Here, the restriction $|\mu|/M_S \le 3$ was not imposed, and the main reason for this observation was a coupling suppression of the charged Higgs contribution to the branching fraction of the $B$ meson decay $B \to X_s \gamma$, thus yielding phenomenologically acceptable values despite the presence of a very light charged Higgs boson. Lastly, a word of caution is in order: the numerical results displayed in this work are based on the \emph{exact} alignment limit, whereas in the global fit studies of the MSSM parameter space in the non-decoupling regime, the parameter points only need to be \emph{near} the alignment limit in order to be phenomenologically viable. 
In particular, Ref.~\cite{Bechtle:2016kui} quantified the maximal values of $|Z_6|/Z_1$ for the parameter points allowed at the $2\sigma$ level in the light Higgs (with low $M_A$) and heavy Higgs interpretation, resulting in $\sim 0.3$ and $\sim 0.2$, respectively. Such values indicate that these parameter regions are not yet parametrically fine-tuned, and non-negligible deviations from the alignment limit are still allowed by the current data.\footnote{For instance, in the global fit of Ref.~\cite{Bechtle:2016kui} employing the heavy Higgs interpretation, $M_A$ values up to around $180\gev$ were found to be viable, whereas in the \emph{exact} alignment limit we find $M_{A,c} \le 125\gev$ in the corresponding parameter region, as shown in the top right panel of Fig.~\ref{Fig:alignment_numerical_MAc}.}

\afterpage{\clearpage}

\section{SM-like Higgs branching ratios in the alignment limit}
\label{sec:Hhh}

In the exact alignment limit, the tree-level couplings of the SM-like Higgs boson are precisely those of the Higgs boson of the SM.  
Nevertheless, in the case of alignment without decoupling,
deviations from SM Higgs boson properties can arise because the effective
theory at the electroweak scale contains additional fields beyond the fields
of the SM.  In this work, we have employed the framework of the MSSM under the assumption that the SUSY-breaking scale $M_S\gg M_Z$, $M_{H^\pm}$.
Thus the effective electroweak theory at energy scales below $M_S$ is the 2HDM.   Moreover, SUSY-breaking effects can generate so-called wrong-Higgs couplings with coefficients that in some cases are $\tan\beta$-enhanced (see footnote~\ref{fn}).  Thus, we are led to consider the 2HDM with the most general Higgs-fermion Yukawa interactions as the effective theory below $M_S$.   In particular, the masses of the additional scalar states are assumed to be of the same order as the scale of electroweak symmetry breaking.
In the exact alignment limit, deviations of the SM-like Higgs boson branching ratios from the corresponding SM predictions can arise due to two possible effects: (i) new loop-contributions due to the exchange of non-SM Higgs scalars that modify partial decay rates, and (ii) new decay channels in which the SM-like Higgs boson decays into a pair of lighter scalars, if kinematically allowed.

If new tree-level Higgs decays are present, these will typically yield the dominant contributions to the deviations of the Higgs branching ratios from their SM values.  In particular, we expect that any additional deviations that arise from the exchange of non-SM Higgs scalars (which compete with the SM loop corrections) would result only in small shifts of the Higgs decay rates away from their corresponding SM predictions, and will be difficult to isolate experimentally.
In contrast, consider the loop-induced Higgs couplings to $\gamma\gamma$ and $Z\gamma$, which have no tree-level counterpart.  In this case,
new loop corrections due to charged Higgs exchange can compete with the corresponding SM loop contributions, since by assumption $M_{H^\pm}$
does not differ appreciably from the mass of the SM-like Higgs boson~\cite{Bernon:2015wef}.
In practice, due to the domination of the $W$-loop contribution to the loop-induced Higgs couplings to $\gamma\gamma$ and $Z\gamma$ relative to the fermion and scalar loop contributions, the shift in the loop-induced Higgs couplings from their SM values due to the contribution of charged Higgs exchange will typically be small.

The most significant deviation from SM Higgs branching ratios in the alignment limit without decoupling arises if new decay channels are present in which 
the SM-like Higgs boson decays into a pair of lighter scalars.  In
Ref.~\cite{Bechtle:2016kui}, we demonstrated that regions of the MSSM
parameter space in which the heavier of the CP-even scalars, $H$, is SM-like
and $m_h< m_H/2$ are still allowed after taking into account the experimental
constraints from SUSY particle searches and the measurement of Higgs boson
properties at the LHC.  In such a scenario, the decay mode $H\to hh$ is
kinematically allowed, which has an impact on the predicted SM Higgs branching ratios.   

At tree-level, the $Hhh$ coupling survives in the exact alignment limit where $s_{\beta-\alpha}=0$.  Indeed, 
when expressed in terms of the coefficients of the scalar potential in the Higgs basis, the tree-level $Hhh$ coupling is given by~\cite{Gunion:2002zf,Bernon:2015wef,Bernon:2015qea}
\beq
 g\ls{Hhh} = -{3v}\bigl[
   Z_1\cbma\sbmaii
   +Z_{345}\cbma\left(\tfrac{1}{3}-\sbmaii\right)-Z_6\sbma(1-3\cbmaii)
    -Z_7\cbmaii\sbma\bigr]\,, 
    \eeq
where
\beq \label{z345}
Z_{345} \equiv Z_3 + Z_4 + Z_5\,.
\eeq    
Thus, in the alignment limit, $g\ls{Hhh} \rightarrow  -v Z_{345}$.

In the one-loop corrected MSSM in the limit of $M_Z$, $M_A\ll M_S$, we make use of the results of Ref.~\cite{Haber:1993an}
to obtain,\footnote{One can obtain \eq{zeethreefourfive} from the radiatively corrected expressions for $Z_3$, $Z_4$ and $Z_5$ given in Appendix A of
Ref.~\cite{Carena:2015moc} after setting $\lambda=0$.}
\beq \label{zeethreefourfive}
Z_{345}v^2=M_Z^2(3s_{2\beta}^2-1)+\frac{9m_t^4\cot^2\beta}{2\pi^2 v^2}\left[\ln\left(\frac{M_S^2}{m_t^2}\right)+\frac{X_t^2+Y_t^2+4X_t Y_t}{6M_S^2}-\frac{X_t^2 Y_t^2}{12M_S^4}\right]\,.
\eeq
Including the approximate leading $\order{\alpha_s m_t^2 h_t^2}$ corrections we obtain
\beq \label{z345twoloop}
Z_{345} v^2  =   M_Z^2 ( 3s_{2\beta}^2 - 1) + \frac{s_{2\beta}^2}{4s_\beta^4} C \left[ 3L(1- 2\overline{\alpha}_s L  +\overline{\alpha}_s ) + (2 X_{34} + X_5)(1 -4\overline{\alpha}_s L +\tfrac{4}{3}\overline{\alpha}_s x_t) \right],
\eeq
where $X_{34} \equiv \tfrac{1}{4}(x_t + y_t)^2 - \tfrac{1}{12} x_t^2 y_t^2$ and $X_5 \equiv x_t y_t (1 - \tfrac{1}{12} x_t y_t)$ 
and the other relevant quantities have been defined in \refeq{Eq:definequantities}. 

\begin{figure}[t!]
\centering
\includegraphics[width=0.48\textwidth]{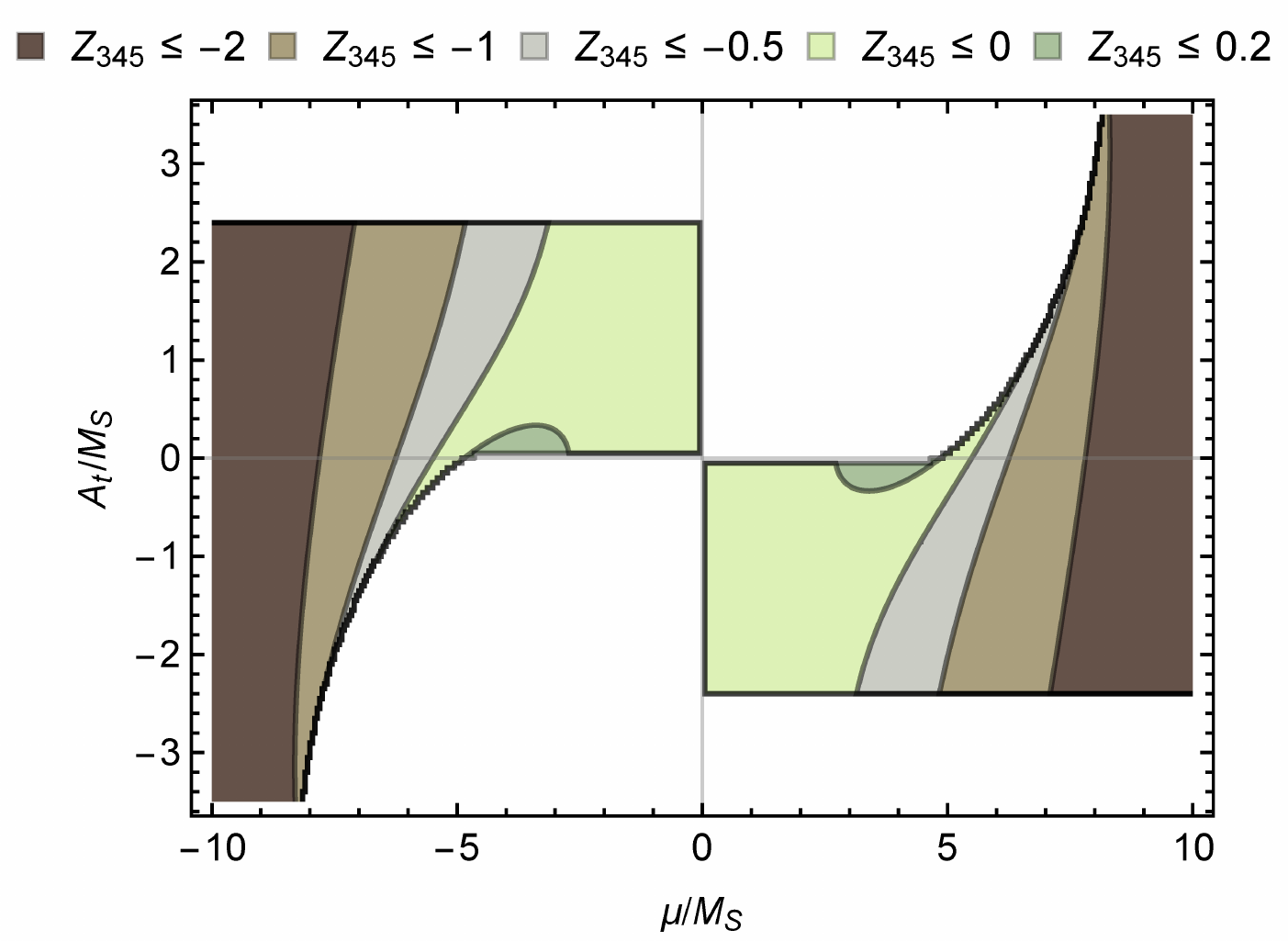}\hfill
\includegraphics[width=0.48\textwidth]{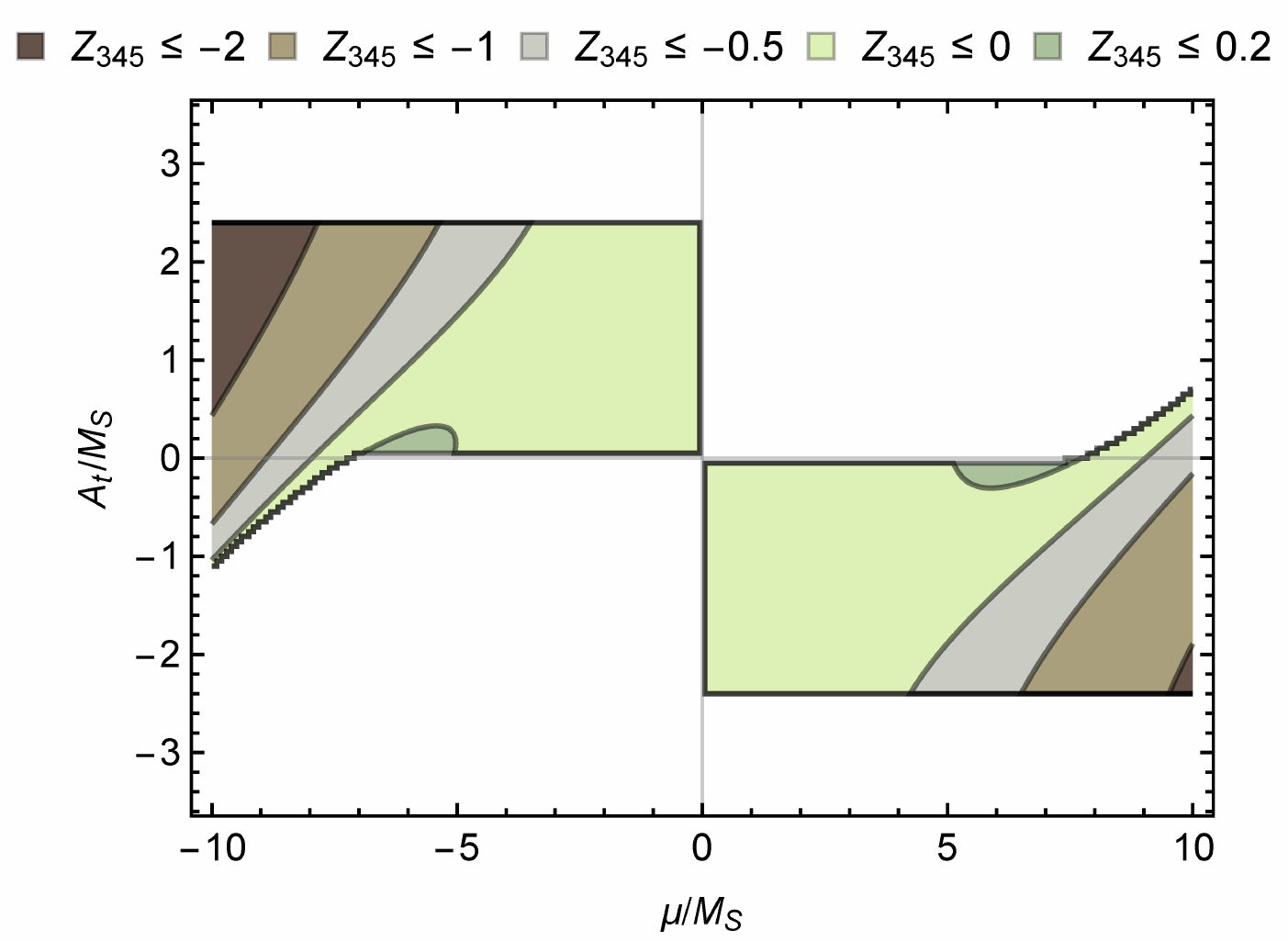}
\caption{Contours of $Z_{345}$ in the exact alignment, corresponding to the strength of the $Hhh$ coupling in the heavy Higgs interpretation. These results correspond to the first alignment solution displayed in the top panels of \figref{Fig:alignment_numerical_TB} in the $(\mu/M_S, A_t/M_S)$ plane. Left: Approximate one-loop result; Right: Two-loop improved result.}
\label{Fig:alignment_numerical_Z345}
\end{figure}

In the left and right panel of \figref{Fig:alignment_numerical_Z345} we show the contours of $Z_{345}$ for the first alignment solution [shown in the top panels in \figref{Fig:alignment_numerical_TB}] derived in the one-loop and two-loop  description, respectively. For most of the parameter space $Z_{345}$ is negative, with large negative values found for large $|\widehat{\mu}|$ values. A small parameter region with small $|\widehat{A}_t| \lesssim 0.4$ and $|\widehat{\mu}|$ between 2.5 and 5 in the one-loop [5 and 7 in the two-loop] description exhibits small positive $Z_{345}$ values, shown by the dark green color in \figref{Fig:alignment_numerical_Z345}. At the boundary between the light and dark green region the coupling $Z_{345}$ vanishes. Thus, in these regions the decay $H\to hh$ becomes coupling-suppressed and the branching fraction can even become zero, irrespective of the available phase-space. This feature has been numerically observed in Ref.~\cite{Bechtle:2016kui} and in particular exploited in Ref.~\cite{Profumo:2016zxo} in the definition of low mass light Higgs boson benchmark scenarios for dark matter studies.

The other two alignment solutions [middle and bottom panels in \figref{Fig:alignment_numerical_TB}] do not exhibit phenomenologically relevant parameter regions where the coupling $Z_{345}$ vanishes, and thus we do not exhibit them here.\footnote{In the second solution [middle panels in \figref{Fig:alignment_numerical_TB}], 
one can achieve a vanishing $Z_{345}$ in a small parameter strip in the second solution [middle panels in \figref{Fig:alignment_numerical_TB}], which we have discarded as phenomenologically irrelevant.} 
For the second alignment solution [shown in the middle panels in \figref{Fig:alignment_numerical_TB}], the $Z_{345}$ values are positive, whereas in the third alignment solution [shown in the bottom panels in \figref{Fig:alignment_numerical_TB}], the $Z_{345}$ values are negative, with larger magnitudes found at larger values of $|\widehat{\mu}|$.

\begin{figure}[t!]
\centering
\includegraphics[width=0.44\textwidth]{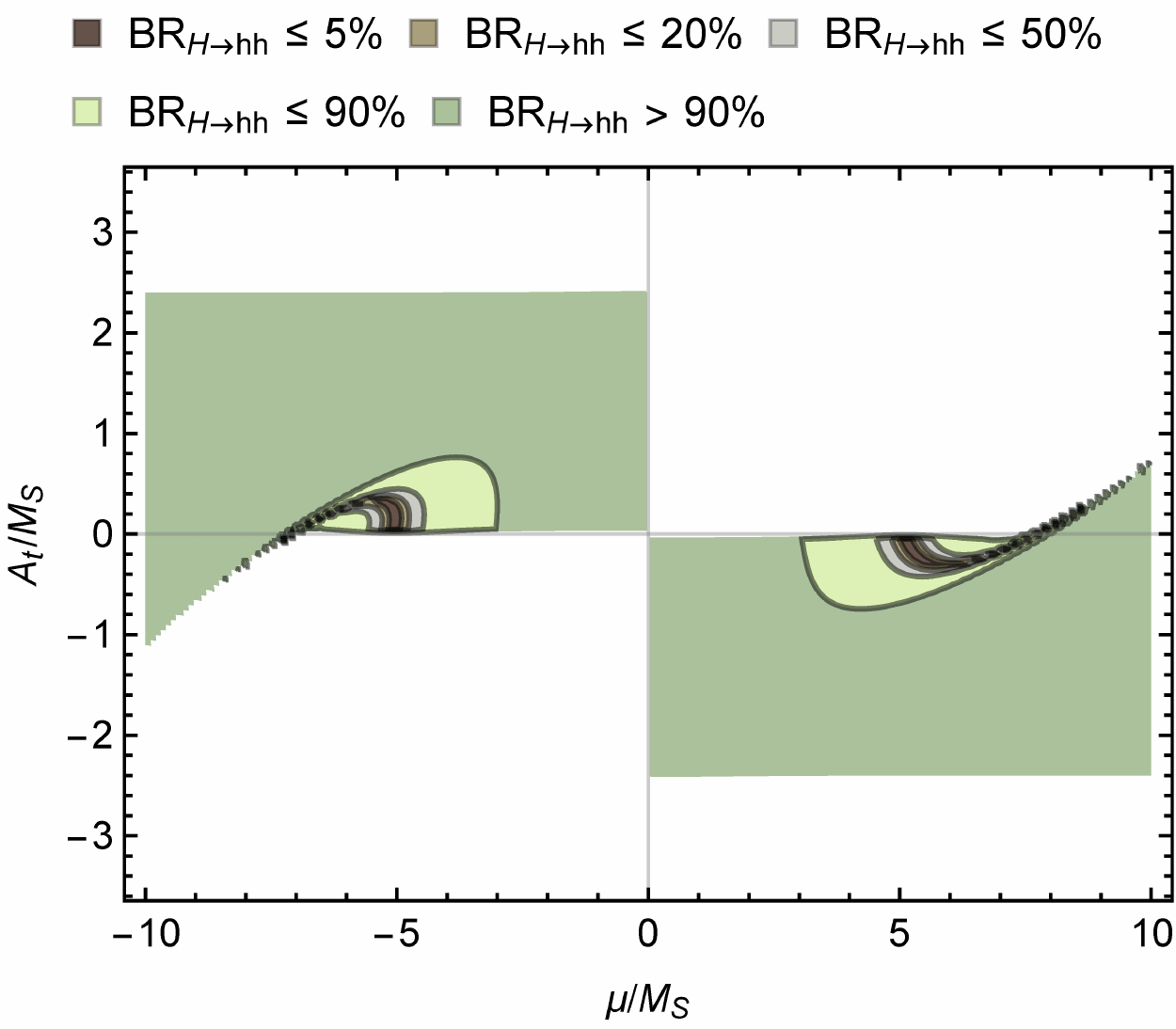}\hfill
\includegraphics[width=0.44\textwidth]{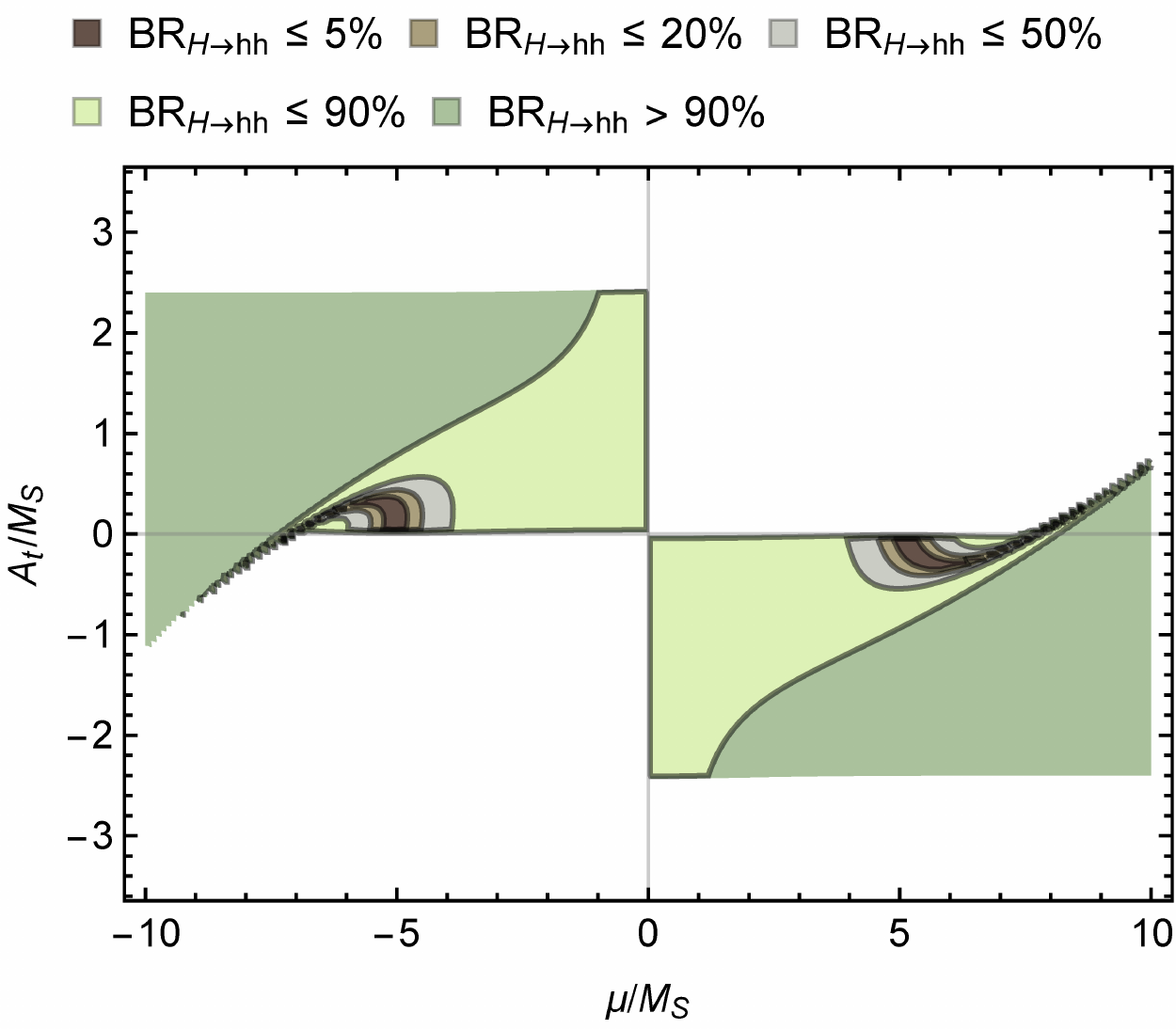}
\caption{Contours of the branching fraction $\mathrm{BR}(H\to hh)$ in the exact alignment in the heavy Higgs interpretation, for the first alignment solution (top panels in~\figref{Fig:alignment_numerical_TB}) in the $(\mu/M_S, A_t/M_S)$ plane. We exhibit only the two-loop improved result here, and assume the light Higgs mass to be $ M_h = 10\gev$ (\emph{left})
 and $60\gev$ (\emph{right}).}
\label{Fig:alignment_numerical_BRHhh}
\end{figure}

In order to further illustrate the feature of a vanishing branching fraction $\mathrm{BR}(H\to hh)$ in the first alignment solution, as mentioned above, we show $\mathrm{BR}(H\to hh)$ for two choices of the light Higgs mass, $M_h = 10\gev$ and $60\gev$, in the left and right panel of \figref{Fig:alignment_numerical_BRHhh}. Here, we exhibit the alignment solution in the two-loop description, and calculate the branching ratio,
\beq
\mathrm{BR}(H\to hh) = \frac{\Gamma (H\to hh)}{\Gamma_\text{tot}^\text{SM} + \Gamma (H\to hh)}\,,
\eeq
where $\Gamma_\text{tot}^\text{SM} = 4.1~\mathrm{MeV}$ is the SM Higgs boson
total decay width~\cite{deFlorian:2016spz}, and%
\footnote{Employing \eq{z345twoloop} for $Z_{345}$ in \eq{hhhdecay} incorporates some of the leading one and two-loop corrections to the $H\to hh$ decay rate.  A more complete one-loop computation can be found in
 \citere{Williams:2011bu}.}
\beqa \label{hhhdecay}
\Gamma (H\to hh) = \frac{Z_{345}^2 v^2}{32 \pi M_H} \left( 1 - \frac{4M_h^2}{M_H^2}\right)^{1/2}\,.
\eeqa

By comparing \figref{Fig:alignment_numerical_BRHhh} with \figref{Fig:alignment_numerical_Z345} one can clearly observe that the branching fraction vanishes in the region where $Z_{345} = 0$. Nevertheless, while such an ``accidental'' parameter constellation leading to a vanishing coupling $Z_{345}$ can occur in this alignment solution, it should be noted that generically the $Hhh$ coupling does not vanish in the alignment limit in the scenario where the heavy CP-even Higgs boson is identified with the observed Higgs boson at $125\gev$. Thus, precision measurements of the properties of the observed Higgs boson can be employed to further constrained the heavy SM-like Higgs scenario.


\section{Conclusions and Outlook}
\label{conclusions}

Given the current precision of Higgs boson measurements at the LHC, the observed state with mass 125 GeV is consistent with a Higgs boson that possesses the spin, CP quantum number and coupling properties predicted by the SM. In an extended Higgs sector of a BSM theory, a scalar mass-eigenstate would possess the properties of the SM Higgs boson if it is aligned in field space with the scalar vacuum expectation value responsible for electroweak symmetry breaking.  This defines the so-called \emph{alignment limit}.  Higgs alignment can be achieved due to the \textit{decoupling} of the non-SM Higgs states, under the assumption that these scalars are considerably heavier than
the SM-like Higgs boson, or via the suppression of
the mixing between the aligned scalar state and the other scalar states of the Higgs sector. In the latter scenario, the 
possibility of \textit{alignment without decoupling} arises if the masses of the non-SM-like Higgs states 
are of the same order of magnitude as the mass of the SM-like Higgs boson.

In the MSSM, the simplest way to achieve approximate Higgs alignment is in the decoupling limit in which $M_A\gg M_h$ and $h$ is identified as the observed SM-like Higgs boson.  In this paper, we have addressed the possibility of achieving approximate Higgs alignment \textit{without} decoupling, which can arise due to
an accidental cancellation of tree-level and loop-level effects, independently of the mass scale of the other Higgs states. Under the assumption that the SUSY-breaking scale, $M_S$, is significantly larger than the masses of the non-SM-like Higgs scalars, the properties of the Higgs sector are well described by an effective two-Higgs doublet extension of the SM (2HDM).
In previous work, alignment without decoupling was achieved due to the cancellation of tree-level and one-loop contributions to the effective 2HDM Lagrangian.\footnote{The analysis of the leading two-loop contributions to the alignment without decoupling scenario given in Section~\ref{Sec:alignment2L} has already been employed in our previous numerical study of the allowed MSSM parameter space~\cite{Bechtle:2016kui}.} The present work goes beyond previous studies and provides a detailed analysis of the leading two-loop corrections and their impact on the parameter regions that exhibit an exact realization of the alignment without decoupling scenario. In particular, we assessed the leading radiative corrections proportional to the strong coupling constant $\alpha_s$, which first enters at the two-loop level,
by employing an approximation scheme developed in Refs.~\cite{Haber:1996fp,Carena:2000dp}.  This scheme employs an optimal choice of the renormalization scale for the running top quark mass, which captures the leading logs at the two-loop level as well as a significant part of the two-loop corrections proportional to the stop mixing parameter $X_t$.

Taking the observed Higgs boson mass of $125\gev$ as an additional constraint, the alignment condition (i.e., the equation corresponding to exact alignment, independent of the value of $M_A$) can only be fulfilled for a specific value of $\tanb$ and $M_S$ that depends on the location 
in the $(\widehat{\mu}\equiv{\mu}/{M_S},\,\widehat{A}_t\equiv {A_t}/{M_S})$ parameter plane. We discussed all physical solutions of the alignment condition both at the one-loop and two-loop level. We found that up to three physical solutions exist simultaneously, out of which at most two appear to be phenomenologically relevant. Comparing the one- and two-loop approximations we found some significant differences in the number of physical solutions in the $(\widehat{\mu},\,\widehat{A}_t)$ parameter plane.  Nevertheless, the gross qualitative features of the one-loop solutions are maintained in the two-loop improved results.

We presented a detailed numerical comparison of the resulting $\tb$ and $M_S$ values obtained in the one-loop and two-loop approximation in the exact alignment limit.
We found that the two-loop corrections are sizable and lead to significant changes of the phenomenology. In particular, the $\tb$ values are corrected towards larger values, and the SUSY mass scale $M_S$ is corrected towards smaller values, with respect to the corresponding values obtained in the one-loop approximation. Because $\tb$ is a parameter that significantly influences the collider phenomenology of the non-SM Higgs bosons (in particular the CP-odd Higgs boson $A$), the two-loop corrections to the alignment condition cannot be neglected in a detailed phenomenological study of the viability of the alignment without decoupling scenario in the MSSM~\cite{Bechtle:2016kui}. We found that the SUSY-breaking mass scale $M_S$ varies in the $(\widehat{\mu}, \widehat{A}_t)$ plane from below $500 \gev$ up to values in the multi-TeV range.

We furthermore defined a critical mass of the CP-odd Higgs boson, $M_{A,c}$. For parameter points with $M_A$ below this value, the heavy CP-even Higgs boson plays the role of the SM-like Higgs boson at $125\gev$, whereas parameter points with $M_A > M_{A,c}$ feature a SM-like light CP-even Higgs boson. We exhibited numerical results for $M_{A,c}$ for all viable solutions to the alignment condition in the one-loop and two-loop approximations. Again, we noted a significant impact of the two-loop corrections, which in general lead to a substantial downward shift of $M_{A,c}$, thus narrowing the parameter space that exhibits a SM-like heavy Higgs boson~$H$.

In the heavy Higgs interpretation, i.e.~the scenario where the heavy CP-even Higgs boson plays the role of the SM-like Higgs boson at $125\gev$, a new decay mode $H\to hh$ is possible if $m_H>2m_h$. We discussed the magnitude of the relevant triple Higgs coupling and the resulting branching fraction $\mathrm{BR}(H\to hh)$ for two choices of the light Higgs mass, $M_h = 10$ and $60\gev$, in the $(\widehat{\mu}, \widehat{A}_t)$ plane. We find that generically the relevant coupling is unsuppressed in the limit of alignment without decoupling, thus leading to a value of $\mathrm{BR}(H\to hh)$ that is in conflict with the LHC Higgs data. However, in one of the solutions to the alignment condition, the responsible triple Higgs coupling (accidentally) vanishes in certain regions of the parameter space. These regions are found at small $|\widehat{A}_t| \lesssim 0.4$ and large $|\widehat{\mu}|$ values around $3$ to $5$ ($5$ to $7$) in the one-loop (two-loop) description. Parameter points exhibiting this accidental suppression of $\mathrm{BR}(H\to hh)$ in the heavy Higgs interpretation have previously been observed numerically in Refs.~\cite{Bechtle:2016kui,Profumo:2016zxo}.

In the effective 2HDM Lagrangian, exact alignment corresponds to setting the effective Higgs basis parameter $Z_6$ to zero.  Given that \textit{exact} Higgs alignment in the MSSM is achieved by an accidental cancellation between tree-level and loop-level contributions to $Z_6$, the astute reader may object that this scenario is of no interest as it represents a set of measure zero of the MSSM parameter space.  To address this concern, we first note that the present Higgs data implies that the observed state at 125 GeV is consistent with that of the SM Higgs boson with an accuracy that is roughly 20--$30\%$.  Consequently, as long as the parameters of the MSSM Higgs sector yield a result close to the alignment limit, such MSSM parameter regions are presently not ruled out by the Higgs data. 
Indeed, in Ref.~\cite{Bechtle:2016kui}, a detailed numerical scan of the parameter space of a phenomenological MSSM governed by eight parameters revealed the existence of regions in which approximate Higgs alignment without decoupling is satisfied.  In the preferred region, some points with values of $Z_6$ as large as  $|Z_6/Z_1|\sim 0.3$ were within two standard deviations of the best fit point.  Thus, the present Higgs data does not require excessive fine-tuning of the MSSM parameters to achieve approximate Higgs alignment without decoupling. 

The analysis of the exact alignment limit given in this paper provides an understanding of the regions of the MSSM parameters where Higgs alignment without decoupling can occur, and the impact of including or neglecting the leading two-loop effects.  Our analytic approximations include the leading effects proportional to the fourth power of the top quark Yukawa coupling and include leading logarithmic terms (sensitive to the mass scale of SUSY-breaking, $M_S$, arising from the top squark sector), and the leading threshold effects at $M_S$ due to top squark mixing.   However, subdominant effects proportional to the square of the top quark Yukawa coupling, the bottom quark and tau lepton Yukawa couplings, and the electroweak gauge couplings have been neglected, as well as non-leading logarithmic terms and non-leading threshold effects due to bottom squark mixing.  It is straightforward to include such effects analytically (see, e.g. Ref~\cite{Haber:1996fp}).  
Additional corrections not treated in this work include effects arising from higher dimensional operators (ultimately arising from integrating out the heavy SUSY sector) as well as genuine electroweak radiative corrections to the low-energy effective 2HDM. 
Nevertheless, the impact of including all such corrections, while modifying some of the precise details of the cancellation between tree-level and loop-level contributions in achieving exact Higgs alignment, will not change the overall qualitative understanding of the MSSM parameter regime that yields the \emph{approximate} alignment limit without decoupling. 

Further experimental Higgs studies at the LHC will improve the precision of the properties of the 125 GeV Higgs boson, while further constraining or discovering the existence of new scalar states of the extended Higgs sector.   Both endeavors will be critical for providing a more fundamental understanding as to why the observed 125 GeV scalar resembles the SM Higgs boson.


\section*{Acknowledgments}

We are especially grateful to Philip Bechtle, Georg Weiglein and Lisa Zeune for their collaboration at an earlier stage of this work, and their helpful comments on the present manuscript.
The work of H.E.H and T.S. is partly funded by the US Department of Energy, grant number
DE-SC0010107. TS is furthermore supported by a  Feodor-Lynen research fellowship sponsored by the Alexander von
Humboldt foundation.
The work of S.H.\ is supported in part by CICYT (Grant FPA 2013-40715-P),
in part by the MEINCOP Spain under contract FPA2016-78022-P, in part by the
``Spanish Agencia Estatal de Investigaci\'on'' (AEI) and the EU
``Fondo Europeo de Desarrollo Regional'' (FEDER) through the project
FPA2016-78645-P, in part  by the AEI through the grant IFT Centro de Excelencia Severo Ochoa SEV-2016-0597,
and by the Spanish MICINN's Consolider-Ingenio 2010 Program
under Grant MultiDark CSD2009-00064. 

\bibliography{MSSMalignment}

\providecommand{\href}[2]{#2}\begingroup\raggedright\begin{thebibliography}{10}

\bibitem{ATLASDiscovery}
{\bf ATLAS} Collaboration, G.~Aad et~al. {\em Phys.~Lett.} {\bf B716} (2012)
  1--29, [\href{http://arxiv.org/abs/1207.7214}{{\tt arXiv:1207.7214}}].

\bibitem{CMSDiscovery}
{\bf CMS} Collaboration, S.~Chatrchyan et~al. {\em Phys.~Lett.} {\bf B716}
  (2012) 30--61, [\href{http://arxiv.org/abs/1207.7235}{{\tt
  arXiv:1207.7235}}].

\bibitem{Aad:2015zhl}
{\bf ATLAS, CMS} Collaboration, G.~Aad et~al. {\em Phys. Rev. Lett.} {\bf 114}
  (2015) 191803, [\href{http://arxiv.org/abs/1503.07589}{{\tt
  arXiv:1503.07589}}].

\bibitem{Gunion:2002zf}
J.~F. Gunion and H.~E. Haber {\em Phys. Rev.} {\bf D67} (2003) 075019,
  [\href{http://arxiv.org/abs/hep-ph/0207010}{{\tt hep-ph/0207010}}].

\bibitem{Craig:2013hca}
N.~Craig, J.~Galloway, and S.~Thomas \href{http://arxiv.org/abs/1305.2424}{{\tt
  arXiv:1305.2424}}.

\bibitem{Haber:2013mia}
H.~E. Haber in {\em {1st Toyama International Workshop on Higgs as a Probe of
  New Physics 2013 (HPNP2013) Toyama, Japan, February 13-16, 2013}}, 2013.
\newblock \href{http://arxiv.org/abs/1401.0152}{{\tt arXiv:1401.0152}}.

\bibitem{Asner:2013psa}
D.~Asner, T.~Barklow, C.~Calancha, K.~Fujii, N.~Graf, et~al.
  \href{http://arxiv.org/abs/1310.0763}{{\tt arXiv:1310.0763}}. See Chapter
  1.3.

\bibitem{Carena:2013ooa}
M.~Carena, I.~Low, N.~R. Shah, and C.~E.~M. Wagner {\em JHEP} {\bf 04} (2014)
  015, [\href{http://arxiv.org/abs/1310.2248}{{\tt arXiv:1310.2248}}].

\bibitem{Dev:2014yca}
P.~S. Bhupal~Dev and A.~Pilaftsis {\em JHEP} {\bf 12} (2014) 024,
  [\href{http://arxiv.org/abs/1408.3405}{{\tt arXiv:1408.3405}}]. [Erratum:
  JHEP11,147(2015)].

\bibitem{Pilaftsis:2016erj}
A.~Pilaftsis {\em Phys. Rev.} {\bf D93} (2016) 075012,
  [\href{http://arxiv.org/abs/1602.02017}{{\tt arXiv:1602.02017}}].

\bibitem{Haber:1989xc}
H.~E. Haber and Y.~Nir {\em Nucl. Phys.} {\bf B335} (1990) 363--394.

\bibitem{Carena:2014nza}
M.~Carena, H.~E. Haber, I.~Low, N.~R. Shah, and C.~E.~M. Wagner {\em Phys.
  Rev.} {\bf D91} (2015) 035003, [\href{http://arxiv.org/abs/1410.4969}{{\tt
  arXiv:1410.4969}}].

\bibitem{Bernon:2015qea}
J.~Bernon, J.~F. Gunion, H.~E. Haber, Y.~Jiang, and S.~Kraml {\em Phys. Rev.}
  {\bf D92} (2015) 075004, [\href{http://arxiv.org/abs/1507.00933}{{\tt
  arXiv:1507.00933}}].

\bibitem{Bernon:2015wef}
J.~Bernon, J.~F. Gunion, H.~E. Haber, Y.~Jiang, and S.~Kraml {\em Phys. Rev.}
  {\bf D93} (2016) 035027, [\href{http://arxiv.org/abs/1511.03682}{{\tt
  arXiv:1511.03682}}].

\bibitem{Nilles:1983ge}
H.~P. Nilles {\em Phys.~Rept.} {\bf 110} (1984) 1--162.

\bibitem{Barbieri:1987xf}
R.~Barbieri {\em Riv.~Nuovo Cim.} {\bf 11N4} (1988) 1--45.

\bibitem{Haber:1984rc}
H.~E. Haber and G.~L. Kane {\em Phys.~Rept.} {\bf 117} (1985) 75--263.

\bibitem{Gunion:1984yn}
J.~F. Gunion and H.~E. Haber {\em Nucl. Phys.} {\bf B272} (1986) 1. [Erratum:
  Nucl. Phys.B402,567(1993)].

\bibitem{Giusti:1998gz}
L.~Giusti, A.~Romanino, and A.~Strumia {\em Nucl. Phys.} {\bf B550} (1999)
  3--31, [\href{http://arxiv.org/abs/hep-ph/9811386}{{\tt hep-ph/9811386}}].

\bibitem{Cheng:2003ju}
H.-C. Cheng and I.~Low {\em JHEP} {\bf 09} (2003) 051,
  [\href{http://arxiv.org/abs/hep-ph/0308199}{{\tt hep-ph/0308199}}].

\bibitem{Harnik:2003rs}
R.~Harnik, G.~D. Kribs, D.~T. Larson, and H.~Murayama {\em Phys. Rev.} {\bf
  D70} (2004) 015002, [\href{http://arxiv.org/abs/hep-ph/0311349}{{\tt
  hep-ph/0311349}}].

\bibitem{Cheng:2004yc}
H.-C. Cheng and I.~Low {\em JHEP} {\bf 08} (2004) 061,
  [\href{http://arxiv.org/abs/hep-ph/0405243}{{\tt hep-ph/0405243}}].

\bibitem{ATLAS-SUSY}
 See: {\tt
  https://twiki.cern.ch/twiki/bin/view/AtlasPublic/SupersymmetryPublicResults}.

\bibitem{CMS-SUSY}
 See: {\tt https://twiki.cern.ch/twiki/bin/view/CMSPublic/PhysicsResultsSUS}.

\bibitem{Heinemeyer:2004gx}
S.~Heinemeyer, W.~Hollik, and G.~Weiglein {\em Phys.~Rept.} {\bf 425} (2006)
  265--368, [\href{http://arxiv.org/abs/hep-ph/0412214}{{\tt hep-ph/0412214}}].

\bibitem{Carena:2013qia}
M.~Carena, S.~Heinemeyer, O.~St{\aa}l, C.~Wagner, and G.~Weiglein {\em
  Eur.Phys.J.} {\bf C73} (2013) 2552,
  [\href{http://arxiv.org/abs/1302.7033}{{\tt arXiv:1302.7033}}].

\bibitem{Profumo:2016zxo}
S.~Profumo and T.~Stefaniak {\em Phys. Rev.} {\bf D94} (2016) 095020,
  [\href{http://arxiv.org/abs/1608.06945}{{\tt arXiv:1608.06945}}].

\bibitem{Bechtle:2016kui}
P.~Bechtle, H.~E. Haber, S.~Heinemeyer, O.~Stål, T.~Stefaniak, G.~Weiglein,
  and L.~Zeune {\em Eur. Phys. J.} {\bf C77} (2017) 67,
  [\href{http://arxiv.org/abs/1608.00638}{{\tt arXiv:1608.00638}}].

\bibitem{Haber:1996fp}
H.~E. Haber, R.~Hempfling, and A.~H. Hoang {\em Z. Phys.} {\bf C75} (1997) 539,
  [\href{http://arxiv.org/abs/hep-ph/9609331}{{\tt hep-ph/9609331}}].

\bibitem{Carena:2000dp}
M.~Carena, H.~E. Haber, S.~Heinemeyer, W.~Hollik, C.~E.~M. Wagner, and
  G.~Weiglein {\em Nucl. Phys.} {\bf B580} (2000) 29--57,
  [\href{http://arxiv.org/abs/hep-ph/0001002}{{\tt hep-ph/0001002}}].

\bibitem{Georgi:1978ri}
H.~Georgi and D.~V. Nanopoulos {\em Phys. Lett.} {\bf B82} (1979) 95--96.

\bibitem{Branco:1999fs}
G.~C. Branco, L.~Lavoura, and J.~P. Silva, {\em {CP Violation}}.
\newblock Oxford University Press, Oxford, UK, 1999.

\bibitem{Davidson:2005cw}
S.~Davidson and H.~E. Haber {\em Phys.~Rev.} {\bf D72} (2005) 035004,
  [\href{http://arxiv.org/abs/hep-ph/0504050}{{\tt hep-ph/0504050}}]. {\it
  Erratum:} Phys.~Rev. {\bf D72} 099902 (2005).

\bibitem{Haber:1990aw}
H.~E. Haber and R.~Hempfling {\em Phys.~Rev.~Lett.} {\bf 66} (1991) 1815--1818.

\bibitem{Okada:1990vk}
Y.~Okada, M.~Yamaguchi, and T.~Yanagida {\em Prog.~Theor.~Phys.} {\bf 85}
  (1991) 1--6.

\bibitem{Ellis:1990nz}
J.~R. Ellis, G.~Ridolfi, and F.~Zwirner {\em Phys.~Lett.} {\bf B257} (1991)
  83--91.

\bibitem{Djouadi:2005gj}
A.~Djouadi {\em Phys.~Rept.} {\bf 459} (2008) 1--241,
  [\href{http://arxiv.org/abs/hep-ph/0503173}{{\tt hep-ph/0503173}}].

\bibitem{Heinemeyer:2004ms}
S.~Heinemeyer {\em Int.~J.~Mod.~Phys.} {\bf A21} (2006) 2659--2772,
  [\href{http://arxiv.org/abs/hep-ph/0407244}{{\tt hep-ph/0407244}}].

\bibitem{Draper:2016pys}
P.~Draper and H.~Rzehak {\em Phys. Rept.} {\bf 619} (2016) 1--24,
  [\href{http://arxiv.org/abs/1601.01890}{{\tt arXiv:1601.01890}}].

\bibitem{Heinemeyer:2011aa}
S.~Heinemeyer, O.~St{\aa}l, and G.~Weiglein {\em Phys. Lett.} {\bf B710} (2014)
  201--206, [\href{http://arxiv.org/abs/1112.3026}{{\tt arXiv:1112.3026}}].

\bibitem{Haber:1993an}
H.~E. Haber and R.~Hempfling {\em Phys. Rev.} {\bf D48} (1993) 4280--4309,
  [\href{http://arxiv.org/abs/hep-ph/9307201}{{\tt hep-ph/9307201}}].

\bibitem{Dobrescu:2010mk}
B.~A. Dobrescu and P.~J. Fox {\em Eur. Phys. J.} {\bf C70} (2010) 263,
  [\href{http://arxiv.org/abs/1001.3147}{{\tt arXiv:1001.3147}}].

\bibitem{Frere:1983ag}
J.~M. Frere, D.~R.~T. Jones, and S.~Raby {\em Nucl. Phys.} {\bf B222} (1983)
  11--19.

\bibitem{Claudson:1983et}
M.~Claudson, L.~J. Hall, and I.~Hinchliffe {\em Nucl. Phys.} {\bf B228} (1983)
  501--528.

\bibitem{Kounnas:1983td}
C.~Kounnas, A.~B. Lahanas, D.~V. Nanopoulos, and M.~Quiros {\em Nucl. Phys.}
  {\bf B236} (1984) 438--466.

\bibitem{Gunion:1987qv}
J.~F. Gunion, H.~E. Haber, and M.~Sher {\em Nucl. Phys.} {\bf B306} (1988)
  1--13.

\bibitem{Casas:1995pd}
J.~A. Casas, A.~Lleyda, and C.~Munoz {\em Nucl. Phys.} {\bf B471} (1996) 3--58,
  [\href{http://arxiv.org/abs/hep-ph/9507294}{{\tt hep-ph/9507294}}].

\bibitem{Langacker:1994bc}
P.~Langacker and N.~Polonsky {\em Phys. Rev.} {\bf D50} (1994) 2199--2217,
  [\href{http://arxiv.org/abs/hep-ph/9403306}{{\tt hep-ph/9403306}}].

\bibitem{Strumia:1996pr}
A.~Strumia {\em Nucl. Phys.} {\bf B482} (1996) 24--38,
  [\href{http://arxiv.org/abs/hep-ph/9604417}{{\tt hep-ph/9604417}}].

\bibitem{Chowdhury:2013dka}
D.~Chowdhury, R.~M. Godbole, K.~A. Mohan, and S.~K. Vempati {\em JHEP} {\bf 02}
  (2014) 110, [\href{http://arxiv.org/abs/1310.1932}{{\tt arXiv:1310.1932}}].

\bibitem{Bagnaschi:2015pwa}
E.~Bagnaschi, F.~Br{\"u}mmer, W.~Buchm{\"u}ller, A.~Voigt, and G.~Weiglein {\em
  JHEP} {\bf 03} (2016) 158, [\href{http://arxiv.org/abs/1512.07761}{{\tt
  arXiv:1512.07761}}].

\bibitem{Hollik:2016dcm}
W.~G. Hollik {\em JHEP} {\bf 08} (2016) 126,
  [\href{http://arxiv.org/abs/1606.08356}{{\tt arXiv:1606.08356}}].

\bibitem{Arbey:2017gmh}
A.~Arbey, F.~Mahmoudi, O.~St{\aa}l, and T.~Stefaniak
  \href{http://arxiv.org/abs/1706.07414}{{\tt arXiv:1706.07414}}.

\bibitem{Marquard:2015qpa}
P.~Marquard, A.~V. Smirnov, V.~A. Smirnov, and M.~Steinhauser {\em Phys. Rev.
  Lett.} {\bf 114} (2015) 142002, [\href{http://arxiv.org/abs/1502.01030}{{\tt
  arXiv:1502.01030}}].

\bibitem{Haber:2007dj}
H.~E. Haber and J.~D. Mason {\em Phys. Rev.} {\bf D77} (2008) 115011,
  [\href{http://arxiv.org/abs/0711.2890}{{\tt arXiv:0711.2890}}].

\bibitem{Carena:2002es}
M.~Carena and H.~E. Haber {\em Prog. Part. Nucl. Phys.} {\bf 50} (2003)
  63--152, [\href{http://arxiv.org/abs/hep-ph/0208209}{{\tt hep-ph/0208209}}].

\bibitem{Dine:2007xi}
M.~Dine, N.~Seiberg, and S.~Thomas {\em Phys. Rev.} {\bf D76} (2007) 095004,
  [\href{http://arxiv.org/abs/0707.0005}{{\tt arXiv:0707.0005}}].

\bibitem{Bechtle:2015pma}
P.~Bechtle, S.~Heinemeyer, O.~St{\aa}l, T.~Stefaniak, and G.~Weiglein {\em Eur.
  Phys. J.} {\bf C75} (2015) 421, [\href{http://arxiv.org/abs/1507.06706}{{\tt
  arXiv:1507.06706}}].

\bibitem{CMSMSSMHiggs}
{\bf CMS} Collaboration \href{http://arxiv.org/abs/CMS-PAS-HIG-16-037}{{\tt
  CMS-PAS-HIG-16-037}}.

\bibitem{ATLASMSSMHiggs}
{\bf ATLAS} Collaboration \href{http://arxiv.org/abs/ATLAS-CONF-2017-050}{{\tt
  ATLAS-CONF-2017-050}}.

\bibitem{Carena:2015moc}
M.~Carena, H.~E. Haber, I.~Low, N.~R. Shah, and C.~E.~M. Wagner {\em Phys.
  Rev.} {\bf D93} (2016) 035013, [\href{http://arxiv.org/abs/1510.09137}{{\tt
  arXiv:1510.09137}}].

\bibitem{deFlorian:2016spz}
{\bf LHC Higgs Cross Section Working Group} Collaboration, D.~de~Florian et~al.
  \href{http://arxiv.org/abs/1610.07922}{{\tt arXiv:1610.07922}}.

\bibitem{Williams:2011bu}
K.~E. Williams, H.~Rzehak, and G.~Weiglein {\em Eur.~Phys.~J.} {\bf C71} (2011)
  1669, [\href{http://arxiv.org/abs/1103.1335}{{\tt arXiv:1103.1335}}].

\end{thebibliography}\endgroup
\bibliographystyle{JHEP}

\end{document}